\documentclass[11pt]{article}
\usepackage{graphicx} 
\usepackage{import}
\usepackage{basic}
\newtheorem*{mainquestion}{Main Question}
\newcounter{introthm}

\usepackage{pict2e}
    \usepackage{color-edits}
    \usepackage[normalem]{ulem}

    \addauthor{MB}{red}

    \addauthor{AS}{orange}

    \addauthor{NN}{blue}

\usepackage{pgfplots}
\pgfplotsset{compat=1.18}
\usetikzlibrary{tikzmark,calc}

\newif\iflocalnotes
\InputIfFileExists{local-settings}{}{}

\title{The Power and Limits of Money Burning in Bilateral Trade\thanks{This research was supported by the Israel Science Foundation (grant numbers 301/24 and 505/23). Moshe Babaioff's research is also supported by a Golda Meir Fellowship.}}
\author{Moshe Babaioff \thanks{Hebrew University of Jerusalem. Email: \texttt{moshe.babaioff@mail.huji.ac.il}} \and Noam Nisan \thanks{Hebrew University of Jerusalem. Email: \texttt{noam@cs.huji.ac.il}} \and
 Ariel Shaulker \thanks{Stanford University. Email: \texttt{arielsha@stanford.edu}}}

\begin{document}

\maketitle

\begin{abstract}

We study the power of money burning in bilateral trade. In this problem, a seller owns an indivisible item, a buyer would like to acquire it, and each of these two traders has a private value for the item and the values are sampled from independent distributions.
We ask whether the traders can benefit from a mechanism that burns money, that is, one that charges the buyer more than it pays the seller, so that neither trader enjoys the difference (money burnt is a loss for the traders). \emph{Is it possible that the traders will ever be better off if the trade mechanism can burn money?}

For Bayesian incentive-compatible mechanisms and for \emph{deterministic} dominant-strategy incentive-compatible (DSIC) mechanisms, it is known that money burning is always a waste and 
can only hurt the traders.
Maybe surprisingly, we show that once a DSIC mechanism can use randomization, money burning can significantly benefit the traders, and the gain of each of them can be arbitrarily large. In contrast, we show that the power of DSIC 
randomized mechanisms that burn money is still very limited in the sense that they cannot guarantee any constant fraction of the first-best gains-from-trade.

\end{abstract}

\thispagestyle{empty}
\newpage
\pagenumbering{arabic}

\section{Introduction}

We study the power of money burning to improve efficiency in the classical
bilateral-trade problem.  In the bilateral-trade problem, a seller owns an
indivisible item, and a buyer would like to acquire it.  
The 
seller and buyer
privately know their respective non-negative values, $\vs$ and $\vb$, for the
item; these values are drawn from independent distributions.
A
mechanism elicits reports of these values, 
{decides the probability with which} the item 
 is traded,
and determines the price that the buyer pays the seller when there is trade.

Traders are assumed to act strategically, aiming to maximize their expected utilities.
Because the traders are strategic, analyzing the outcome of a mechanism
requires understanding the traders' behavior.  
The literature
considers two approaches.  In the first, the traders are assumed to play a Bayes--Nash
equilibrium, and by the revelation principle it is without loss to consider
direct mechanisms in which truthful reporting is a Bayes--Nash equilibrium (\emph{Bayesian incentive-compatible}{, or BIC,} mechanisms)~\cite{Myerson-Satterthwaite, BothWorlds,Bru, MM23, GDTCA}. The second considers
\emph{dominant-strategy incentive-compatible} mechanisms \cite{BD14, italians, BD21, KPV, CW-STOC, CSTOC, improved-fp}, which impose a
stronger requirement than Bayes--Nash equilibrium.  We start with a discussion of Bayesian
incentive-compatible mechanisms, and move to dominant-strategy
incentive-compatible mechanisms later in this introduction.
As usual
in mechanism design, traders' participation is voluntary, so a  mechanism must be individually rational (IR), ensuring traders are better off participating. As all mechanisms we consider are required to be IR, we sometimes omit stating this requirement explicitly.

If the traders' values were public, efficiency (or social-welfare maximization) would require trade whenever
$\vb>\vs$.  This efficient allocation rule is called the \emph{first-best}.
Myerson and Satterthwaite~\cite{Myerson-Satterthwaite} showed that, under mild
conditions, no BIC,
interim 
IR 
mechanism implements the first-best.\footnote{Note that in the problem as described so far, the buyer pays the seller, so no outside subsidy is allowed. With an outside subsidy, the impossibility result of~\cite{Myerson-Satterthwaite} no longer holds, and the VCG mechanism implements the first-best.} 
 
The first-best cannot be implemented, but can truthful mechanisms ``come close'' to it? To measure that, the literature considers two benchmarks:
social welfare and gains-from-trade. For a valuation profile, the \emph{social welfare} is the value of the trader that ends up with the item: the seller's value if he keeps the item, and
the buyer's value if the item is traded.  
The \emph{gains-from-trade} objective measures the
\emph{improvement} in social welfare relative to the no-trade outcome (in which the seller keeps the item): trade creates gains of $\vb-\vs$, while no trade creates no gains.  The social welfare and gains-from-trade of a truthful mechanism are
the expectations of these quantities over the traders' values and the mechanism's
internal randomness. Given the Myerson and Satterthwaite~\cite{Myerson-Satterthwaite} impossibility result, the literature focuses on approximating these benchmarks: maximizing the mechanism's guaranteed fraction of the corresponding expectations
under the first-best.  

Approximating gains-from-trade is harder than approximating social welfare: any
approximation to gains-from-trade implies the same approximation to
social welfare, while any impossibility for social welfare
implies the same impossibility for gains-from-trade.  For gains-from-trade,
BIC 
 mechanisms obtain a
constant approximation to the first-best~\cite{GDTCA,IGFT}.
These results, as well as the  line of work in computer science on the
approximation guarantees of bilateral-trade mechanisms{~\cite{BD14,BM,italians,BD21,KPV,CW-STOC,CSTOC,improved-fp,JGC26,LSDKLYZ26},
which we discuss later in this section}, consider mechanisms in
which the buyer's entire
payment is transferred to the seller.

The bilateral-trade problem as presented so far, and as defined by Myerson and Satterthwaite~\cite{Myerson-Satterthwaite}, inherently assumes this property: the amount of money the buyer pays is exactly equal to the amount of money the seller receives. This property is called \emph{strong budget balance} (SBB), and in the literature it is also often referred to simply as \emph{budget balance}. However, nothing in the problem itself requires this property. 
The requirement which captures the inability of the mechanism to finance the trade is the weaker condition of ``no deficit'': 
the amount of money the buyer pays must be at least the amount of money the seller receives, so that no outside subsidy is needed. This weaker requirement is called \emph{weak budget balance} (WBB), or \emph{no deficit}.

Under WBB, the buyer's payment may exceed the amount the seller
receives.  Neither trader enjoys this difference: the buyer pays it, but the
seller does not receive it.  Hence, when only the traders are counted, as is
common in economics, this difference is lost, and we view it as money burned by
the mechanism~\cite{DK15,MM23}. The sum of the buyer's and seller's utilities is called
the \emph{traders' utility}; in the literature on bilateral trade, this
objective is also called \emph{aggregate welfare} or \emph{total
welfare}~\cite{DK15,MM23}.
Its improvement relative to no trade is called the \emph{traders' surplus}.  Thus,
traders' utility is social welfare minus the money burned, and traders' surplus
is gains-from-trade minus the money burned.  
One can alternatively consider the benchmarks in which the mechanism
itself is considered part of the society (so the money it collects is counted). This gives rise to the social-welfare and gains-from-trade benchmarks, defined above.
Under SBB, no money is burned, so the two pairs of benchmarks coincide.

When the goal is the first-best itself, SBB is without loss: the first-best leaves no room for waste, since any part of the buyer's payment that the seller does not receive is lost to both traders. Thus, a mechanism that attains the first-best in terms of the traders' utility must be SBB. 
This argument no longer applies when aiming to \emph{approximate} the  first-best, rather than getting it exactly. 
{Burning money seems inherently wasteful, as the money burnt is not given to the traders. Can the traders ever be better off when the mechanism can burn money? If so, how much better can the approximation be?}

Is the restriction to SBB then benign?  
For BIC and interim IR
mechanisms with independent values, it is: a  
reduction~\cite{BN09,Bru,MM23} converts any
ex-ante WBB mechanism into an ex-post SBB
one without reducing the traders' expected utility.\footnote{\label{fn:bic-pointwise}For BIC mechanisms, the correct notions of WBB and SBB are pointwise, that is, for every pair of values. Note that the ex-ante notions, in expectation over both values, trivially coincide: the expected extra money can be returned to the traders as a fixed payment.} {The reduction keeps the allocation rule and only changes the payments, so money burning gives BIC
mechanisms no new allocation rules.
}  
Thus, for 
{BIC mechanisms, the picture} is
essentially complete: money burning {is always a waste}. Additionally, strongly
budget-balanced 
{BIC mechanisms already} provide a constant approximation to the
first-best gains-from-trade~\cite{GDTCA,IGFT}, and therefore to the
first-best social welfare as well.

In this paper we focus on the power of money burning for dominant-strategy incentive-compatible (DSIC) bilateral trade mechanisms.  In such a mechanism, truthful reporting maximizes each trader's expected utility,
regardless of the other trader's report {(the expectation is only over the mechanism's internal randomness)}.  This makes these mechanisms more
robust, but this robustness comes at a cost.
No \emph{deterministic} SBB DSIC mechanism
guarantees any constant approximation~\cite{BD14} to the first-best
gains-from-trade. It was shown that every randomized mechanism with these properties is a distribution over posted prices~\cite{HR87,CP16}%
, and for any distributions of values, some single posted price in its support performs at least as well. Hence, from an approximation point of view, randomization does not help such mechanisms, and this impossibility extends to randomized mechanisms with the same properties. For social welfare, deterministic,
{DSIC 
mechanisms do}
provide constant approximations~\cite{BD14,italians,BD21,KPV,CW-STOC,CSTOC,improved-fp,LSDKLYZ26,JGC26}. Here as well, all these mechanisms are SBB.\footnote{
{Some works on regret minimization post a higher price to the buyer than to the seller~\cite{regret1,regret2}. This helps learning across rounds, 
yet does not (and cannot) improve the outcome of each round 
(see Section~\ref{sec:related-work}).}\label{fn:intro-regret-prices}}
{For DSIC mechanisms, 
however, no}
reduction like that of~\cite{BN09,Bru,MM23} 
{for BIC mechanisms is known}, and whether money burning can
help these mechanisms has remained open.
In this work, we study the power of 
{money burning in DSIC mechanisms,}
focusing on the traders' utility and the traders' surplus (although we give
results for all four benchmarks). We focus on the following questions.

\begin{mainquestion}\phantomsection\label{question:intro:money-burning}
For dominant-strategy incentive-compatible trade mechanisms with independent values, can money burning improve the traders' surplus and the traders' utility?  If so,
to what extent? Can the improvement be so large as to approximate the first-best?
\end{mainquestion}


As discussed above, money burning is 
{useless for BIC mechanisms.} 
Moreover, for DSIC mechanisms that are restricted to be deterministic, money burning is also useless: the optimal WBB mechanism for maximizing the traders' utility can always be chosen to be SBB~\cite {DK15}.
As money burning is useless for deterministic mechanisms, it is also useless for any randomized mechanism that is a distribution over deterministic mechanisms that each individually satisfies these properties (DSIC and WBB).

This observation might suggest that money burning also cannot help randomized
mechanisms.  However, not 
every randomized 
WBB DSIC mechanism 
is necessarily
a distribution over deterministic
mechanisms that each satisfies these properties.
In algorithmic game theory,
this is the distinction between universally truthful and truthful-in-expectation.  Under SBB, the characterization
in~\cite{HR87,CP16} implies that every truthful-in-expectation mechanism can be implemented as a universally truthful one, with SBB and IR holding ex-post, so it is natural
to ask whether the same holds under WBB. If that is the case, then money burning is useless even for randomized DSIC mechanisms.

All prior work that we have discussed above seems to indicate that, for independent distributions, money burning is useless.  {Our \hyperref[question:intro:money-burning]{Main Question} asks whether this is indeed the case for DSIC (randomized) mechanisms.} 

\subsection{Our Results}

Our work answers {the \hyperref[question:intro:money-burning]{Main Question}}. First, we show that for randomized dominant-strategy incentive-compatible mechanisms, 
money burning can substantially improve the traders' outcome: 
mechanisms that burn money can outperform strongly budget-balanced mechanisms in maximizing the traders'
surplus, and their advantage can be arbitrarily large, even for each trader separately. Motivated by this gap, we also consider correlated values, and strengthen the result of~\cite{SW18} by showing an unbounded separation not only for the traders' surplus, but even
for the traders' utility. These positive results are presented in  Section~\ref{sec:intro-money-pos}. On the negative side, we present impossibility results for first-best approximations, showing that even with money burning, DSIC mechanisms are not very powerful.
Specifically, they cannot guarantee a constant approximation to the first-best traders' surplus, and, for correlated values, to the first-best traders' utility.
These two impossibility results hold even if we ignore the fact that some money is burnt by the mechanism, and count it towards the gains-from-trade or the social welfare.
These negative results are presented in  Section~\ref{sec:intro-limits}.



\subsubsection{The Advantages of Money Burning}\label{sec:intro-money-pos}

We demonstrate the power of money burning for 
{randomized DSIC mechanisms}.  We prove that the traders' surplus (the improvement in the traders'
utilities relative to no trade) of money burning mechanisms can be arbitrarily larger than that 
of any strongly budget-balanced mechanism, even for each trader separately.
\paragraph*{Notation.}
For a truthful mechanism $\mechanism$ and independent seller and buyer distributions $\distseller$ and $\distbuyer$, we write $\swelfare_\mechanism(\distseller\times\distbuyer)$ and $\sgft_\mechanism(\distseller\times\distbuyer)$ for the expected traders' utility and traders' surplus, respectively, under mechanism $\mechanism$ when the agents' values are sampled according to these distributions. 
These quantities are equal to the expected social welfare and gains-from-trade,
respectively, minus the expected money burned (the expected difference between the buyer's payment and the amount the seller receives).

\medskip
\noindent\refstepcounter{introthm}\label{thm:intro-advantages-ind}\textbf{Theorem~\theintrothm.}
For every $\varepsilon>0$, there are independent seller and buyer distributions
    $\distseller$ and $\distbuyer$ and a dominant-strategy incentive-compatible,
    individually rational, and weakly budget-balanced mechanism $\wbbmechanism$ such that,
    for every dominant-strategy incentive-compatible, individually rational, and strongly
    budget-balanced mechanism $\sbbmechanism$, it holds that
    \[
        \varepsilon\cdot\sgft_{\wbbmechanism}(\distseller\times\distbuyer)
        > \sgft_{\sbbmechanism}(\distseller\times\distbuyer).
    \]
    {Moreover, the same gap holds for each trader separately: $\varepsilon$ times the expected surplus of each trader under $\wbbmechanism$ is larger than the expected surplus of that trader under $\sbbmechanism$.}
\medskip

Note that this strong separation holds even though we 
{do not credit the WBB mechanism} for the money burnt. {The separation for each trader separately follows since our construction is symmetric between the two traders} (Remark~\ref{rem:both-traders-benefit}). 
Thus, each trader might be arbitrarily better off by money burning (the gain for one is not due to a loss for the other).
Observe that a separation for independent distributions immediately implies the same separation for correlated values.

We next move to consider the traders' utility.
An unbounded separation does not hold 
for the traders' utility
under independent values, because
fixed-price mechanisms already guarantee a constant approximation to the first-%
best 
\cite{BD14,italians,BD21,KPV,CW-STOC,CSTOC,improved-fp}.
Does such a constant approximation hold for correlated distributions, or is there a strong separation for this general case? We show that for correlated values, there is a strong separation for the traders' utility as well.

For a joint distribution $\distjoint$ over the agents' values, we write $\swelfare_\mechanism(\distjoint)$ for the expected traders' utility under mechanism $\mechanism$ when the agents' values are sampled according to $\distjoint$.

\medskip
\noindent\refstepcounter{introthm}\label{thm:intro-advantages-correlated}\textbf{Theorem~\theintrothm.}
For every $\varepsilon>0$, there is a \emph{joint distribution} $\distjoint$
    and a dominant-strategy incentive-compatible, individually rational, and weakly
    budget-balanced mechanism $\wbbmechanism$ such that, for every
    dominant-strategy incentive-compatible, individually rational, and strongly
    budget-balanced mechanism $\sbbmechanism$, it holds that
    \[
        \varepsilon\cdot\swelfare_{\wbbmechanism}(\distjoint)
        > \swelfare_{\sbbmechanism}(\distjoint).
    \]
\medskip

{Table~\ref{tab:intro-money-burning-gain} summarizes how much money burning can help the traders, for each class of mechanisms and each type of prior.}

\begin{table}[t]
\centering
\caption[How Much Can Money Burning Help the Traders?]{How Much Can Money Burning Help the Traders? ($\frac{\text{WBB}}{\text{SBB}}$)}
\label{tab:intro-money-burning-gain}
\renewcommand{\arraystretch}{1.4}
\setlength{\tabcolsep}{4pt}

\begin{tabular}{@{}lcccc@{}}
\toprule
& \multicolumn{2}{c}{\textbf{Traders' Utility}}
& \multicolumn{2}{c}{\textbf{Traders' Surplus}} \\
\cmidrule(lr){2-3} \cmidrule(lr){4-5}

\begin{tikzpicture}[baseline={([yshift=-0.5ex]current bounding box.center)}]
    \node[anchor=south west,inner sep=0pt] (mech)
        at (0,0) {\textbf{Mechanism}};
    \node[anchor=north east,inner sep=0pt] (vals)
        at (2.9,0.9) {\textbf{Values}};
    \draw[thin] (0.2,0.85) -- (2.7,0.15);
    \path (0,-0.15);
\end{tikzpicture}
& \textbf{Independent}
& \textbf{Correlated}
& \textbf{Independent}
& \textbf{Correlated} \\
\midrule


\begin{tabular}[c]{@{}l@{}}\textbf{Deterministic DSIC}\\ \cite{DK15}\end{tabular}
& $1$ & $1$ & $1$ & $1$ \\

\addlinespace[3pt]

\textbf{General DSIC}
& \makebox[3.4em][l]{$>1$\rlap{\ \tikzmarknode{gUI}{(new)}}}
& $\infty$ (new)
& \tikzmarknode{gSI}{$\infty$ (new)}
& \tikzmarknode{gSC}{$\infty$ (new)} \\[5pt]

& \tikzmarknode{gUB}{\makebox[3.4em][l]{$\leq 1.355$}}
& & & \\[-2pt]

& {\footnotesize\cite{JGC26,LSDKLYZ26}}
& & & \\

\bottomrule
\end{tabular}
\begin{tikzpicture}[remember picture,overlay,>=stealth]

    \draw[dashed,thin] ([xshift=-2pt,yshift=7pt]gUB.north west) -- ([xshift=2pt,yshift=7pt]gUB.north east);

    \draw[->,thin]
        ([xshift=4pt]gSI.east)
        --
        ([xshift=-4pt]gSC.west);

    \draw[->,thin,shorten <=2pt,shorten >=2pt]
        ([xshift=-3pt,yshift=1pt]gSI.north west)
        to[out=165,in=15]
        ([xshift=3pt,yshift=1pt]gUI.north east);

\end{tikzpicture}

\caption*{Table~\thetable: Each entry is the largest possible ratio between the best
WBB mechanism and the best SBB
mechanism.
Entries marked (new) are our results, and arrows point from
a result to an entry that follows from it.
For deterministic DSIC mechanisms, the entries for correlated values follow from Proposition~\ref{prop:prelim:dk-dominance}. 
For general DSIC mechanisms, the unbounded gap in traders' surplus
for correlated values follows from the unbounded gap for independent values.
The gap in traders' surplus for independent values also implies
a gap in traders' utility, since traders' utility is the
traders' surplus plus the seller's expected value, which is
the same under both mechanisms (see Appendix~\ref{app:ts-to-tu} for more details).
This gap can be much smaller, but is strictly larger than one.
The upper bound of $1.355$ holds since a fixed price obtains
approximately a {$0.738024 \approx \frac{1}{1.355}$} fraction of the first-best social welfare. Finding the exact ratio is an open problem.
}
\end{table}

The intuition behind this immense gap lies in the structural rigidity of SBB mechanisms. Every SBB DSIC mechanism is a distribution over posted prices. A posted price requires both agents to face the 
{same price, which depends on neither report, limiting} the mechanism's ability to trade the item at disjoint valuation profiles. WBB decouples the buyer's and seller's prices, allowing the mechanism to trade at multiple disjoint profiles with high probability (at the cost of burning money at intermediate profiles). {Randomization is necessary for this: if a deterministic mechanism trades at two disjoint profiles, it also trades at the profile with the lower seller value and the higher buyer value, and there the buyer pays less than the seller receives, so WBB fails (see Proposition~\ref{prop:prelim:dk-dominance}). Randomized mechanisms can avoid this deficit, at the cost of trading with probability less than 1 at the two disjoint profiles, as we explain in Section~\ref{sec:intro-technique}.} We show how to 
{chain such profiles to generate unbounded relative gains} in expected traders' surplus. 

\subsubsection{The Limitations of Weak Budget Balance Mechanisms}
\label{sec:intro-limits}

We have seen that DSIC mechanisms that burn money can be unboundedly more powerful than strongly budget-balanced mechanisms. {Moreover, on the distributions used in Theorem~\ref{thm:intro-advantages-ind} and Theorem~\ref{thm:intro-advantages-correlated}, our WBB mechanism obtains a constant fraction of the first-best.} This raises the natural question of whether 
money burning is powerful enough to 
%
overcome  the known barriers to approximating the first-best 
{with SBB DSIC 
 mechanisms.} Our next results show a sharp limit: even \emph{weakly} budget-balanced DSIC mechanisms cannot guarantee a constant approximation to the first-best traders' surplus (for independent priors) or the first-best traders' utility (for correlated priors). These impossibilities hold even if we ignore the fact that some money is burnt by the mechanism, and count it towards the gains-from-trade or the social welfare.


\paragraph*{Notation.}
For a truthful mechanism $\mechanism$ and a prior $\distjoint$ over the agents' values, we write $\welfare_\mechanism(\distjoint)$ and $\gft_\mechanism(\distjoint)$ for the expected social welfare and gains-from-trade, respectively, under mechanism $\mechanism$ when the agents' values are sampled according to $\distjoint$.  Unlike the traders’ utility and surplus, these quantities do not subtract the money
burned by the mechanism, and thus depend only on the allocation rule.

We write $\fbwelfare(\distjoint)$ and $\fbgft(\distjoint)$ for the expected social welfare and gains-from-trade of the {first-best}, which trades whenever $\vb>\vs$.

\medskip
\noindent\refstepcounter{introthm}\label{thm:intro-limitations}\textbf{Theorem~\theintrothm\ (limitations of weak budget balance).}
The following statements hold:
\begin{enumerate}
    \item For every $\varepsilon>0$, there are independent seller and buyer distributions
    $\distseller$ and $\distbuyer$ such that every dominant-strategy
    incentive-compatible, individually rational, and weakly budget-balanced
    mechanism $\wbbmechanism$ satisfies
    \[ 
    \sgft_{\wbbmechanism}(\distseller\times\distbuyer)
         \leq \gft_{\wbbmechanism}(\distseller\times\distbuyer)
        < \varepsilon\cdot\fbgft(\distseller\times\distbuyer).
    \]
    \item For every $\varepsilon>0$, there is a joint distribution $\distjoint$
    such that every dominant-strategy incentive-compatible, individually
    rational, and weakly budget-balanced mechanism $\wbbmechanism$ satisfies
    \[
         \swelfare_{\wbbmechanism}(\distjoint)
        \leq \welfare_{\wbbmechanism}(\distjoint)
        < \varepsilon\cdot\fbwelfare(\distjoint). \] 
    
\end{enumerate}
\medskip


For independent values, no 
SBB DSIC mechanism guarantees
a constant fraction of the first-best gains-from-trade~\cite{BD14}.  The first
statement extends this impossibility to 
WBB DSIC mechanisms.

For correlated values, no SBB mechanism in which both traders have a dominant
strategy guarantees a constant approximation to first-best social
welfare~\cite{DS24}.  The second statement shows that
allowing 
the 
{DSIC mechanism to be only} WBB
does not overcome this barrier.
{Table~\ref{tab:intro-approximation} places these results in the context of the literature.}


\begin{table}[t]
\centering
\caption{Can the first-best be approximated to within a constant factor?}
\label{tab:intro-approximation}
\renewcommand{\arraystretch}{1.4}
\setlength{\tabcolsep}{6pt}

\begin{tabular}{@{}lcccc@{}}
\toprule
& \multicolumn{2}{c}{\textbf{Social Welfare}}
& \multicolumn{2}{c}{\textbf{Gains-from-Trade}} \\
\cmidrule(lr){2-3} \cmidrule(lr){4-5}

\begin{tikzpicture}[baseline={([yshift=-0.5ex]current bounding box.center)}]
    \node[anchor=south west,inner sep=0pt] (mech)
        at (0,0) {\textbf{Mechanism}};
    \node[anchor=north east,inner sep=0pt] (vals)
        at (2.9,0.9) {\textbf{Values}};
    \draw[thin] (0.2,0.85) -- (2.7,0.15);
    \path (0,-0.15);
\end{tikzpicture}
& \textbf{Independent}
& \textbf{Correlated}
& \textbf{Independent}
& \textbf{Correlated} \\
\midrule

\textbf{WBB BIC}
& \tikzmarknode{aBWI}{Yes} & \tikzmarknode{aBWC}{Yes} & \tikzmarknode{aBWG}{Yes} & ? \\[22pt]

\textbf{SBB BIC}
& \tikzmarknode{aBSI}{Yes} & \tikzmarknode{aBSC}{Yes~\cite{DS24}} & \tikzmarknode{aBSG}{Yes~\cite{GDTCA}} & No~\cite{RW-private} \\

\textbf{WBB DSIC}
& \tikzmarknode{aWSI}{Yes}
& No (new)
& \tikzmarknode{aWGI}{No (new)}
& \tikzmarknode{aWGC}{No} \\

\textbf{SBB DSIC}
& \tikzmarknode{aSSI}{Yes~\cite{BD14}}
& No~\cite{DS24}
& \tikzmarknode{aSGI}{No~\cite{BD14}}
& \tikzmarknode{aSGC}{No} \\

\bottomrule
\end{tabular}

\begin{tikzpicture}[remember picture,overlay,>=stealth]

    \draw[->,thin] ([yshift=1pt]aSSI.north) -- ([yshift=-1pt]aWSI.south);
    \draw[<->,thin,shorten <=7pt,shorten >=7pt] ([yshift=1pt]aBSI.north) -- node[right,inner sep=2pt,font=\footnotesize,align=left] {$\star$} ([yshift=-1pt]aBWI.south);
    \draw[->,thin,shorten <=7pt,shorten >=7pt] ([yshift=1pt]aBSC.north) -- ([yshift=-1pt]aBWC.south);
    \draw[<->,thin,shorten <=7pt,shorten >=7pt] ([yshift=1pt]aBSG.north) -- node[right,inner sep=2pt,font=\footnotesize,align=left] {$\star$} ([yshift=-1pt]aBWG.south);

    \draw[->,thin] ([xshift=-1pt]aSSI.west) to[out=150,in=210] ([xshift=-1pt]aBSI.west);

    \draw[->,thin] ([xshift=4pt,yshift=0.5ex]aWGI.base east) -- ([xshift=-4pt,yshift=0.5ex]aWGC.base west);
    \draw[->,thin] ([xshift=4pt,yshift=0.5ex]aSGI.base east) -- ([xshift=-4pt,yshift=0.5ex]aSGC.base west);

\end{tikzpicture}

\caption*{Table~\thetable: Each entry says whether the mechanisms in that row can guarantee a constant fraction of the first-best.
Entries marked (new) are our results, and arrows point from a result to an entry that follows from it.
Every SBB mechanism is also WBB, and every DSIC mechanism is also BIC, so the positive result for SBB DSIC mechanisms with independent values also holds for WBB DSIC and BIC mechanisms.
For BIC mechanisms, the IR requirement is of  interim IR, and WBB and SBB are pointwise (see Footnote~\ref{fn:bic-pointwise}). A two-sided arrow means that each of the two entries implies the other, and $\star$ means that this is due to the reduction of~\cite{BN09,Bru,MM23}.
An impossibility for independent values also holds for correlated values.
For the positive results with independent values, we cite the first constant approximation, later works improved the constants (see Section~\ref{sec:related-work}).
In this work our focus is on DSIC mechanisms.  For BIC mechanisms, SBB mechanisms already guarantee a constant fraction of the first-best in all entries except for the gains-from-trade with correlated values.  Whether money burning can help to approximate the first-best in this case is an interesting open question.
}
\end{table}

These impossibilities clarify the scope of our positive results.  Money burning DSIC mechanisms can be arbitrarily more powerful
{than SBB DSIC mechanisms}, but allowing the mechanism to be only WBB is not enough to 
overcome the fundamental gap between DSIC
mechanisms and the
first-best, even if we ignore the fact that some money is burnt by the mechanism, and count it towards the social welfare or the gains-from-trade.


\subsection{Our Technique: The Advantages of Money Burning}\label{sec:intro-technique}


We now describe the main ideas behind Theorem~\ref{thm:intro-advantages-ind} and Theorem~\ref{thm:intro-advantages-correlated}, 
which show that DSIC mechanisms that are weakly
budget-balanced can obtain unboundedly better outcomes
than strongly
budget-balanced ones: arbitrarily more traders' surplus when the values are independent,
and arbitrarily more traders' utility when the values may be correlated.  We first
identify a limitation of SBB DSIC
mechanisms, {explain why randomization is needed to overcome it,} and then
describe a WBB DSIC
mechanism that overcomes it.  We then
explain why this mechanism also leaves the traders a large surplus, and finally
describe the distributions on which the separations hold.  The full details
are in Section~\ref{sec:separation-results}.

\paragraph{A limitation of strongly budget-balanced mechanisms.}
{It was shown that every} 
SBB DSIC mechanism is a distribution over posted
prices~\cite{HR87,CP16}: at a profile $(\vs,\vb)$, it trades only if its random
price falls in $[\vs,\vb]$.  Hence, if $\price$ is its random price and
$\sbballocation$ is its allocation rule, then for any profiles
$(\vs^{(1)},\vb^{(1)})$, \ldots, $(\vs^{(m)},\vb^{(m)})$ whose intervals
$[\vs^{(j)},\vb^{(j)}]$ are pairwise disjoint,
\begin{align*}
    \sum_{j=1}^{m}\sbballocation(\vs^{(j)},\vb^{(j)})
    \le\sum_{j=1}^{m}\Pr\left[\price\in[\vs^{(j)},\vb^{(j)}]\right]
    \le 1.
\end{align*}
Thus, if almost all of the first-best gains-from-trade comes from these $m$ profiles and each profile
contributes the same amount, a strongly
budget-balanced mechanism obtains at most about a $1/m$ fraction of the first-best
gains-from-trade.  
{Our WBB mechanism trades with}
probability more than $1/2$ at many such profiles simultaneously, and
therefore obtains at least about half of the first-best gains-from-trade on such
distributions.

\paragraph{Why randomization is needed.}\phantomsection\label{par:intro-why-randomization}
WBB lets the buyer's price and the seller's price differ, and each can depend on the other trader's report. The question is whether this lets a mechanism trade with substantial probability at many profiles whose intervals are disjoint. For deterministic mechanisms it does not (Proposition~\ref{prop:prelim:dk-dominance}, essentially~\cite{DK15}), and the following example shows why. Take two agents' profiles $(s_1,b_1)$ and $(s_2,b_2)$ with $b_1<s_2$, so no posted price trades at both, and suppose 
a deterministic WBB DSIC mechanism trades at both. By monotonicity, it also trades at $(s_1,b_2)$. There, the buyer faces the same price as at $(s_1,b_1)$, since the seller's report is the same, so this price is at most $b_1$. Similarly, the seller faces the same price as at $(s_2,b_2)$, so it is at least $s_2$. Since $b_1<s_2$, the buyer pays less than the seller receives, which violates WBB.

Randomization resolves this. To see how, suppose the mechanism trades with probability more than $1/2$ at both $(s_1,b_1)$ and $(s_2,b_2)$, which no lottery over posted prices can do, since its trade probabilities at these two profiles sum to at most one. For simplicity of the argument, take both probabilities to be the same $x>1/2$.\footnote{The argument extends to different probabilities at the two profiles.} Suppose also that it trades with a higher probability $y>x$ at $(s_1,b_2)$. Think of the trade at $(s_1,b_2)$ as two parts: a part with probability $x$, which the buyer would also get at $(s_1,b_1)$ and the seller would also get at $(s_2,b_2)$, and an extra part with probability $y-x$. By Myerson's lemma, the buyer pays price $b_1$ on the first part and $b_2$ on the extra part, while the seller receives $s_2$ on the first part and $s_1$ on the extra part. The first part still runs a deficit of $(s_2-b_1)x$, but the extra part brings in $(b_2-s_1)(y-x)$, which covers it when the gap $s_2-b_1$ is small compared to $b_2-s_1$. A deterministic mechanism cannot do this, since it already trades with probability one at $(s_1,b_1)$ and $(s_2,b_2)$, so $x=1$, and since $y$ cannot exceed one, there is no extra part to cover the deficit. This is exactly the violation of WBB in the deterministic example above.

This two-profile argument is the basic building block of our construction. 
We build a grid and apply this argument repeatedly along the profiles between consecutive grid points: these profiles have disjoint intervals, every two adjacent ones play the roles of $(s_1,b_1)$ and $(s_2,b_2)$ as described above, and the profile with the seller value of the first and the buyer value of the second plays the role of $(s_1,b_2)$. Each step requires a slightly higher trade probability than the previous one, and if each increase is small, the trade probability does not exceed 1, even after many steps. We next describe the grid and the mechanism in detail. 

\paragraph{The construction.}
The mechanism is defined by a grid $0=\qpoint{0}<\cdots<\qpoint{n}=1$ and a
small $\slack>0$.  Each interior grid point $\qpoint{i}$ defines a seller
threshold $\sellerendpoint{i}=\qpoint{i}+\slack$ slightly above it and a buyer
threshold $\buyerendpoint{i}=\qpoint{i}-\slack$ slightly below it.  The
mechanism rounds the seller's report up to the nearest seller threshold and
the buyer's report down to the nearest buyer threshold, and it trades only if
the buyer's rounded index is strictly above the seller's rounded index.  We call
the difference between the two rounded indices the \emph{level} of the
profile.  The trade probability depends only on the level: at level $d\ge1$ it
is $\growth^{-(n-d)}$ for a \emph{growth factor} $\growth>1$, so moving up one
level, by increasing the buyer's rounded index or decreasing the seller's
rounded index by one, multiplies the trade probability by $\growth$, and at the
top level the trade probability is $1$ (see
Figure~\ref{fig:wbb-allocation-buyer-view}).  For the purpose of this discussion, we denote this
allocation rule by $\wbballocation$.  It is nondecreasing in the buyer's value and
nonincreasing in the seller's value, so by Myerson's lemma there are payments
that make the mechanism 
DSIC.  For a large enough growth factor, the mechanism is also
WBB, and we fix such a $\growth$.

The smallest positive trade probability is at level~$1$, and it is
$\growth^{-(n-1)}$.  The growth factor depends on the grid, and when we apply
the construction, we carefully choose the grid so that $\growth^{n-1}<2$ 
({Definition~\ref{def:separation-grid}} gives a sufficient
condition), and hence every positive trade probability is more than $1/2$.  In
particular, consider the profiles
\begin{align*}
    (\sellerendpoint{0},\buyerendpoint{1}),\,
    (\sellerendpoint{1},\buyerendpoint{2}),\,\ldots,\,
    (\sellerendpoint{n-1},\buyerendpoint{n}).
\end{align*}
At the profile $(\sellerendpoint{i-1},\buyerendpoint{i})$, the seller's
rounded index is $i-1$ and the buyer's rounded index is $i$, so all these
profiles are at level~$1$, and for a small enough $\slack$, their intervals
$[\sellerendpoint{i-1},\buyerendpoint{i}]$ are pairwise disjoint.  Hence, the
trade probabilities of 
{every SBB DSIC mechanism at these}
profiles sum to at most one, while our WBB DSIC mechanism trades at each of them with
probability more than $1/2$.

\paragraph{From trade probability to traders' surplus.}
If we only cared about the gains from trade, which do
not depend on the payments, we could stop here: as explained above, for
such distributions, the difference in the trade probabilities
gives an unbounded separation between the gains from trade of
WBB and SBB DSIC mechanisms.  
However, at the profile
$(\sellerendpoint{i-1},\buyerendpoint{i})$, the payments leave the traders
with zero surplus.  The payments of Myerson's lemma charge each trader the
values at which his or her trade probability jumps, and at this profile these
are the thresholds themselves: the buyer pays
$\buyerendpoint{i}\cdot\wbballocation(\sellerendpoint{i-1},\buyerendpoint{i})$
and the seller receives
$\sellerendpoint{i-1}\cdot\wbballocation(\sellerendpoint{i-1},\buyerendpoint{i})$,
so the entire gain from trade is burned. 
To obtain a separation for the traders' surplus, we need a more nuanced argument, which uses how the construction
relates the trade probabilities at different levels.

By Myerson's lemma, WBB at a profile $(\vs,\vb)$ with
$\wbballocation(\vs,\vb)>0$ is equivalent to
\begin{align*}
    \int_{\vs}^{\vb}\wbballocation(\vs,z)\d z
    +\int_{\vs}^{\vb}\wbballocation(z,\vb)\d z
    \le
    (\vb-\vs)\,\wbballocation(\vs,\vb).
\end{align*}
The left-hand side is the traders' surplus: the first integral is the buyer's
utility and the second is the seller's surplus.  The right-hand side is the
expected gain from trade at this profile.  Thus, WBB says that the traders'
surplus is at most this gain, and the difference is the
money burned.  Apart from the profiles whose values are rounded to the same
thresholds as $(\vs,\vb)$, both integrals only involve profiles at lower levels,
or profiles with zero trade probability (see
Figure~\ref{fig:wbb-level-dependence}).  We therefore construct
the allocation rule level by level, from the bottom up.  At the level-$1$
profiles $(\sellerendpoint{i-1},\buyerendpoint{i})$, both integrals are zero,
so the trade probabilities at level~$1$ can be chosen freely, and the traders'
surplus is zero.  Now consider the level-$2$ profile
$(\sellerendpoint{i-1},\buyerendpoint{i+1})$.
{It plays the role of $(s_1,b_2)$ in the argument above (see the paragraph ``\hyperref[par:intro-why-randomization]{Why randomization is needed}''), with the level-$1$ profiles $(\sellerendpoint{i-1},\buyerendpoint{i})$ and $(\sellerendpoint{i},\buyerendpoint{i+1})$ in the roles of $(s_1,b_1)$ and $(s_2,b_2)$.} 
  The buyer's integral covers the
trade probability at level~$1$ for buyer values between $\buyerendpoint{i}$
and $\buyerendpoint{i+1}$, and the seller's integral covers the trade
probability at level~$1$ for seller values between $\sellerendpoint{i-1}$ and
$\sellerendpoint{i}$.  These two ranges together are slightly longer than
$\buyerendpoint{i+1}-\sellerendpoint{i-1}$, so the trade probability at
level~$2$ must be higher than at level~$1$, which is where the growth factor
comes from.  On the other hand, since the trade probability at level~$1$ is
more than $1/2$, the traders' surplus at this profile is more than
$(\buyerendpoint{i+1}-\sellerendpoint{i-1})/2$, that is, more than half of the
first-best gain from trade at this profile.  The same holds at every higher
level.

{An SBB DSIC mechanism, in contrast, trades} at the profile
$(\sellerendpoint{i-1},\buyerendpoint{i+1})$ only if its price falls in
$[\sellerendpoint{i-1},\buyerendpoint{i+1}]\subseteq[\qpoint{i-1},\qpoint{i+1}]$,
so it leaves the traders at most the gain from trade times the probability of
this event ({Claim~\ref{clm:level-two-comparison}}).  Unlike the level-$1$
intervals, the intervals $[\qpoint{i-1},\qpoint{i+1}]$ overlap, but every
price lies in at most three of them, so the probabilities that the price falls
in them sum to at most three.  Lemma~\ref{lem:separation} describes the
distributions for which this gives an unbounded gap: if most of the first-best
gains-from-trade comes from such level-$2$ profiles, and each $i$ contributes
the same amount, then our WBB DSIC mechanism obtains more than half of every
contribution, while
{every SBB DSIC mechanism obtains, in}
total, only a constant number of them.

\paragraph{The distributions.}
It remains to construct distributions that satisfy the conditions of
Lemma~\ref{lem:separation}.  
{For independent values, the seller's value is likely to be near a high grid point and the buyer's value is likely to be near a low grid point. Hence, a profile in which the buyer's grid point is more than two steps above the seller's is unlikely, and a profile in which it is below the seller's has no gain from trade. Thus, all but a small fraction of the first-best gains-from-trade comes from level-$2$ profiles.}
{Figure~\ref{fig:ind-heatmap-probability} in Section~\ref{sec:proof:wbb-sbb-unbounded-gap-ind-gft} shows the probabilities of the resulting independent distribution, and Figure~\ref{fig:ind-heatmap-gft} shows how they translate into contributions to the first-best gains-from-trade, almost all of which come from the level-$2$ profiles.} 
For correlated values, we draw an index $i$ and
then a seller value near $\qpoint{i-1}$ and a buyer value near
$\qpoint{i+1}$, with probability inversely proportional to the gain from trade
between these values, so that each $i$ contributes approximately the same
amount to the first-best gains-from-trade.  Since the gaps of the grid grow quickly, the seller's values
are small compared to the gains from trade, so the traders' utility is close
to the traders' surplus.

\subsection{Our Technique: The Limitations of Money Burning}

We now describe the main ideas behind Theorem~\ref{thm:intro-limitations}, which shows that although DSIC mechanisms that are only \emph{weakly} budget-balanced can be arbitrarily better for the agents than those that are strongly so, 
WBB DSIC mechanisms still cannot provide, in the worst case, a constant fraction of the first-best gains-from-trade when the values are independent, or of the first-best social welfare when the values may be correlated.  Previous impossibility results~\cite{BD14,DS24} rely on the fact that every SBB DSIC mechanism is a distribution over posted prices~\cite{HR87,CP16}.  This fails for WBB DSIC mechanisms, so we need a different approach.


We show that WBB DSIC mechanisms are inherently limited: they fail on some distributions in which almost all of the contribution to the first-best benchmark comes from the accumulated contributions of many profiles spread out over the value space, each with small gain from trade. We show that these mechanisms fail by proving that the average trade probability of every such mechanism at these profiles must be close to zero. 
To see why, note that to obtain a constant fraction of the first-best on such distributions, a mechanism has to trade at these profiles with substantial average probability.  For WBB to hold, this forces a higher average trade probability at profiles with a larger gain from trade.  After many such steps, the average trade probability exceeds $1$, which is impossible (Lemma~\ref{lem:lb-trade-mass-bound}).  For the social welfare with correlated values, a similar argument applies to profiles on a geometric grid (Lemma~\ref{lem:geometric-staircase-bound}).

\paragraph*{Organization of the paper.}
Section~\ref{sec:related-work} discusses additional related work.
Section~\ref{sec:preliminaries} presents the model, the budget-balance requirements,
the allocation and traders' objectives, and the structural results used
throughout.
Section~\ref{sec:separation-results} explains the intuition behind our
construction of WBB mechanisms, presents the construction
and a separation lemma based on it, and uses them to
prove the unbounded separations for the traders' surplus under independent
values and for the traders' utility under correlated values.
Section~\ref{sec:wbb:limitations} shows that WBB
mechanisms cannot guarantee a constant fraction of the first-best
gains-from-trade under independent values, or of the first-best social welfare
under correlated values.  Omitted proofs are deferred to
Appendix~\ref{app:proofs}.

\section{{Additional} Related Work}\label{sec:related-work}

\paragraph{Approximately efficient mechanisms.}
Most of the results in this literature are for independent values, and we discuss these first.
Myerson and Satterthwaite~\cite{Myerson-Satterthwaite} showed that no
BIC and no-deficit 
mechanism can implement the first-best under mild conditions.  This impossibility motivates the study of approximately
efficient mechanisms.
{For gains-from-trade, \cite{mcaffee} showed that when the median
of the buyer's value exceeds the median of the seller's value, any fixed
price between the two medians obtains at least half of the first-best
gains-from-trade.  However, \cite{BD14} showed that
without such a condition,} no
deterministic, 
SBB mechanism in
which both traders have a dominant strategy guarantees a constant
approximation to the first-best.
This impossibility extends to randomized mechanisms with these
properties, since by \cite{HR87,CP16} every such randomized mechanism is a distribution over posted
prices.\footnote{Every posted price is a deterministic mechanism
with these properties.  Hence, for each of the objectives we consider, such as
the gains-from-trade in the result of~\cite{BD14}, the performance of the
randomized mechanism is the average of the performances of the posted prices
in its support, so at least one of these posted prices performs at least as
well.  The same argument extends every impossibility for deterministic,
SBB DSIC mechanisms to randomized ones.}
To overcome this impossibility,
{BIC mechanisms were considered.}  Blumrosen and
Mizrahi~\cite{BM} showed that the seller-offering mechanism
provides a constant approximation for distributions satisfying a certain
condition.
For general independent distributions, \cite{Bru} showed that the
random-offerer mechanism, which with probability $1/2$ runs the
seller-offering mechanism and with probability $1/2$ the symmetric
buyer-offering mechanism, obtains at least half of the second-best
gains-from-trade, that is, of the gains-from-trade of the optimal
SBB BIC
mechanism.  In a breakthrough, \cite{GDTCA}
proved that this mechanism also provides a constant approximation to the first-%
best.  Subsequent works improved the approximation factor and analyzed the
limitations of the random-offerer mechanism~\cite{IGFT,Jo26,BWZ26,DKB,CGMZ21,CGLM26,SC26}.
Recently, \cite{LQRW26} showed that the second-best mechanism always obtains
at least half of the first-best gains-from-trade, and that this is tight.

For social welfare, however, fixed prices suffice to give a constant
approximation under independent values.  A long line of work has therefore
studied how well a fixed price can perform~\cite{BD14,italians,BD21,KPV,CW-STOC,CSTOC,improved-fp}, showing that the best fixed
price always obtains approximately a $0.738024$ fraction of the first-best
welfare, and that this is tight~\cite{JGC26,LSDKLYZ26}.
By the posted-price characterization~\cite{HR87,CP16}, no
{randomized,} 
SBB DSIC mechanism obtains a better ratio. {When incentive compatibility is relaxed so that only
one trader has a dominant strategy and the other is 
{BIC, a better ratio} is possible: the buyer-offering mechanism
with a reserve price obtains a $0.746$ fraction of the first-best
welfare~\cite{DS26}.} {When incentive compatibility is
relaxed further, so that 
{both traders are only BIC,}
the second-best mechanism
obtains approximately a $0.888$ fraction of the first-best welfare, and this
is tight~\cite{LQW26welfare}.}

With
correlated values, no SBB mechanism in which both traders have a dominant strategy
guarantees a constant approximation to the first-best social welfare, while the
buyer-offering mechanism{, in which only the seller has a dominant
strategy, obtains a $(1-1/e)$ fraction of the first-best
welfare}~\cite{DS24}.
For gains-from-trade, no SBB BIC mechanism guarantees a constant approximation to the first-best~\cite{RW-private}.
Welfare approximation was also studied for traders with interdependent
values~\cite{DEGST25}.

\paragraph{Weakly budget-balanced mechanisms in bilateral trade.}
For 
{\emph{BIC} and interim IR mechanisms with independent values, money burning does not help:}
{every 
{ex-ante WBB mechanism can be converted into an ex-post SBB one} without reducing the
traders' expected utility~\cite{BN09,Bru,MM23}.}
For 
{\emph{deterministic DSIC} mechanisms}, money
burning does not help either: the results of~\cite{DK15} imply that the allocation rule of every
deterministic, 
WBB DSIC mechanism is pointwise dominated by that of a
posted price (Proposition~\ref{prop:prelim:dk-dominance}).

The work most closely related to our positive results shows that for correlated values, a weakly
budget-balanced mechanism can obtain more traders' surplus (which they call gains-from-trade) than every
posted-price mechanism~\cite{SW18}.
They suggest that correlation is crucial for this gain, since with independent values they found no case where their mechanisms beat posted prices, and they ask whether the optimal mechanism is indeed SBB in that case.
Our results strengthen this separation in
two ways.  First,
our separation for traders' surplus holds even for independent distributions,
while theirs uses correlated values.  Second, the advantage in our separation
is unbounded, also for the traders' utility when the values are correlated, while theirs establishes a constant separation. 

Manea and Maskin~\cite{MM23} also consider withholding or damaging the item in
Bayesian bilateral trade with independent values.  For discrete distributions,
neither withholding the item from both traders nor damaging it for the buyer
improves welfare.  They show, however, that damaging it for the seller can
sometimes be useful.  Their
results extend to continuous distributions under additional assumptions.


\paragraph{Extensions of bilateral trade.}

Bilateral trade has been extended to double auctions with many buyers and
sellers.  McAfee's trade reduction mechanism~\cite{TradeReduction} is
a WBB DSIC double auction.  It
was extended to supply chains~\cite{BabaioffNisan04,BabaioffWalsh2005}
and to spatially distributed markets~\cite{BNP09}.
Segal-Halevi, Hassidim, and Aumann~\cite{SHA16} extended it to 
{an SBB DSIC double auction}.  {Brustle, Cai, Wu, and Zhao}~\cite{Bru} obtained a
constant approximation to the second-best gains-from-trade in double auctions.
{Babaioff, Cai, Gonczarowski, and Zhao}~\cite{BothWorlds} combined trade reduction with the
random-offerer mechanism, which reduces double auctions to bilateral trade
whenever the first-best has a single trade.
Together with~\cite{GDTCA}, this
gives a constant approximation to the first-best gains-from-trade in double
auctions.  For the social welfare, constant approximations are known as
well~\cite{italians}.  The random-offerer
mechanism was extended to matching markets, where it obtains a constant
approximation to the first-best gains-from-trade~\cite{BRTW26,BWZ26}, and
the guarantees of the second-best mechanism were extended to matching
markets as well~\cite{BLWZ26,LQW26welfare}.  Bilateral trade was also
extended to traders with multi-dimensional values~\cite{CGMZ21,RTZ26}, to
markets with one seller and two buyers~\cite{BFN24}, and to
trades that involve more than two agents~\cite{EGR26}.

\paragraph{No-regret learning in bilateral trade.}
Bilateral trade was also studied in an online setting, where in every round
the mechanism posts prices to a new seller and buyer, and the goal is to
minimize the regret with respect to the best fixed price in
hindsight~\cite{CCCFL24MOR,regret2,regret1,BCCF24,CJLZ26,SublinearBudgetViolation25,PerturbedMarkets26,BCCC24fair,BCC24brokerage,BCC25contextual,GBCCP25,CC25volume,ABDF24,LCM25}.
Some of these works post a higher price to the buyer than to the seller, or
allow the mechanism to violate budget balance in some rounds.  There, this
helps learning by providing more informative feedback
across rounds. Each of these two prices is set independently of both traders' reports, and each trader only decides whether to accept the price he faces. Without learning, if the buyer's price is higher, offering both traders any common price between these two prices can only increase the probability of trade (without any impact on incentives), so posting such different prices can only be harmful.

\paragraph{Money burning in other settings.}
Money burning was previously
studied in settings with one-sided private
information: only the buyers have private values, while the designer, who is also the seller and owns the items, has none.  The goal is to maximize the buyers' surplus, that
is, the value the buyers get from the allocation minus their payments, while
the seller's revenue does not count.
This objective makes
sense in settings where payments are unavailable or undesirable, such as the
allocation of healthcare.  The study of money-burning mechanisms for
this objective, also known as the \emph{residual surplus}, was introduced
in~\cite{HartlineRoughgarden08}, which characterized the optimal mechanisms in
single-parameter auction 
settings.  Money burning in general domains was studied
in~\cite{FTTZ16}.  For unit-demand buyers with i.i.d.\ values,
\cite{GoldnerLundy25} give mechanisms whose surplus
approximates the optimal welfare, with tight bounds.  A framework for
designing mechanisms that maximize the surplus was given in~\cite{ESS25}, and
applied to unit-demand bidders, multi-unit auctions, and divisible goods,
obtaining mechanisms with optimal approximation guarantees with respect to the
social welfare.  In general multidimensional settings, a prior-free mechanism
obtains a $1/H_n$ fraction of the optimal welfare, and this is
tight~\cite{Ezra26}.  Unit-demand buyers with independent values were studied
in~\cite{GLT26}, and an online setting in~\cite{GMT26}.
The allocation of items when agents can signal their values only through
wasteful costs was also studied in
economics~\cite{Condorelli12,Yoon11,ChakravartyKaplan13}.
{Money burning was also studied for the allocation of a single item between
two buyers who can exchange money, with the goal of maximizing the buyers'
surplus~\cite{ShaoZhou16}.  Finally, money burning also appears in delegation
settings~\cite{AmadorBagwell20}.}

\paragraph{Redistribution of VCG payments.}
A closely related line of work studies the same objective as in the
previous paragraph, the agents' total utility, but focuses on the VCG mechanism.  In allocation problems, the VCG mechanism is 
{efficient and DSIC, but} the agents' payments go to the designer and reduce
the agents' total utility.  These works redistribute as much of these payments as
possible back to the agents without violating incentive compatibility, and
any payment that is not
redistributed is lost~\cite{Bailey97,Cavallo06,GC09,Moulin09}.  Guo and
Conitzer~\cite{GC08} show that allocating inefficiently, by withholding some
units, can increase the agents' total utility, and de Clippel et
al.~\cite{deClippel14} show that destroying items can have a similar effect.




\section{Preliminaries}\label{sec:preliminaries}

\subsection{The Bilateral Trade Problem}

In bilateral trade there are two agents:
a seller, who initially owns one indivisible
item, and a buyer.  The buyer's value for receiving the item is
$\vb\in\suppb$, and the seller's value for keeping the item is
$\vs\in\supps$, where the value spaces $\suppb, \supps\subseteq \mathbb{R}_{\geq 0}$ are (possibly unbounded)  intervals, satisfying $\suppb= \supps=\supp$. {Although both value spaces are equal, we keep separate notation for them to make clear which agent's value each argument refers to.}

The valuation profile $(\vs,\vb)$ is drawn from a known joint distribution $\distjoint$ over $\supps \times \suppb$. The values may be correlated. We denote the seller and buyer marginals of $\distjoint$ by $\distseller$ and $\distbuyer$, respectively.  When values are
independent, $\distjoint=\distseller\times\distbuyer$.

A \emph{randomized (direct revelation) mechanism} 
$\mechanism=(\allocation,\paymentB,\paymentS)$ consists of three measurable functions:
\begin{itemize}
    \item an \emph{allocation rule} $\allocation:\supps\times\suppb\to[0,1]$,
    where $\allocation(\vs,\vb)$ is the probability of trade; and 
    \item \emph{payment rules}\footnote{Although the payment rules are a priori allowed to take negative
    values, individual rationality and weak budget balance together imply that
    both payments are nonnegative; see the proof of
    Lemma~\ref{lem:wbb-allocation-inequality} for a formal derivation.} $\paymentB,\paymentS:\supps\times\suppb\to\mathbb{R}$,
    where $\paymentB(\vs,\vb)$ is the expected payment made by the buyer, and
    $\paymentS(\vs,\vb)$ is the expected payment that the seller receives.
\end{itemize}
{A direct revelation mechanism is \emph{deterministic} if it uses no internal randomness (its allocation and payments are deterministically determined). In particular, its allocation function $\allocation$ takes values in $\{0,1\}$.}\footnote{{It suffices to consider mechanisms whose allocation function takes values in $\{0,1\}$: since $\paymentB(\vs,\vb)$ and $\paymentS(\vs,\vb)$ are the expected payments at each report profile, charging the buyer exactly $\paymentB(\vs,\vb)$ and paying the seller exactly $\paymentS(\vs,\vb)$ implements the same allocation and payment functions with no internal randomness, and hence preserves incentive compatibility, individual rationality, and budget balance.}}
We assume that both the seller and buyer have quasi-linear utilities and are risk
neutral.  That is, in a mechanism
$\mechanism=(\allocation,\paymentB,\paymentS)$, for a valuation profile
$(\vs,\vb)\in\supps\times\suppb$ and reports $\vs'\in\supps$ and
$\vb'\in\suppb$, the agents' utilities are:
\begin{align*}
    \utilb(\vb,(\vs',\vb'))
    &= \vb\cdot\allocation(\vs',\vb')-\paymentB(\vs',\vb'), \\
    \utils(\vs, (\vs',\vb'))
    &= \paymentS(\vs',\vb')+\vs\cdot
    \bigl(1-\allocation(\vs',\vb')\bigr).
\end{align*}

\begin{definition}\label{def:randomized-dsic}
    A randomized mechanism $\mechanism=(\allocation,\paymentB,\paymentS)$ is
    \emph{dominant-strategy incentive compatible (DSIC)} if, for every valuation profile $(\vs,\vb)\in\supps\times\suppb$, each agent maximizes his expected utility by reporting his actual value, regardless of the other agent's report, where the expectation is over the mechanism's internal randomness. That is, for every
    $\vb,\vb'\in\suppb$ and $\vs,\vs'\in\supps$,
    \begin{align}\label{equ:DSIC-condition-buyer}
        \utilb(\vb, (\vs', \vb)) & \geq \utilb(\vb, (\vs', \vb')) \\       \label{equ:DSIC-condition-seller}
        \utils(\vs, (\vs, \vb')) &\geq \utils(\vs, (\vs', \vb'))
    \end{align}
\end{definition}

In this paper, we only consider DSIC mechanisms, and thus by the revelation principle, we can always assume that the agents report truthfully{, that is, $\vs'=\vs$ and $\vb'=\vb$, and for such mechanisms we drop the reports from the utility notation:}
\begin{align}
    \utilb(\vs,\vb)
    &= \vb\cdot\allocation(\vs,\vb)-\paymentB(\vs,\vb),
    \label{equ:truthful-utility-buyer}\\
    \utils(\vs,\vb)
    &= \paymentS(\vs,\vb)+\vs\cdot
    \bigl(1-\allocation(\vs,\vb)\bigr).
    \label{equ:truthful-utility-seller}
\end{align}

\noindent Following the standard terminology from the algorithmic
mechanism-design literature, we sometimes distinguish between two classes of randomized
DSIC mechanisms.  The first class is of \emph{universally truthful}
mechanisms; such a mechanism is a distribution over deterministic DSIC mechanisms.  Here,
truthful reporting maximizes each agent's utility for every realization of the
mechanism's internal randomness.  The second class is \emph{truthful in expectation} mechanisms.  In such a mechanism truthful reporting maximizes each agent's expected
utility, where the expectation is over the mechanism's internal randomness.  This is consistent with Definition~\ref{def:randomized-dsic}.
Thus, every universally truthful mechanism is truthful in expectation, but not necessarily vice-versa.

\begin{definition}
A  randomized mechanism is \emph{individually rational (IR)} if, for every valuation
profile $(\vs,\vb)\in\supps\times\suppb$, each agent's expected utility, where
the expectation is only over the mechanism's internal randomness, is at least as large as his utility when not participating in the mechanism, under truthful reporting.  Equivalently,
\begin{align}
    \utilb(\vs,\vb)  & \ge 0,
    \label{equ:ir-buyer}\\
    \utils(\vs,\vb) &\ge \vs.
    \label{equ:ir-seller}
\end{align}
\end{definition}

As we focus on DSIC mechanisms, individual rationality requires the above inequalities to hold for \emph{every valuation profile} separately: the expectation is only over the mechanism's internal randomness.
Unlike interim individual rationality in the Bayesian setting, it does
\emph{not} average over the other agent's value.

Throughout this paper, unless stated otherwise, any mechanism we consider must be both individually rational (IR) and dominant-strategy incentive compatible (DSIC). Trade mechanisms are also required not to lose money:

\begin{definition}
A randomized mechanism is \emph{weakly budget balanced (WBB)} if, for every valuation profile
$(\vs,\vb)\in\supps\times\suppb$, the expected payment of the buyer is at least the expected payment of the seller, where the expectation is only over the mechanism's internal randomness.  Equivalently,
\begin{align}\label{equ:def-wbb}
    \paymentB(\vs,\vb)\ge \paymentS(\vs,\vb).
\end{align}
The mechanism is called \emph{strongly budget-balanced (SBB)} if this inequality holds as equality.
\end{definition}

We also consider ex-post versions of these notions. A randomized mechanism satisfies \emph{\expirname} if, for every valuation profile $(\vs,\vb)\in\supps\times\suppb$, the individual rationality conditions in Equations~\eqref{equ:ir-buyer} and \eqref{equ:ir-seller} hold for every realization of the mechanism’s internal randomness, rather than merely in expectation over this randomness. Similarly, a randomized mechanism is \emph{ex-post weakly budget balanced (ex-post WBB)} or \emph{ex-post strongly budget-balanced (ex-post SBB)} if the corresponding payment condition (inequality \eqref{equ:def-wbb}) holds for every valuation profile and every realization of the mechanism’s internal randomness {(with equality, for ex-post SBB)}.




Throughout the paper, individual rationality, strong budget balance, and weak
budget balance refer to expectations over the mechanism's internal randomness,
at every valuation profile.  We explicitly use the qualifier ``ex-post'' when
requiring these properties for every realization of this randomness.

\begin{remark}
For deterministic mechanisms, individual rationality, strong budget balance,
and weak budget balance coincide with their ex-post versions.  Thus, a
universally truthful mechanism is ex-post individually rational, ex-post
strongly budget balanced, or ex-post weakly budget balanced whenever every
deterministic mechanism in its support satisfies the corresponding property.
Universal truthfulness alone does not make individual rationality or budget
balance hold ex-post.
\end{remark}

The following remark is a simple useful fact, also used in the literature
before.

\begin{remark}\label{remark:moving-from-expectation-to-ex-post}
Any individually rational, strongly budget-balanced (respectively, weakly
budget-balanced) truthful-in-expectation mechanism can be converted into a truthful-in-expectation
mechanism that satisfies ex-post individual rationality and ex-post strong
budget balance (respectively, ex-post weak budget balance), without changing its
allocation rule or expected payments.  At every valuation profile for which
$\allocation(\vs,\vb)>0$, charge the buyer
$\paymentB(\vs,\vb)/\allocation(\vs,\vb)$ and pay the seller
$\paymentS(\vs,\vb)/\allocation(\vs,\vb)$ when trade occurs, and make no
payments otherwise.  Individual rationality and weak budget balance imply
$\vs\cdot\allocation(\vs,\vb)\leq\paymentS(\vs,\vb)
\leq\paymentB(\vs,\vb)\leq\vb\cdot\allocation(\vs,\vb)$.
Thus, conditional on trade, the seller receives at least $\vs$, the buyer
pays at most $\vb$, and the required budget-balance condition holds.  If
$\allocation(\vs,\vb)=0$, both expected payments are zero, so we make no
payments.  This construction
preserves the agents' expected utilities and therefore truthfulness in
expectation, while ensuring ex-post individual rationality and the relevant
ex-post budget-balance condition in every realization.
\end{remark}


\subsection{Basic Properties}
\label{sec:basic-properties}

We repeatedly use the following simple consequence of dominant-strategy incentive compatibility,
individual rationality, and weak budget balance, which follows from  Myerson's lemma.

\begin{lemma}[Allocation inequality from dominant-strategy incentive compatibility, individual rationality, and weak budget balance]
\label{lem:wbb-allocation-inequality}
Let $\mechanism=(\allocation,\paymentB,\paymentS)$ be a dominant-strategy incentive-compatible,
individually rational, and weakly budget-balanced
mechanism. For every
$\vs\in \supps,\vb\in \suppb$ such that $\vs\le \vb$,
\begin{equation}\label{ineq:wbb-tie}
    \int_{\vs}^{\vb} \allocation(z,\vb)\d{z}
    +
    \int_{\vs}^{\vb} \allocation(\vs,z)\d{z}
    \le
    (\vb-\vs)\cdot \allocation(\vs,\vb).
\end{equation}
\end{lemma}
\begin{proof}[Proof of Lemma~\ref{lem:wbb-allocation-inequality}]
Note that, for every valuation profile $(\vs,\vb)$, WBB and
IR imply
\begin{align}\label{ineq:wbb-ir}
    \vs\cdot\allocation(\vs,\vb)
    \le \paymentS(\vs,\vb)
    \le \paymentB(\vs,\vb)
    \le \vb\cdot\allocation(\vs,\vb).
\end{align}
This has two consequences.  First, if there is no trade, that is,
$\allocation(\vs,\vb)=0$, then both payments are zero:
$\paymentB(\vs,\vb)=\paymentS(\vs,\vb)=0$.  Second, if $\vs>\vb$, then
Equation~\eqref{ineq:wbb-ir} forces $\allocation(\vs,\vb)=0$, and hence both payments
are zero.  If $\vs=\vb$, all the inequalities in Equation~\eqref{ineq:wbb-ir} hold with
equality, so
\begin{equation}
\label{eq:equal-value-payments}
    \paymentB(\vs,\vs)
    =
    \paymentS(\vs,\vs)
    =
    \vs\cdot\allocation(\vs,\vs).
\end{equation}
Applying Myerson's lemma, for $\vs\le\vb$ we obtain
\begin{align*}
    \paymentB(\vs,\vb)
    &=
    \vb\cdot \allocation(\vs,\vb)
    -
    \int_{\vs}^{\vb}\allocation(\vs,z)\d{z}
    +
    \paymentB(\vs,\vs)
    -
    \vs\cdot\allocation(\vs,\vs)\\
    &\stackrel{\eqref{eq:equal-value-payments}}{=}
    \vb\cdot \allocation(\vs,\vb)
    -
    \int_{\vs}^{\vb}\allocation(\vs,z)\d{z}
\end{align*}
and
\begin{align*}
    \paymentS(\vs,\vb)
    &=
    \vs\cdot \allocation(\vs,\vb)
    +
    \int_{\vs}^{\vb}\allocation(z,\vb)\d{z}
    +
    \paymentS(\vb,\vb)
    -
    \vb\cdot\allocation(\vb,\vb)\\
    &\stackrel{\eqref{eq:equal-value-payments}}{=}
    \vs\cdot \allocation(\vs,\vb)
    +
    \int_{\vs}^{\vb}\allocation(z,\vb)\d{z}.
\end{align*}
WBB gives
$\paymentB(\vs,\vb)\ge \paymentS(\vs,\vb)$.  Substituting the two payment
identities into this inequality and rearranging gives Equation~\eqref{ineq:wbb-tie}.
\end{proof}

Drexl and Kleiner~\cite{DK15} show that, for every pair of independent
distributions over the agents' values, an optimal deterministic, WBB DSIC and IR
mechanism for the sum of the agents' utilities can be chosen to be SBB.
Their proof implies the following stronger, pointwise statement about the
allocation rule.

\begin{proposition}[Essentially~\cite{DK15}]
\label{prop:prelim:dk-dominance}
For every deterministic, dominant-strategy incentive-compatible, individually rational, and weakly budget-balanced bilateral-trade mechanism with
allocation rule $\allocation$, there is a deterministic, dominant-strategy incentive-compatible, individually rational, and strongly budget-balanced
mechanism with allocation rule $\allocation'$ such that
\begin{align*}
    \allocation(\vs,\vb)\leq \allocation'(\vs,\vb)
\end{align*}
for every valuation profile $(\vs,\vb)$.
\end{proposition}
The proof of Proposition~\ref{prop:prelim:dk-dominance} is deferred to Appendix~\ref{app:proof-dk-dominance}.

\subsection{Objectives}
\label{sec:objectives}

{Our two main objectives are the \emph{traders' utility}, which is the sum
of the buyer's and seller's utilities, and the \emph{traders' surplus}, which
is the improvement in this sum relative to no trade (the seller keeps the
item).  
Under WBB, the buyer
may pay more than the seller receives.  The \emph{mechanism's surplus},
denoted by $u_{\mechanism}(\vs,\vb)$, is
the expectation of
the difference between the payment collected from the buyer and the payment made to the seller.
Under
WBB it is nonnegative, and under SBB it is
zero.  We take the traders' perspective and interpret the mechanism's surplus as money burned: it is paid by the buyer but not received by the seller.

Adding the mechanism's surplus to the traders' utility gives the standard
notion of \emph{social welfare}: the total utility of the buyer, the seller,
and the mechanism.  Since payments cancel in this sum, it equals the expected
value of the agent who holds the item 
at the end of the mechanism's execution.
Similarly,
adding the mechanism's surplus to the traders' surplus gives the
\emph{gains-from-trade}: the improvement in social welfare relative to no
trade.  These objectives depend only on the allocation, so we call them
\emph{allocation-based} objectives.  Equivalently, the traders' utility and the traders'
surplus are obtained from social welfare and gains-from-trade by removing the
mechanism's surplus.  Thus, under WBB, each traders' objective
is at most its allocation-based counterpart; under SBB, the
corresponding objectives coincide.}

Formally, for every mechanism $\mechanism=(\allocation,\paymentB,\paymentS)$ and valuation profile
$(\vs,\vb)\in\supps\times\suppb$, the \emph{buyer's surplus} is his utility, $\sgft^{b}_{\mechanism}(\vs,\vb)=\vb\cdot\allocation(\vs,\vb)-\paymentB(\vs,\vb)$, and the \emph{seller's surplus} is the improvement in her utility relative to keeping the item, $\sgft^{s}_{\mechanism}(\vs,\vb)=\paymentS(\vs,\vb)-\vs\cdot\allocation(\vs,\vb)$. By individual rationality, both are nonnegative. The mechanism's surplus is
\begin{equation}\label{eq:mechanism-surplus}
    u_{\mechanism}(\vs,\vb)
    = \paymentB(\vs,\vb)-\paymentS(\vs,\vb).
\end{equation}
The objectives that include this surplus are
\begin{align}
    \welfare_{\mechanism}(\vs,\vb)
    &= \utilb(\vs,\vb)+\utils(\vs,\vb)+u_{\mechanism}(\vs,\vb)\notag\\
    &= \vb\cdot\allocation(\vs,\vb)
       +\vs\cdot\bigl(1-\allocation(\vs,\vb)\bigr),
    \label{equ:def-welfare}\\
    \gft_{\mechanism}(\vs,\vb)
    &= \welfare_{\mechanism}(\vs,\vb)-\vs
     = (\vb-\vs)\cdot\allocation(\vs,\vb).
    \label{equ:def-gft}
\end{align}
The objectives that count only the traders' utilities are
\begin{align}
    \swelfare_{\mechanism}(\vs,\vb)
    &= \utilb(\vs,\vb)+\utils(\vs,\vb)\notag\\
    &= \welfare_{\mechanism}(\vs,\vb)-u_{\mechanism}(\vs,\vb),
    \label{eq:welfare-with-payments-objective}\\
    \sgft_{\mechanism}(\vs,\vb)
    &= \sgft^{b}_{\mechanism}(\vs,\vb)+\sgft^{s}_{\mechanism}(\vs,\vb)\notag\\
    &= \swelfare_{\mechanism}(\vs,\vb)-\vs\notag\\
    &= \gft_{\mechanism}(\vs,\vb)-u_{\mechanism}(\vs,\vb).
    \label{eq:residual-surplus-objective}
\end{align}

For a prior $\distjoint$, we evaluate each objective in expectation over the
distribution of traders' values.  That is,
\begin{align*}
    \welfare_{\mechanism}(\distjoint)
    & =
    \Exp_{(\vs,\vb)\sim\distjoint}
    \left[\welfare_{\mechanism}(\vs,\vb)\right],\\
    \gft_{\mechanism}(\distjoint)
    & =
    \Exp_{(\vs,\vb)\sim\distjoint}
    \left[\gft_{\mechanism}(\vs,\vb)\right]{=\welfare_{\mechanism}(\distjoint)-\Exp_{(\vs,\vb)\sim\distjoint}\left[\vs\right]},\\
        \swelfare_{\mechanism}(\distjoint)
    & =
    \Exp_{(\vs,\vb)\sim\distjoint}
    \left[\swelfare_{\mechanism}(\vs,\vb)\right]{=\welfare_{\mechanism}(\distjoint)-\Exp_{(\vs,\vb)\sim\distjoint}\left[u_{\mechanism}(\vs,\vb)\right]},\\
    \sgft_{\mechanism}(\distjoint) 
    & =
    \Exp_{(\vs,\vb)\sim\distjoint} 
    \left[\sgft_{\mechanism}(\vs,\vb)\right]{=\gft_{\mechanism}(\distjoint)-\Exp_{(\vs,\vb)\sim\distjoint}\left[u_{\mechanism}(\vs,\vb)\right]}.
\end{align*}

Similarly, for a prior $\distjoint$, the buyer's and the seller's expected surpluses are
\begin{align*}
    \sgft^{b}_{\mechanism}(\distjoint) &= \Exp_{(\vs,\vb)\sim\distjoint}\left[\sgft^{b}_{\mechanism}(\vs,\vb)\right],\\
    \sgft^{s}_{\mechanism}(\distjoint) &= \Exp_{(\vs,\vb)\sim\distjoint}\left[\sgft^{s}_{\mechanism}(\vs,\vb)\right].
\end{align*}

For a distribution $\distjoint$ over the traders' values, we denote by
$\fbwelfare(\distjoint)$ and $\fbgft(\distjoint)$ 
the optimal expected social
welfare and gains-from-trade, respectively, if the traders' private values were
public.  These are the \emph{first-best} benchmarks{:
\begin{align*}
    \fbwelfare(\distjoint)
    = \Exp_{(\vs,\vb)\sim\distjoint}\left[\max\{\vs,\vb\}\right],
    \qquad
    \fbgft(\distjoint)
    = \Exp_{(\vs,\vb)\sim\distjoint}\left[\max\{\vb-\vs,0\}\right].
\end{align*}}
We denote by $\fbswelfare(\distjoint)$ and $\fbsgft(\distjoint)$ the traders' utility and traders' surplus of the first-best. Since the first-best burns no money, $\fbswelfare(\distjoint)=\fbwelfare(\distjoint)$ and $\fbsgft(\distjoint)=\fbgft(\distjoint)$.

\section{Money Burning Mechanisms are Arbitrarily Better }\label{sec:separation-results}\label{sec:seperation-results}

%
%
In this section we present our main result, showing that for randomized dominant-strategy incentive-compatible and
individually rational mechanisms, 
the traders can be arbitrarily better off when the mechanism burns money than when it does not. That is, weakly budget-balanced mechanisms can perform
arbitrarily better than every strongly budget-balanced mechanism. 
We show this in two
settings: first, for traders' surplus under independent distributions, and second, for traders' utility under
correlated distributions. We first state the two results. Both rely on a general
construction of WBB DSIC and IR mechanisms, which we present in
Section~\ref{sec:wbb:construction}, followed by the proofs of the two results in
Sections~\ref{sec:proof:wbb-sbb-unbounded-gap-ind-gft} and~\ref{sec:proof-wbb-sbb-seperation-welfare}, respectively. 

For traders' surplus, SBB DSIC mechanisms cannot
provide a constant approximation to first-best gains-from-trade,
even under independent distributions~\cite{BD14}\footnote{Note that the impossibility result of \cite{BD14} applies to deterministic mechanisms. However, it extends to randomized ones by the characterization of \cite{HR87,CP16}.}.  
{We show that randomized WBB DSIC mechanisms can do arbitrarily better
than SBB ones (Theorem~\ref{thm:wbb-sbb-unbounded-gap-ind-gft}).
Yet, in Section~\ref{sec:wbb:limitations} we show that they still cannot guarantee a
constant approximation to the first-best gains-from-trade
(Theorem~\ref{thm:tie-wbb-no-constant-ind-gft}).}

\begin{theorem}\label{thm:wbb-sbb-unbounded-gap-ind-gft}
For every $\varepsilon>0$, there are independent seller and buyer distributions $\distseller$ and $\distbuyer$ and a randomized dominant-strategy incentive-compatible, individually rational, and weakly budget-balanced mechanism $\wbbmechanism$ such that, for every randomized dominant-strategy incentive-compatible, individually rational, and strongly budget-balanced mechanism $\sbbmechanism$,
\[
    \varepsilon\cdot\sgft_{\wbbmechanism}(\distseller\times\distbuyer)
    > \sgft_{\sbbmechanism}(\distseller\times\distbuyer)
    = \gft_{\sbbmechanism}(\distseller\times\distbuyer).
\]
\end{theorem}
{Moreover, the construction is symmetric, and so \emph{each} of the traders is arbitrarily better off when moving from SBB mechanisms to  using money burning (see Remark~\ref{rem:both-traders-benefit}).}

For traders' utility, we cannot obtain an unbounded separation under
independent distributions, since SBB mechanisms provide a
constant approximation to first-best social
welfare~\cite{BD14,italians,BD21,KPV,CSTOC,CW-STOC,improved-fp,JGC26,LSDKLYZ26}.  For correlated
distributions, however, no SBB mechanism provides such a
guarantee~\cite{DS24}.  We therefore show an unbounded separation between
WBB and SBB mechanisms for traders' utility under
correlated distributions
(Theorem~\ref{thm:wbb-sbb-unbounded-gap-correlated-residual-welfare}).

\begin{theorem}
\label{thm:wbb-sbb-unbounded-gap-correlated-residual-welfare}
For every $\varepsilon>0$, there is a joint distribution $\distjoint$ and a dominant-strategy incentive-compatible, individually rational, and weakly budget-balanced mechanism $\wbbmechanism$ such that, for every dominant-strategy incentive-compatible, individually rational, and strongly budget-balanced mechanism $\sbbmechanism$,
\[
    \varepsilon\cdot\swelfare_{\wbbmechanism}(\distjoint)
    > \swelfare_{\sbbmechanism}(\distjoint)
    = \welfare_{\sbbmechanism}(\distjoint).
\]
\end{theorem}

{In both theorems, the money burned by the WBB mechanism is not
counted in its objective: the traders' surplus and the traders' utility 
objectives only count
the two traders. 
For a WBB
mechanism,
counting the burnt money can only increase these two objectives, 
while for an SBB mechanism, traders' surplus coincides with gains-from-trade, and traders' utility coincides with social welfare.

Hence, both theorems also give the corresponding separations for gains-from-trade and
social welfare:
\begin{align*}
    \varepsilon\cdot\gft_{\wbbmechanism}(\distseller\times\distbuyer)
    &\ge \varepsilon\cdot\sgft_{\wbbmechanism}(\distseller\times\distbuyer)
    > \gft_{\sbbmechanism}(\distseller\times\distbuyer)
    \tag*{(by Theorem~\ref{thm:wbb-sbb-unbounded-gap-ind-gft}),}\\
    \varepsilon\cdot\welfare_{\wbbmechanism}(\distjoint)
    &\ge \varepsilon\cdot\swelfare_{\wbbmechanism}(\distjoint)
    > \welfare_{\sbbmechanism}(\distjoint)
    \tag*{(by Theorem~\ref{thm:wbb-sbb-unbounded-gap-correlated-residual-welfare}).}
\end{align*}}


\subsection{A Construction of Weakly Budget-Balanced DSIC Mechanisms}\label{sec:wbb:construction}

In this section, we present a general construction of dominant-strategy incentive-compatible,
individually rational, and weakly budget-balanced mechanisms that we use
to establish our two separation results
(Theorems~\ref{thm:wbb-sbb-unbounded-gap-ind-gft}
and~\ref{thm:wbb-sbb-unbounded-gap-correlated-residual-welfare}).  
We first provide
intuition for the construction and explain its main features.  The formal
construction and its proof follow in
Lemma~\ref{lem:general-wbb-step-construction}. 

\paragraph{Intuition and high-level description.}
The purpose of this construction is to obtain
WBB DSIC and IR mechanisms that can outperform SBB DSIC and IR mechanisms. We begin by identifying a structural limitation of strongly budget-balanced mechanisms, and then we construct a
 WBB mechanism that circumvents this limitation.

Recall that since we consider DSIC mechanisms, the incentive constraints do not involve the distribution of the values: truthful reporting must be optimal for each agent regardless of the other agent's report. Hence, when constructing our mechanism, we only need to consider the agents' value spaces $\supps$ and $\suppb$; the distribution plays a role only when we measure the performance of the mechanism. For the purpose of this discussion, let $\supps=\suppb=[0,1]$.\footnote{We state the construction for this case; it can easily be extended to any interval.}

{It was shown that every} randomized SBB DSIC and IR
mechanism can be represented as a distribution over posted prices~\cite{HR87,CP16}. Consider such a mechanism, let $\allocation$ denote its allocation function, and let $\price$ be the random posted price drawn from a distribution $\mu$. At the
agents' profile $(\vs,\vb)$, the allocation probability is at most the probability that the posted price lies in $[\vs,\vb]$, that is, $\allocation(\vs,\vb) \le \Pr_{\price\sim \mu}[\price \in [\vs,\vb]]$. Consequently, for any profiles $(\vs^{(1)},\vb^{(1)})$, \ldots, $(\vs^{(m)},\vb^{(m)})$ whose intervals $[\vs^{(j)},\vb^{(j)}]$ are pairwise disjoint,
\begin{align*}
\sum_{j=1}^{m} \allocation(\vs^{(j)},\vb^{(j)})
\le \sum_{j=1}^{m} \Pr_{\price \sim \mu}\left[\price\in[\vs^{(j)},\vb^{(j)}]\right]
\le 1.
\end{align*}
Thus, an SBB mechanism cannot assign substantial allocation probability to all these profiles simultaneously: at least one of them has allocation probability at most $1/m$. Our WBB construction overcomes precisely this limitation: it assigns high allocation probability to all
 such profiles simultaneously (in the specific constructions we use, at least $1/2$).

We now describe the construction. Its allocation rule is a step function in each agent's report: fixing the buyer's report, the allocation probability is a step function of the seller's report, and vice versa. The thresholds at which the steps occur are derived from a grid on the value space $[0,1]$ that is common to the buyer and the seller. This grid is the input of the construction: an increasing sequence $\qset$ of $n+1$ values
\[
0=\qpoint{0}<\cdots<\qpoint{n}=1.
\]

Each interior grid point $\qpoint{i}$ gives rise to two nearby thresholds: a seller threshold $\sellerendpoint{i}=\qpoint{i}+\slack$ slightly above $\qpoint{i}$ and a buyer threshold $\buyerendpoint{i}=\qpoint{i}-\slack$ slightly below $\qpoint{i}$, for a small $\slack>0$.\footnote{{For the endpoints of the value space, both thresholds coincide with the grid values, that is, $\sellerendpoint{0}=\buyerendpoint{0}=\qpoint{0}=0$ and $\sellerendpoint{n}=\buyerendpoint{n}=\qpoint{n}=1$.}} Thus, the buyer and seller thresholds are derived from the same grid, and the slight separation between them distinguishes the two sides of each grid point. In particular, $\sellerendpoint{i}>\buyerendpoint{i}$ for every interior grid point, while for $\slack$ that is small enough, $\sellerendpoint{i-1}<\buyerendpoint{i}$ for every $i\in[n]$. Hence, for any small enough $\slack>0$, the intervals
\begin{align*}
    K_i=[\sellerendpoint{i-1},\buyerendpoint{i}], \qquad i\in[n],
\end{align*}
that is, $[\sellerendpoint{0},\buyerendpoint{1}],[\sellerendpoint{1},\buyerendpoint{2}],\ldots,[\sellerendpoint{n-1},\buyerendpoint{n}]$, are nonempty and pairwise disjoint. These are the intervals on which we exhibit the gap between the two classes of mechanisms: by the discussion above, the allocation probabilities of an SBB mechanism at the profiles $(\sellerendpoint{i-1},\buyerendpoint{i})$ sum to at most $1$, whereas {in the specific constructions we use, this sum is} {at least $n/2$}.

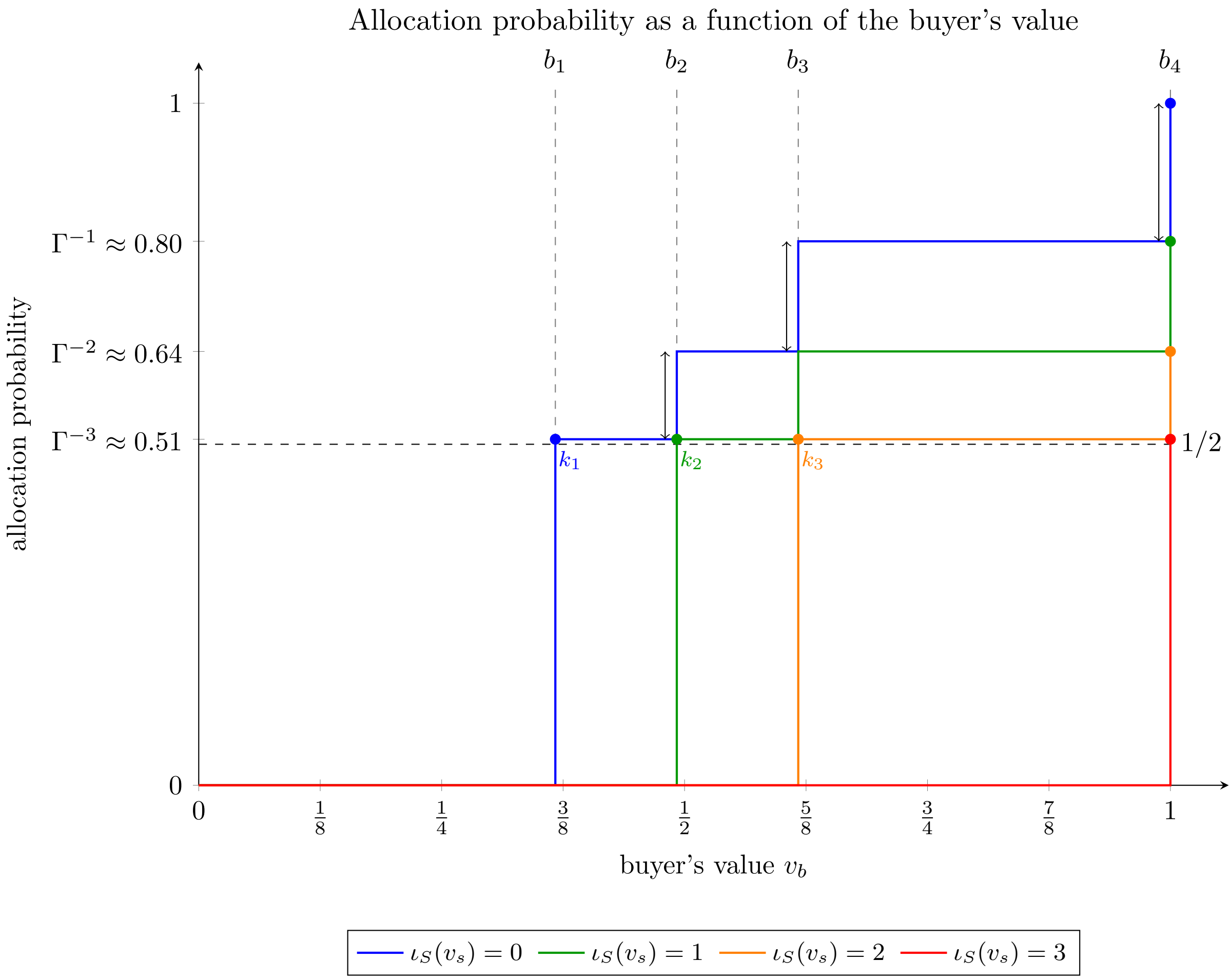
\begin{figure}[!t]
    \centering
    \begin{tikzpicture}
        \pgfmathsetmacro{\growthval}{1.254}
        \pgfmathsetmacro{\gpowA}{\growthval^(-1)}
        \pgfmathsetmacro{\gpowB}{\growthval^(-2)}
        \pgfmathsetmacro{\gpowC}{\growthval^(-3)}
        \def\bOne{0.367}
        \def\bTwo{0.492}
        \def\bThree{0.617}
        \def\bFour{1}
        \begin{axis}[
            width=0.9\textwidth,
            height=0.66\textwidth,
            axis lines=left,
            xmin=0,
            xmax=1.06,
            ymin=0,
            ymax=1.06,
            title={Allocation probability as a function of the buyer's value},
            xlabel={buyer's value $\vb$},
            ylabel={allocation probability},
            xtick={0,0.125,...,1},
            xticklabels={
                $0$, $\frac{1}{8}$, $\frac{1}{4}$, $\frac{3}{8}$, $\frac{1}{2}$,
                $\frac{5}{8}$, $\frac{3}{4}$, $\frac{7}{8}$, $1$
            },
            ytick={0,{\gpowC},{\gpowB},{\gpowA},1},
            yticklabels={
                $0$,
                $\growth^{-3}\approx0.51$,
                $\growth^{-2}\approx0.64$,
                $\growth^{-1}\approx0.80$,
                $1$
            },
            tick label style={font=\small},
            label style={font=\small},
            legend style={
                at={(0.5,-0.2)},
                anchor=north,
                legend columns=-1,
                /tikz/every even column/.append style={column sep=0.3em},
                draw=black,
                fill=white,
                font=\footnotesize
            },
            every axis plot/.append style={thick},
            clip=false
        ]

        \foreach \x in {\bOne,\bTwo,\bThree,\bFour} {
            \addplot[gray, dashed, thin, forget plot]
                coordinates {(\x,0) (\x,1.03)};
        }
        \node[anchor=south, font=\small] at (axis cs:\bOne,1.03) {$\buyerendpoint{1}$};
        \node[anchor=south, font=\small] at (axis cs:\bTwo,1.03) {$\buyerendpoint{2}$};
        \node[anchor=south, font=\small] at (axis cs:\bThree,1.03) {$\buyerendpoint{3}$};
        \node[anchor=south, font=\small] at (axis cs:\bFour,1.03) {$\buyerendpoint{4}$};

        \addplot[black, dashed, thin, forget plot]
            coordinates {(0,0.5) (\bFour,0.5)};
        \node[anchor=west, font=\small] at (axis cs:\bFour,0.5) {$1/2$};

        \addplot[const plot, blue] coordinates {
            (0,0) (\bOne,{\gpowC}) (\bTwo,{\gpowB}) (\bThree,{\gpowA}) (\bFour,1)
        };
        \addlegendentry{$\sellerround{\vs}=0$}

        \addplot[const plot, green!60!black] coordinates {
            (0,0) (\bOne,0) (\bTwo,{\gpowC}) (\bThree,{\gpowB}) (\bFour,{\gpowA})
        };
        \addlegendentry{$\sellerround{\vs}=1$}

        \addplot[const plot, orange] coordinates {
            (0,0) (\bTwo,0) (\bThree,{\gpowC}) (\bFour,{\gpowB})
        };
        \addlegendentry{$\sellerround{\vs}=2$}

        \addplot[const plot, red] coordinates {
            (0,0) (\bThree,0) (\bFour,{\gpowC})
        };
        \addlegendentry{$\sellerround{\vs}=3$}

        \fill[blue] (axis cs:\bFour,1) circle[radius=2pt];
        \fill[green!60!black] (axis cs:\bFour,{\gpowA}) circle[radius=2pt];
        \fill[orange] (axis cs:\bFour,{\gpowB}) circle[radius=2pt];
        \fill[red] (axis cs:\bFour,{\gpowC}) circle[radius=2pt];

        \fill[blue] (axis cs:\bOne,{\gpowC}) circle[radius=2pt];
        \node[font=\scriptsize, text=blue, anchor=north west, inner sep=1pt]
            at (axis cs:\bOne,0.495) {$k_1$};
        \fill[green!60!black] (axis cs:\bTwo,{\gpowC}) circle[radius=2pt];
        \node[font=\scriptsize, text=green!60!black, anchor=north west, inner sep=1pt]
            at (axis cs:\bTwo,0.495) {$k_2$};
        \fill[orange] (axis cs:\bThree,{\gpowC}) circle[radius=2pt];
        \node[font=\scriptsize, text=orange, anchor=north west, inner sep=1pt]
            at (axis cs:\bThree,0.495) {$k_3$};

        \draw[<->, thin] (axis cs:\bTwo-0.012,{\gpowC}) -- (axis cs:\bTwo-0.012,{\gpowB});
        \draw[<->, thin] (axis cs:\bThree-0.012,{\gpowB}) -- (axis cs:\bThree-0.012,{\gpowA});
        \draw[<->, thin] (axis cs:\bFour-0.012,{\gpowA}) -- (axis cs:\bFour-0.012,1);

        \end{axis}
    \end{tikzpicture}
    \caption{
    The allocation rule of Lemma~\ref{lem:general-wbb-step-construction},
    drawn to scale, for $n=4$, the grid
    $(\qpoint{0},\ldots,\qpoint{4})=(0,\frac{3}{8},\frac{1}{2},\frac{5}{8},1)$,
    and $\slack=0.002$, as a function of the buyer's value.  {The dashed
    vertical lines are the buyer thresholds
    $\buyerendpoint{i}=\qpoint{i}-\slack$ for $1\le i<n$, each $\buyerendpoint{i}$ lies slightly below its grid point $\qpoint{i}$.}  For this grid,
    Equation~\eqref{eq:general-wbb-grid-condition} gives the growth factor
    $\growth\approx1.254$.  Each curve corresponds to all seller values that
    are rounded to the same index $\sellerround{\vs}$.  Since
    $\growth^{3}\approx1.97<2$, every positive allocation probability is above
    $1/2$ (dashed line).  The levels are not evenly spaced.  The black arrows
    show the jumps of the curve $\sellerround{\vs}=0$ at $\buyerendpoint{2}$,
    $\buyerendpoint{3}$, and $\buyerendpoint{4}$, of sizes approximately
    $0.13$, $0.16$, and $0.20$: each jump is $\growth$ times the previous one.
    The dot labeled $k_i$ marks the
    allocation probability at the profile
    $(\sellerendpoint{i-1},\buyerendpoint{i})$, whose seller and buyer values
    are the two endpoints of the interval {$K_i=[\sellerendpoint{i-1},\buyerendpoint{i}]$}.
    {For example, if the seller's value is $0.2$ and the buyer's value is
    $0.7$, then the seller's value is rounded up to $\sellerendpoint{1}$, so
    $\sellerround{\vs}=1$ (the green curve), and the buyer's value is rounded
    down to $\buyerendpoint{3}$.  The two rounded indices are two grid steps
    apart, so trade happens with probability $\growth^{-2}\approx0.64$.}
    }
\label{fig:wbb-allocation-buyer-view}
\end{figure}

The allocation rule of the mechanism starts by rounding the reports of the traders by the rounding procedure below. Then there is trade with some probability (that is later specified), if the rounded buyer value is {strictly} larger than the rounded seller value.
The seller's report is rounded upward to the nearest seller threshold, and the
buyer's report is rounded downward to the nearest buyer threshold; we denote the
resulting indices by $\sellerround{\vs}$ and $\buyerround{\vb}$, respectively.
Hence, each profile is associated with a pair of grid indices. 
{The
profiles with seller index $r$ and buyer index $t$ form a \emph{rectangle}: the
product of the set of seller values rounded up to $r$ and the set of buyer values
rounded down to $t$. We say that this rectangle is at \emph{level}~$t-r$, that is,
the level of a profile is the number of grid steps by which the buyer's index
exceeds the seller's.}
The mechanism
trades exactly when the buyer's index is strictly larger than the seller's, and
the allocation probability depends only on the difference between the two
indices: increasing the buyer's index by one, or decreasing the seller's index
by one, multiplies the allocation probability by the same factor $\growth > 1$,
which we call the \emph{growth factor}. Thus, for buyer index that is larger than the seller index by $d>0$, the allocation probability is $\growth^{-n+d}$.
See Figure~\ref{fig:wbb-allocation-buyer-view} for an illustration of the
allocation rule.

The resulting allocation rule is nondecreasing in the buyer's report
and nonincreasing in the seller's report. Thus, by Myerson's lemma, there are
payments that make the mechanism DSIC and
IR; we use the normalized payments defined in the proof of
Lemma~\ref{lem:general-wbb-step-construction}. For any large enough growth
factor $\growth$, the resulting mechanism is also WBB. We
choose $\growth>1$ as in Equation~\eqref{eq:general-wbb-grid-condition}, and
we normalize the allocation probabilities so that they are at most $1$.

The growth factor depends on the grid. 
{When we apply the construction in the proofs of
Theorems~\ref{thm:wbb-sbb-unbounded-gap-ind-gft}
and~\ref{thm:wbb-sbb-unbounded-gap-correlated-residual-welfare}, we carefully pick the
grid points so that $\growth^{n-1}<2$, and hence every positive allocation
probability is greater than $1/2$.}


We now return to the intervals $K_1,\ldots,K_n$ and compare the two classes
of mechanisms on them. Fix $i\in[n]$ and consider the profile
$(\sellerendpoint{i-1},\buyerendpoint{i})$, whose values are the endpoints of
$K_i$. By the definition of the rounding,
$\sellerround{\sellerendpoint{i-1}}=i-1$ and
$\buyerround{\buyerendpoint{i}}=i$, so the two rounded indices are exactly one
grid step apart, {that is, the profile is at level~$1$,} and
\begin{align*}
    \allocation(\sellerendpoint{i-1},\buyerendpoint{i})=\growth^{-(n-1)}.
\end{align*}

These are the profiles marked in
Figure~\ref{fig:wbb-allocation-buyer-view}. {When we apply the construction,
we choose the grid so that $\growth^{n-1}<2$. Hence, in our construction, each
of these allocation probabilities is greater than $1/2$, and their sum is
greater than $n/2$, while for every SBB mechanism this
sum is at most $1$.}


{If we only cared about gains from trade or social welfare, which
do not depend on the payments, we could stop here: the difference in the
allocation probabilities at the profiles
$(\sellerendpoint{i-1},\buyerendpoint{i})$ is enough to obtain an unbounded separation between {the gains from trade (for independent distributions) and the social welfare (for correlated distributions) of} WBB and SBB mechanisms. 
{However, at the exact profile $(\sellerendpoint{i-1},\buyerendpoint{i})$ the
payments leave the traders with zero surplus.  For a step allocation rule, the
payments of Myerson's lemma charge each trader the values at which the
trader's allocation probability jumps.  At this profile, the buyer's allocation
probability, as a function of his value, jumps only at $\buyerendpoint{i}$,
from $0$ to $\allocation(\sellerendpoint{i-1},\buyerendpoint{i})$: lowering his
value below $\buyerendpoint{i}$ moves the profile to a rectangle with no
trade.  Similarly, the seller's allocation probability jumps only at
$\sellerendpoint{i-1}$.  Hence, the buyer pays
$\buyerendpoint{i}\cdot\allocation(\sellerendpoint{i-1},\buyerendpoint{i})$ and
the seller receives
$\sellerendpoint{i-1}\cdot\allocation(\sellerendpoint{i-1},\buyerendpoint{i})$.
These payments equal the traders' values times the allocation probability, so
the entire gain from trade is burned.\footnote{At any other profile
$(\vs,\vb)$ in the same rectangle, the payments are the same, so the traders
keep only the rounding difference
$\left((\vb-\buyerendpoint{i})+(\sellerendpoint{i-1}-\vs)\right)\cdot
\allocation(\sellerendpoint{i-1},\buyerendpoint{i})$.  This surplus is small
when $(\vs,\vb)$ is close to $(\sellerendpoint{i-1},\buyerendpoint{i})$, and it
is not enough for our purposes.}}
To obtain the separation for traders'
surplus and traders' utility, we need a more elaborate construction and more nuanced argument.
}

\begin{figure}[!htbp]
    \centering
    \begin{tikzpicture}[x=0.95cm,y=0.95cm, >=stealth]
        \foreach \ox/\tk/\tl/\ttl in {0/1/3/{(a) a rectangle at level $2$},
                                      7.5/1/4/{(b) a rectangle at level $3$}} {
            \begin{scope}[shift={(\ox,0)}]
                \foreach \k in {0,...,4} {
                    \foreach \l in {1,...,5} {
                        \pgfmathtruncatemacro{\d}{\l-\k}
                        \ifnum\d>0
                            \ifnum\d=1
                                \def\fc{yellow!45}
                            \else
                                \pgfmathtruncatemacro{\pc}{20*\d-25}
                                \def\fc{blue!\pc}
                            \fi
                        \else
                            \def\fc{gray!8}
                        \fi
                        \fill[\fc] (\k,\l-1) rectangle (\k+1,\l);
                        \draw[gray!60] (\k,\l-1) rectangle (\k+1,\l);
                    }
                }
                \foreach \i in {1,...,5} {
                    \node[font=\scriptsize] at (\i-0.5,\i-0.5) {$K_{\i}$};
                }
                \node[font=\scriptsize, gray, align=center] at (3.5,1) {no\\trade};
                \foreach \k in {0,...,4} {
                    \node[font=\scriptsize, below] at (\k+0.5,0) {$\k$};
                }
                \foreach \l in {1,...,5} {
                    \node[font=\scriptsize, left] at (0,\l-0.5) {$\l$};
                }
                \node[font=\small] at (2.5,-0.8) {$\sellerround{\vs}$};
                \node[font=\small] at (-1.0,2.5) {$\buyerround{\vb}$};
                \node[font=\small] at (2.5,5.6) {\ttl};
                \pgfmathsetmacro{\tkr}{\tk+1}
                \pgfmathsetmacro{\tlb}{\tl-1}
                \draw[red!70!black, thick, dashed]
                    (\tk,\tk) rectangle (\tkr,\tlb);
                \draw[red!70!black, thick, dashed]
                    (\tkr,\tlb) rectangle (\tl,\tl);
                \draw[red!70!black, very thick] (\tk,\tlb) rectangle (\tkr,\tl);
                \draw[->, very thick, red!70!black]
                    (\tk+0.8,\tk+0.1) -- (\tk+0.8,\tl-0.8);
                \draw[->, very thick, red!70!black]
                    (\tl-0.1,\tl-0.8) -- (\tk+0.8,\tl-0.8);
            \end{scope}
        }
    \end{tikzpicture}
    \caption{
    Dependence between the levels of the allocation rule, for $n=5$. Each
    cell is the rectangle of profiles with seller index $\sellerround{\vs}$
    (horizontal) and buyer index $\buyerround{\vb}$ (vertical); cells are not
    drawn to scale. Gray cells have no trade. The cells on the diagonal are at
    level~$1$ (yellow): the cell labeled $K_i$ contains the profile
    $(\sellerendpoint{i-1},\buyerendpoint{i})$. Darker cells are at higher
    levels. WBB at the cell with the solid red border involves
    the cells below it in its column and to its right in its row,
    down to level~$1$ (dashed),
    which correspond to the two integrals. (a) The level-$2$ cell above $K_2$
    and $K_3$ is constrained by these two level-$1$ cells. (b) A level-$3$ cell
    is constrained by two cells at level~$1$ and two cells at level~$2$. Hence,
    the allocation rule is determined level by level, starting from
    level~$1$.
    }
    \label{fig:wbb-level-dependence}
\end{figure}
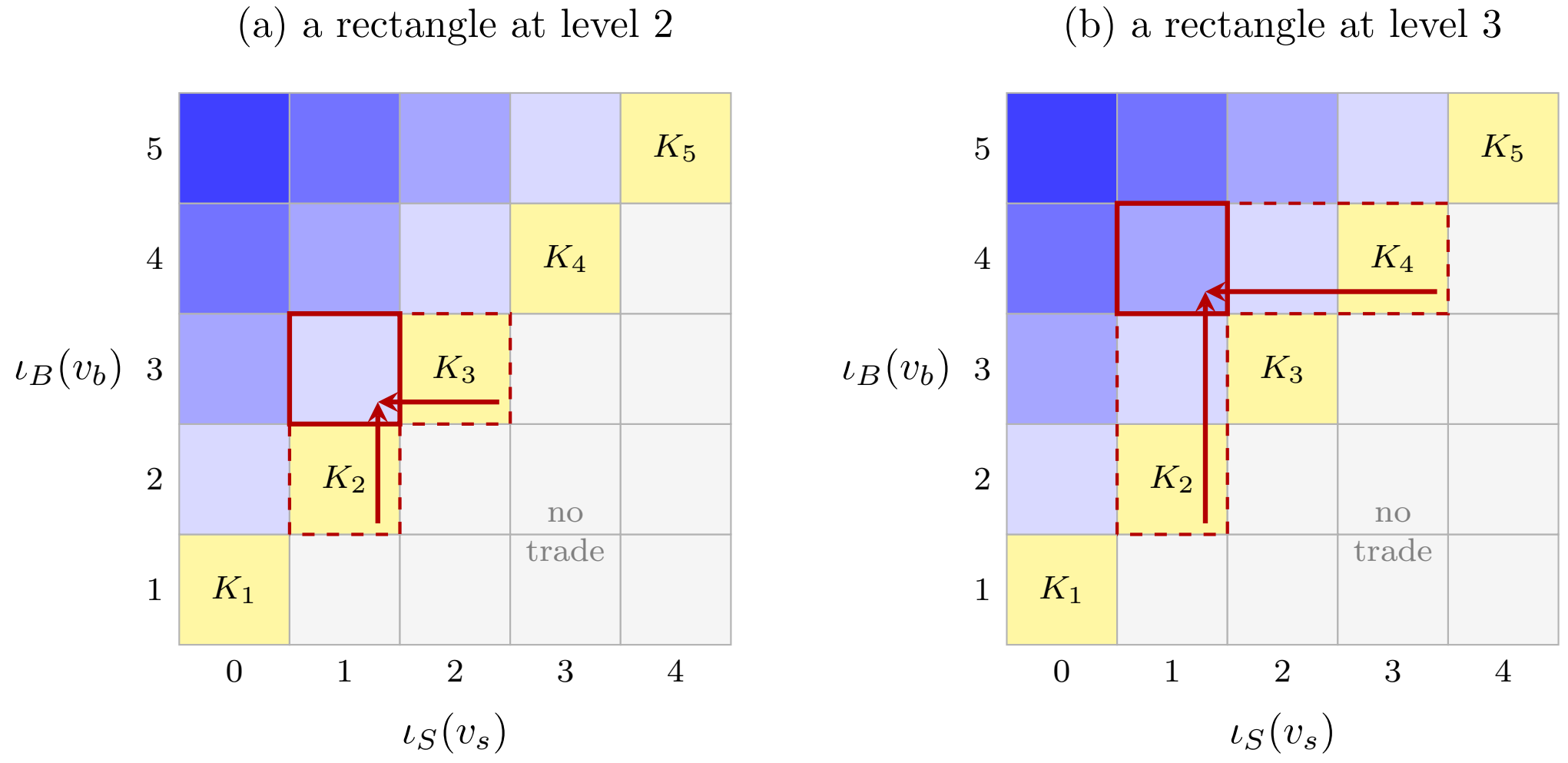

\paragraph{Constructing the allocation rule level by level.}
Once the allocation probabilities at the rectangles of level~$1$,
which contain the profiles $(\sellerendpoint{i-1},\buyerendpoint{i})$, are
fixed, WBB and DSIC
(through monotonicity) impose restrictions on the allocation probabilities at
the rectangles of higher levels. Specifically, the allocation probability at
each rectangle is tied to the allocation probabilities at the rectangles with
the same seller values and lower buyer values (below it in the same column)
and at the rectangles with the same buyer values and higher seller values (to
its right in the same row; see Figure~\ref{fig:wbb-level-dependence}), all of
which are at lower levels. We
therefore construct the allocation rule from the bottom up, one level at a
time. Along the way, we show how the construction leaves the traders a
positive surplus at levels~$2$ and higher.

To see where the dependence comes from, recall that with the payments given by
Myerson's lemma, WBB at a profile $(\vs,\vb)$ with positive
allocation probability is equivalent to
\begin{align*}
    (\vb-\vs)\,\allocation(\vs,\vb)
    \ge
    \int_{\vs}^{\vb} \allocation(\vs,z)\d z
    +
    \int_{\vs}^{\vb} \allocation(z,\vb)\d z
\end{align*}
(see the proof of Lemma~\ref{lem:general-wbb-step-construction}). The first
integral is over buyer values between $\vs$ and $\vb$ with the seller value fixed,
and the second is over seller values between $\vs$ and $\vb$ with the buyer value
fixed. Apart from the rectangle of $(\vs,\vb)$ itself, both integrals only involve
rectangles at lower levels, or profiles with zero allocation probability.
Moreover, the right-hand side is exactly the traders' surplus at
$(\vs,\vb)$: the first integral is the buyer's utility and the second is the
seller's surplus (the allocation probability is zero for buyer values below
$\vs$ and for seller values above $\vb$). The left-hand side is the gain from
trade realized at $(\vs,\vb)$. Thus, WBB says that the traders'
surplus is at most the realized gain from trade, and the difference is the
money burned.

\emph{Level $1$.} Consider the rectangle with seller index $i-1$ and buyer
index $i$, which contains the profile
$(\sellerendpoint{i-1},\buyerendpoint{i})$. All the other profiles in the two
integrals have zero allocation probability, and the inequality reduces to
$\buyerendpoint{i}\ge\sellerendpoint{i-1}$, which holds since $K_i$ is
nonempty. Hence, the allocation probabilities at level~$1$ can be chosen
freely, and we set all of them to the same value. At the profile
$(\sellerendpoint{i-1},\buyerendpoint{i})$ both integrals are zero, so, as
noted above, the traders' surplus there is zero.

\emph{Level $2$.} Consider the rectangle with seller index $i-1$ and buyer
index $i+1$, which lies above the level-$1$ rectangles of $K_i$ and $K_{i+1}$
(see Figure~\ref{fig:wbb-level-dependence}(a)). Let $x_i$ and $x_{i+1}$ be the
allocation probabilities at these two level-$1$ rectangles, and let $y$ be the
allocation probability at the level-$2$ rectangle. Computing the two integrals,
WBB at this rectangle becomes
\begin{align*}
    (\buyerendpoint{i+1}-\sellerendpoint{i-1})\,y
    \ge
    (\buyerendpoint{i+1}-\buyerendpoint{i})\,x_i
    +
    (\sellerendpoint{i}-\sellerendpoint{i-1})\,x_{i+1}.
\end{align*}
Thus, $y$ is bounded from below by a combination of $x_i$ and $x_{i+1}$ whose
weights sum to
\begin{align*}
    \frac{(\buyerendpoint{i+1}-\buyerendpoint{i})+(\sellerendpoint{i}-\sellerendpoint{i-1})}
         {\buyerendpoint{i+1}-\sellerendpoint{i-1}}
    =
    1+\frac{\sellerendpoint{i}-\buyerendpoint{i}}
           {\buyerendpoint{i+1}-\sellerendpoint{i-1}}
    >1.
\end{align*}
In particular, if $x_i=x_{i+1}$, the allocation probability must grow by a
factor of at least
$1+(\sellerendpoint{i}-\buyerendpoint{i})/(\buyerendpoint{i+1}-\sellerendpoint{i-1})$
from level~$1$ to level~$2$. In addition, monotonicity requires
$y\ge\max\{x_i,x_{i+1}\}$.
The right-hand side is the traders' surplus at the profile
$(\sellerendpoint{i-1},\buyerendpoint{i+1})$: the buyer's utility comes from
the level-$1$ rectangle of $K_i$, and the seller's utility comes from the
level-$1$ rectangle of $K_{i+1}$. Unlike at level~$1$, this surplus is
positive. If $x_i=x_{i+1}=x$ and $y=\growth x$, the traders' surplus is at least
$(\buyerendpoint{i+1}-\sellerendpoint{i-1})\,x$, so the traders keep at least
a $1/\growth$ fraction of the realized gain
$(\buyerendpoint{i+1}-\sellerendpoint{i-1})\,y$, and only the rest is burned.

\emph{Higher levels.} In general, the rectangle with seller index $i-1$ and
buyer index $j$ is constrained by
the rectangles below it in its column and to its right in its
row, down to level~$1$ (see Figure~\ref{fig:wbb-level-dependence}): those with
seller index $i-1$ and buyer indices $i,\ldots,j-1$, and those with buyer index
$j$ and seller indices $i,\ldots,j-1$. All of these are at lower levels. Hence,
once the allocation probabilities at levels $1,\ldots,d-1$ are fixed, weak
budget balance and monotonicity give lower bounds on the allocation
probabilities at level~$d$, and we can continue to the next level.

Our construction makes the simplest choice at every level: all rectangles at
the same level have the same allocation probability, and moving one level up
multiplies it by the growth factor $\growth$. Under this choice, the weak
budget balance constraint at level $d$ becomes
inequality~\eqref{eq:general-wbb-reduced-budget} in the proof of
Lemma~\ref{lem:general-wbb-step-construction}, and Equation~\eqref{eq:general-wbb-grid-condition}
chooses a $\growth$ that satisfies these constraints at all levels
simultaneously. Finally, we divide by $\growth^{n-1}$ so that the allocation
probability at the top level is $1$.

As at level~$2$, the traders' surplus is positive at every level
$d\ge2$. At the profile
$(\sellerendpoint{i-1},\buyerendpoint{j})$ with $j\ge i+1$, the buyer's
integral covers $[\buyerendpoint{i},\buyerendpoint{j}]$ and the seller's
integral covers $[\sellerendpoint{i-1},\sellerendpoint{j-1}]$, and the
allocation probability on both is at least $\growth^{-(n-1)}$. Since the total
length of these two intervals is at least
$\buyerendpoint{j}-\sellerendpoint{i-1}$, the traders' surplus is at least
$\growth^{-(n-1)}(\buyerendpoint{j}-\sellerendpoint{i-1})$, which is more than
half of the gain from trade when $\growth^{n-1}<2$. This is the surplus that
our separation results use: in both proofs, the lower bound on the traders' surplus of our mechanism comes from
profiles at level~$2$. 

{For correlated values, it is straightforward to construct a distribution under which most of the first-best gains-from-trade comes from these profiles. We draw an index $i$, and then a seller value slightly above $\qpoint{i}$ and a buyer value slightly below $\qpoint{i+2}$. Since Theorem~\ref{thm:wbb-sbb-unbounded-gap-correlated-residual-welfare} is about the traders' utility, we also make sure that the seller's expected value is negligible compared to the first-best gains-from-trade. For independent values, this is more challenging, since a seller value near $\qpoint{2\ell-1}$ can appear together with a buyer value near $\qpoint{2r+1}$ for any $r$. To circumvent this challenge, we make the seller's value more likely to be near high grid points, and the buyer's value more likely to be near low grid points{, with probabilities that grow (for the seller) and shrink (for the buyer) exponentially along the grid}. If $r<\ell$, the buyer's value is below the seller's value, so there is no gain from trade. If $r>\ell$, there is a gain from trade, but either the seller's value is near an unlikely low point or the buyer's value is near an unlikely high point. {Figure~\ref{fig:ind-heatmap-probability} shows the probabilities of the resulting independent distribution, and Figure~\ref{fig:ind-heatmap-gft} shows that almost all of its first-best gains-from-trade comes from the level-$2$ profiles.} We give the formal distributions in Sections~\ref{sec:proof:wbb-sbb-unbounded-gap-ind-gft} and~\ref{sec:proof-wbb-sbb-seperation-welfare}.}

\begin{lemma}[weak-budget-balance construction]
\label{lem:general-wbb-step-construction}
Let the value spaces be $\supps=\suppb=[0,1]$. Given an integer $n\ge2$ and a sequence 
\begin{align*}
    0=\qpoint{0}<\qpoint{1}<\cdots<\qpoint{n}=1,
\end{align*}
set $\slack$ such that
\begin{align}\label{eq:general-wbb-slack-condition}
    0<\slack<\frac12\min_{i\in[n]}(\qpoint{i}-\qpoint{i-1}).
\end{align}
Let the seller and buyer thresholds be
\begin{align*}
    \sellerendpoint{0}&=0,
    &\sellerendpoint{i}&=\qpoint{i}+\slack &&(1\le i\le n-1),\\
    \sellerendpoint{n}&=1,\\
    \buyerendpoint{0}&=0,
    &\buyerendpoint{i}&=\qpoint{i}-\slack &&(1\le i\le n-1),\\
    \buyerendpoint{n}&=1,
\end{align*}
set the growth factor $\growth$ to
\begin{align}\label{eq:general-wbb-grid-condition}
    \growth
    =1+\max_{1\le i<j\le n}
    \frac{
        \qpoint{j-1}-\qpoint{i}+2\slack
    }{
        \qpoint{j}-\qpoint{i-1}
        -\slack\cdot\mathbbm{1}_{\{i>1\}}
        -\slack\cdot\mathbbm{1}_{\{j<n\}}
    }.
\end{align}

Then there exists a dominant-strategy incentive-compatible, individually rational, and weakly budget-balanced mechanism whose allocation rule is
\begin{align}\label{eq:general-wbb-step-allocation}
    \allocation(\vs,\vb)
    =
    \begin{cases}
        0,
        &\sellerround{\vs}\ge\buyerround{\vb},\\[1mm]
        \displaystyle
        \frac{\growth^{\buyerround{\vb}-\sellerround{\vs}-1}}
             {\growth^{n-1}},
        &\sellerround{\vs}<\buyerround{\vb},
    \end{cases}
\end{align}
where
\begin{align*}
    \sellerround{\vs}
    &=\min\{r\in\{0,\ldots,n\}:\vs\le \sellerendpoint{r}\},\\
    \buyerround{\vb}
    &=\max\{r\in\{0,\ldots,n\}:\vb\ge \buyerendpoint{r}\}.
\end{align*}
In particular, whenever the buyer's rounded index $\buyerround{\vb}$ is strictly larger than the seller's rounded index $\sellerround{\vs}$, the allocation probability is $\allocation(\vs,\vb)\ge\growth^{-(n-1)}$.
\end{lemma}

The proof of Lemma~\ref{lem:general-wbb-step-construction} formalizes the
discussion above, and is deferred to
Appendix~\ref{app:proof-wbb-construction}.

We next present our main lemma,
Lemma~\ref{lem:separation}, that is used in the proofs of both
Theorems~\ref{thm:wbb-sbb-unbounded-gap-ind-gft}
and~\ref{thm:wbb-sbb-unbounded-gap-correlated-residual-welfare}.
Definition~\ref{def:separation-grid} below states the conditions on the grid and the slack that 
{the lemma assumes}, and
{the first step in the proof of the lemma shows that} for such a grid every
positive allocation probability of the mechanism of
Lemma~\ref{lem:general-wbb-step-construction} is greater than $\frac{1}{2}$.
{The second step uses it to compare}, at a profile whose
values are two grid steps apart, our mechanism with an SBB
mechanism.  Our mechanism leaves the traders at least half of the gain from
trade, while an SBB mechanism, which posts a random price,
leaves them at most the probability that the price falls between the two grid
points times the gain.  Since every price falls between only a few such
pairs of grid points, an SBB mechanism does well on only a
few of these profiles, while our mechanism does well on all of them.
Lemma~\ref{lem:separation} uses this to obtain a gap for every distribution of
the agents' values under which most of the first-best gains-from-trade comes
from these profiles.  Thus, to prove
Theorems~\ref{thm:wbb-sbb-unbounded-gap-ind-gft}
and~\ref{thm:wbb-sbb-unbounded-gap-correlated-residual-welfare}, we need to
construct a grid and a distribution of the agents' values that satisfy the
{conditions of this lemma.}  For
Theorem~\ref{thm:wbb-sbb-unbounded-gap-ind-gft}, the seller's and buyer's
values must be independent, which makes it harder to concentrate the
probability on the profiles two grid steps apart.  For
Theorem~\ref{thm:wbb-sbb-unbounded-gap-correlated-residual-welfare}, the values
may be correlated, but since the theorem compares the traders' utility, the
expected seller's value must also be small.

{Lemma~\ref{lem:separation} below assumes} the following conditions on the grid and the slack, and we refer to a grid with slack $\slack$ that satisfies them as a \emph{separation grid}.


\begin{definition}[Separation grid]\label{def:separation-grid}
For $n\geq 4$, fix a grid $0=\qpoint{0}<\cdots<\qpoint{n}=1$, and let $L=\min_{r\in[n]}(\qpoint{r}-\qpoint{r-1})$ be the smallest gap of the grid.
We say that this grid is a \emph{separation grid} 
with slack $\slack$ if there exists $0<\separation\le \frac{1}{100\cdot n}$ such that  $0<\slack\le\frac{\separation\cdot L}{100}$, and it holds that
    \begin{align}\label{eq:grid-condition}
        \qpoint{j-1}-\qpoint{i}\le 4\separation\,(\qpoint{j}-\qpoint{i-1})
        \qquad\text{for every } 1\le i<j\le n.
    \end{align}
\end{definition}
The separation in our result grows with $n$, so we next always assume that $n\geq 4$. For every $1\le i\le n-3$, let $\levelrect{i}$ be the set of profiles $(\vs,\vb)$ in which the seller's value is slightly above $\qpoint{i}$ and the buyer's value is slightly below $\qpoint{i+2}$, that is, $\qpoint{i}\le\vs<\qpoint{i}+\slack$ and $\qpoint{i+2}-\slack<\vb\le\qpoint{i+2}$.
{The following lemma gives the separation.}  Suppose that the distribution of the agents' values is concentrated on
profiles two grid steps apart, that is, on profiles in the sets $\levelrect{i}$, so that each such pair of grid points
contributes the same amount to the first-best gains-from-trade, and the
contribution of all other profiles is arbitrarily small.  That is, most of the
first-best gains-from-trade comes from these profiles.  Then our mechanism obtains much more traders'
surplus than every SBB mechanism.


\begin{lemma}\label{lem:separation} 
Let $0=\qpoint{0}<\cdots<\qpoint{n}=1$ be a separation grid with slack $\slack$, and let $\wbbmechanism$ be the mechanism of Lemma~\ref{lem:general-wbb-step-construction} for this grid and slack. Let $I$ be a nonempty subset of indices $I \subseteq \{ 1, \dots, n-3 \}$, of size $m$, and let $\distjoint$ be a distribution over the agents' values. 
Suppose there exist $\surplusunit>0$ and $\leftover\ge0$ such that the following two conditions hold:
\begin{enumerate}
    \item\label{item:separation-inside} For every $i \in I$, the contribution to the first-best gains-from-trade of the profiles in $\levelrect{i}$ is between $\surplusunit/2$ and $\surplusunit$. That is, 
    \[
    \frac{\surplusunit}{2}\le\Exp[(\vb-\vs)_{+}\cdot\indicator_{(\vs, \vb)\in \levelrect{i}}]\le\surplusunit.
    \]
    \item\label{item:separation-outside} The contribution to the first-best gains-from-trade of the profiles outside $\bigcup_{i \in I}\levelrect{i}$ is at most $\leftover$. That is, 
    \[
    \Exp[(\vb-\vs)_{+}\cdot\indicator_{(\vs, \vb)\notin \bigcup_{i\in I}\levelrect{i}}] \leq \leftover.
    \]
\end{enumerate}

Here $(\vb-\vs)_{+}=\max\{\vb-\vs,0\}$ is the first-best gain from trade at $(\vs,\vb)$.
\
Let $h$ be the maximal number of intervals $[\qpoint{i},\qpoint{i+2}]$ for $i\in I$ that contain a common price. Then $\wbbmechanism$ satisfies
\begin{align*} \sgft_{\wbbmechanism}(\distjoint)>\frac{m\,\cdot \surplusunit}{4}, \end{align*}
and every dominant-strategy incentive-compatible, individually rational, and strongly budget-balanced mechanism $\sbbmechanism$ satisfies 
\begin{align*} \sgft_{\sbbmechanism}(\distjoint)\le\overlap\,\cdot \surplusunit+\leftover. \end{align*}

\end{lemma}

Note that each of the intervals $[\qpoint{i},\qpoint{i+2}]$ for $i\in I$ contains only three grid points, so $\overlap\le 3$ regardless of $n$, and depending on $I$ it can be smaller.
Hence, if $\leftover\le\surplusunit$,
then the 
{ratio} between the traders' surplus of $\wbbmechanism$ and the traders'
surplus of every SBB mechanism becomes unbounded as $m$
grows.

{The proof of Lemma~\ref{lem:separation} is deferred to Appendix~\ref{app:proof-separation}.} {Here we only give a high-level description of the proof.}
{Fix an SBB DSIC and IR mechanism $\sbbmechanism$. The proof compares the traders' surplus of $\wbbmechanism$ and of $\sbbmechanism$ at the profiles in the sets $\levelrect{i}$ for $i\in I$. By Condition~\ref{item:separation-outside} of the lemma, the profiles outside these sets contribute little to the first-best gains-from-trade, and hence to the traders' surplus of every mechanism, so we can focus on the profiles in these sets. We first show that the allocation probability of $\wbbmechanism$ is greater than $\frac{1}{2}$ at every profile in which the rounded value of the seller is below the rounded value of the buyer. We then use Myerson's lemma to show that although $\wbbmechanism$ burns money, its traders' surplus at every profile in $\levelrect{i}$ is more than half of the first-best gain from trade. For $\sbbmechanism$, we use the characterization of~\cite{HR87,CP16}, by which $\sbbmechanism$ posts a price drawn from some distribution $\pricemeasure$, to show that it trades at a profile in $\levelrect{i}$ with probability at most $\pricemeasure([\qpoint{i},\qpoint{i+2}])$. Since every price belongs to at most $\overlap$ of the intervals $[\qpoint{i},\qpoint{i+2}]$ for $i\in I$, these probabilities sum to at most $\overlap$. By Condition~\ref{item:separation-inside}, each of the sets $\levelrect{i}$ contributes roughly the same amount to the first-best gains-from-trade. We conclude that $\wbbmechanism$ obtains a constant fraction of the contribution of each of the $m$ sets, while from all the sets together $\sbbmechanism$ obtains at most the contribution of about $\overlap$ of them
. Finally, the lemma is about the traders' surplus, but Theorem~\ref{thm:wbb-sbb-unbounded-gap-correlated-residual-welfare} is about the traders' utility, which is the traders' surplus plus the seller's expected value $\Exp[\vs]$. In its proof (Section~\ref{sec:proof-wbb-sbb-seperation-welfare}), we construct a distribution that satisfies the conditions of the lemma and under which the seller's expected value is at most $\surplusunit$, so the separation in traders' surplus implies a separation in traders' utility.}

\subsection{Proof of Theorem~\ref{thm:wbb-sbb-unbounded-gap-ind-gft}}\label{sec:proof:wbb-sbb-unbounded-gap-ind-gft}


We prove the theorem by applying Lemma~\ref{lem:separation} to a specific grid
and specific independent distributions.  Throughout this subsection, fix
$\varepsilon>0$, let $k\ge2$ be an integer such that $(k-1)/12\ge1/\varepsilon$,
and let $n=2k$.

\paragraph{The grid.}
Let $\separation=1/(100n)$.  We build the grid symmetrically around $1/2$:
the middle point is $\qpoint{k}=1/2$, the points on the left approach $1/2$
with distances $\separation,\separation^2,\ldots,\separation^{k-1}$, and the
points on the right are their reflections around $1/2$.  Formally,
\begin{align}\label{eq:ind-grid}
    \qpoint{i}= \begin{cases}
        0 & i=0;\\
        \frac{1}{2}-\separation^i & 1\le i<k;\\
        \frac{1}{2} & i=k;\\
        \frac{1}{2}+\separation^{2k-i} & k<i\le 2k-1;\\
        1 & i=n.
    \end{cases}
\end{align}
Thus, the gaps between consecutive points decrease geometrically when moving
from either endpoint toward the middle.  For every $i\in[n]$, let
$\qlength{i}=\qpoint{i}-\qpoint{i-1}$ denote the gap between
$\qpoint{i-1}$ and $\qpoint{i}$.  Then
\begin{align*}
    \qlength{i}= \begin{cases}
        \frac{1}{2}-\separation & i=1 \text{ or } i=n;\\
        (1-\separation)\separation^{i-1} & 2\le i\le k-1;\\
        \separation^{k-1} & i=k \text{ or } i=k+1;\\
        (1-\separation)\separation^{n-i} & k+2\le i\le n-1.
    \end{cases}
\end{align*}
Let $\slack>0$ be small enough for the grid with slack $\slack$ to be a separation grid
(this is possible by Lemma~\ref{lem:ind-grid} below), and let $\wbbmechanism$
be the resulting mechanism of Lemma~\ref{lem:general-wbb-step-construction}.

\paragraph{The distributions.}
The seller's value is concentrated slightly above the grid points
$\qpoint{1},\qpoint{3},\ldots,\qpoint{n-3}$, and the buyer's value is
concentrated slightly below the grid points
$\qpoint{3},\qpoint{5},\ldots,\qpoint{n-1}$.  For every $\ell\in[k-1]$, a
profile with a seller value near $\qpoint{2\ell-1}$ and a buyer value near $\qpoint{2\ell+1}$
lies in $\levelrect{2\ell-1}$
.  Ideally, the seller's value would be near $\qpoint{2\ell-1}$ exactly when
the buyer's value is near $\qpoint{2\ell+1}$, as in a correlated distribution.
Since the values are independent, there are also profiles in which the
seller's value is near $\qpoint{2\ell-1}$ and the buyer's value is near
$\qpoint{2r+1}$ for some $r\neq\ell$.  We choose the probabilities so that these cases contribute
little.  If $r<\ell$, the buyer's value is below the seller's value, so the
first-best gains-from-trade are $0$.  If $r>\ell$, the gains from trade are
positive, but this case is unlikely: the seller's value is more likely to be
near high grid points, and the buyer's value is more likely to be near low grid
points, so either the seller's value is near an unlikely low point or the
buyer's value is near an unlikely high point.  Thus, although the values are
independent, positive gains from trade arise, up to a small error, only when
the seller's value is near $\qpoint{2\ell-1}$ and the buyer's value is near
$\qpoint{2\ell+1}$ for the same $\ell$, that is, in $\levelrect{2\ell-1}$.  Finally, we choose the probabilities
so that each $\ell$ contributes approximately the same amount to the
first-best gains-from-trade: the probability of $\levelrect{2\ell-1}$ is inversely
proportional to $\qpoint{2\ell+1}-\qpoint{2\ell-1}$, which is approximately the
gain from trade in this case.
{Figure~\ref{fig:ind-heatmap-probability} shows the probabilities of the resulting distribution: the seller's value is likely to be near a high grid point, and the buyer's value is likely to be near a low grid point. Figure~\ref{fig:ind-heatmap-gft} shows the contribution of each pair of seller and buyer values to the first-best gains-from-trade: the pairs in which the buyer's value is below the seller's value contribute nothing, each level-$2$ pair contributes about the same amount, and all other pairs contribute little.}

Formally, fix $0<\eta<\slack$.  For every $\ell\in[k-1]$, let
\begin{align*}
    \sellerinterval{\ell}
    =[\qpoint{2\ell-1},\qpoint{2\ell-1}+\eta],
    \qquad
    \buyerinterval{\ell}
    =[\qpoint{2\ell+1}-\eta,\qpoint{2\ell+1}],
    \qquad
    \profilegap{\ell}
    =\qpoint{2\ell+1}-\qpoint{2\ell-1}.
\end{align*}
Let $R>1$ be large enough that
\begin{align}\label{eq:ind-choice-of-R}
    \sum_{1\le\ell<r\le k-1}
    R^{\ell-r}\,
    \frac{\qpoint{2r+1}-\qpoint{2\ell-1}}
         {\sqrt{\profilegap{\ell}\profilegap{r}}}
    \le 1;
\end{align}
such $R$ exists since the sum is finite and each term tends to $0$ as
$R\to\infty$.  For every $\ell\in[k-1]$, let
\begin{align*}
    \sellerweight{\ell}=\frac{R^\ell}{\sqrt{\profilegap{\ell}}},
    \qquad
    \buyerweight{\ell}=\frac{R^{-\ell}}{\sqrt{\profilegap{\ell}}},
    \qquad
    \surplusunit 
    =
    \frac{1}{\left(\sum_r \sellerweight{r}\right)
    \left(\sum_r \buyerweight{r}\right)}.
\end{align*}
To draw the seller's value, we first draw an index $\ell\in[k-1]$ with
probability $\sellerweight{\ell}/\sum_r \sellerweight{r}$, and then draw the
value uniformly from $\sellerinterval{\ell}$.  Similarly, to draw the buyer's
value, we first draw an index $\ell\in[k-1]$ with probability
$\buyerweight{\ell}/\sum_r \buyerweight{r}$, and then draw the value uniformly
from $\buyerinterval{\ell}$.  The two values are drawn independently.  Let
$\distseller$ and $\distbuyer$ denote the resulting seller and buyer
distributions.  Note that, for $(\vs,\vb)\sim\distseller\times\distbuyer$ and every
$\ell,r\in[k-1]$,
\begin{align*}
    \Pr\left[(\vs,\vb)\in\sellerinterval{\ell}\times\buyerinterval{r}\right]
    =\surplusunit\cdot\sellerweight{\ell}\cdot\buyerweight{r}.
\end{align*}
In particular, for $r=\ell\in[k-1]$,
$\Pr\left[(\vs,\vb)\in\sellerinterval{\ell}\times\buyerinterval{\ell}\right]=\surplusunit/\profilegap{\ell}$.
\pgfplotsset{
    colormap={bluered}{rgb255=(49,54,149) rgb255=(116,173,209) rgb255=(224,243,248) rgb255=(254,224,144) rgb255=(244,109,67) rgb255=(165,0,38)},
    support axis/.style={
        width=7cm, height=7cm,
        xmin=0, xmax=1, ymin=0, ymax=1,
        axis on top,
        xtick={0,0.0833,0.25,0.4167,0.5833,0.75,0.9167,1},
        xticklabels={$0$,$\qpoint{1}$,$\qpoint{3}$,$\qpoint{5}$,$\qpoint{7}$,$\qpoint{9}$,$\qpoint{11}$,$1$},
        ytick={0,0.0833,0.25,0.4167,0.5833,0.75,0.9167,1},
        yticklabels={$0$,$\qpoint{1}$,$\qpoint{3}$,$\qpoint{5}$,$\qpoint{7}$,$\qpoint{9}$,$\qpoint{11}$,$1$},
        xlabel={buyer's value $\vb$},
        ylabel={seller's value $\vs$},
        grid=major,
        grid style={gray!20},
        colormap name=bluered,
        colorbar,
        colorbar shift/.style={xshift=0.7cm},
        colorbar style={width=0.25cm},
        scatter/use mapped color={draw=mapped color, fill=mapped color},
    },
    support marks/.style={scatter, only marks, mark=square*, mark size=3.2pt, point meta=explicit},
}

\begin{figure}[!t]
\centering
\begin{tikzpicture}
\begin{axis}[support axis, point meta min=-40, point meta max=0,
    colorbar style={ytick={-40,-30,-20,-10,0}, yticklabels={$10^{-40}$,$10^{-30}$,$10^{-20}$,$10^{-10}$,$1$}, ylabel={probability}}]
\addplot[gray, dashed, domain=0:1] {x};
\addplot[support marks] table[meta=c] {
x y c
0.238 0.095 -20.01
0.238 0.262 -11.93
0.238 0.428 -4.00
0.238 0.595 -1.93
0.238 0.762 -0.01
0.405 0.095 -21.93
0.405 0.262 -13.85
0.405 0.428 -5.92
0.405 0.595 -3.85
0.405 0.762 -1.93
0.572 0.095 -24.00
0.572 0.262 -15.92
0.572 0.428 -7.99
0.572 0.595 -5.92
0.572 0.762 -4.00
0.738 0.095 -31.93
0.738 0.262 -23.85
0.738 0.428 -15.92
0.738 0.595 -13.85
0.738 0.762 -11.93
0.905 0.095 -40.01
0.905 0.262 -31.93
0.905 0.428 -24.00
0.905 0.595 -21.93
0.905 0.762 -20.01
};
\end{axis}
\end{tikzpicture}
\caption{The support of the independent distribution
$\distseller\times\distbuyer$ of
Section~\ref{sec:proof:wbb-sbb-unbounded-gap-ind-gft}, for $k=6$ (so $n=12$)
and $R=10^5$.  The seller's value is slightly above one of
$\qpoint{1},\qpoint{3},\ldots,\qpoint{9}$, and the buyer's value is slightly
below one of $\qpoint{3},\qpoint{5},\ldots,\qpoint{11}$, so the values lie in
the small squares $\sellerinterval{\ell}\times\buyerinterval{r}$, and all
other profiles have probability~$0$.  The color of each square shows its
probability, from blue (low) to red (high).  The probabilities range from
about $10^{-40}$ to almost $1$, so the color scale is logarithmic.  The
seller's value is likely to be near a high grid point and the buyer's value is
likely to be near a low grid point, so the most likely squares are at the
top left, above the dashed diagonal $\vb=\vs$, where $\vb<\vs$ and there is no gain from trade (see Figure~\ref{fig:ind-heatmap-gft}).
  The grid points are drawn evenly spaced, not
to scale.}
\label{fig:ind-heatmap-probability}
\end{figure}
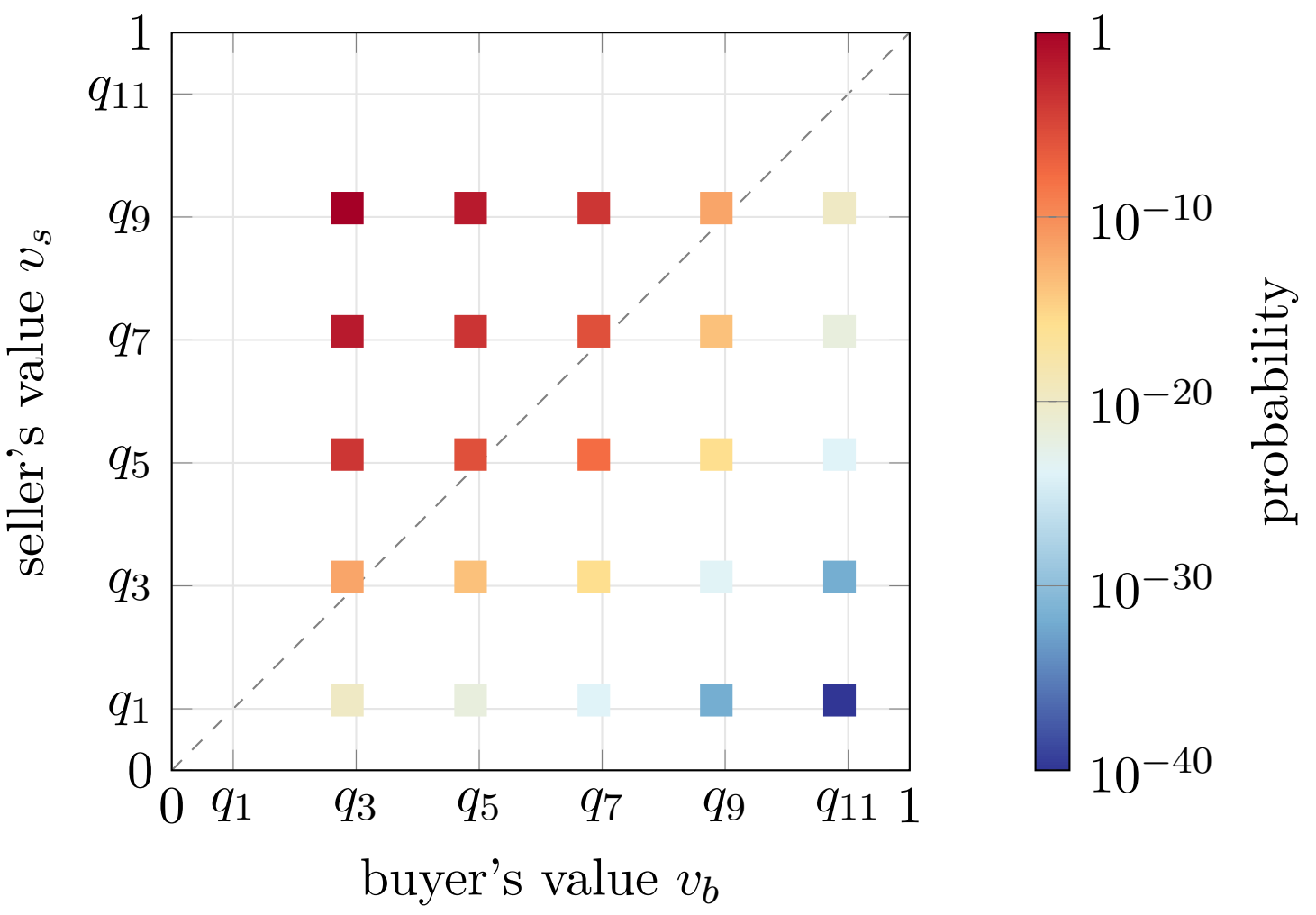

\begin{figure}[!t]
\centering
\begin{tikzpicture}
\begin{axis}[support axis, point meta min=0, point meta max=1,
    colorbar style={ytick={0,0.25,0.5,0.75,1}, ylabel={normalized contribution}}]
\fill[gray!12] (axis cs:0,0) -- (axis cs:1,1) -- (axis cs:0,1) -- cycle;
\addplot[gray, dashed, domain=0:1] {x};
\node[gray, font=\footnotesize, align=center] at (axis cs:0.22,0.88) {no gain\\from trade};
\addplot[only marks, mark=square, mark size=3.2pt, gray] coordinates {
(0.238,0.262) (0.238,0.428) (0.238,0.595) (0.238,0.762)
(0.405,0.428) (0.405,0.595) (0.405,0.762)
(0.572,0.595) (0.572,0.762)
(0.738,0.762)
};
\addplot[support marks] table[meta=c] {
x y c
0.238 0.095 1
0.405 0.095 0.012
0.405 0.262 1
0.572 0.095 0.0001
0.572 0.262 0.0085
0.572 0.428 1
0.738 0.095 0
0.738 0.262 0
0.738 0.428 0.0085
0.738 0.595 1
0.905 0.095 0
0.905 0.262 0
0.905 0.428 0.0001
0.905 0.595 0.012
0.905 0.762 1
};
\end{axis}
\end{tikzpicture}
\caption{The contribution of each square of
Figure~\ref{fig:ind-heatmap-probability} to the first-best gains-from-trade,
that is,
$\Exp[(\vb-\vs)_{+}\cdot\indicator_{(\vs,\vb)\in\sellerinterval{\ell}\times\buyerinterval{r}}]$.
The color of each square shows its normalized contribution, that is, its
contribution divided by $\surplusunit$, so a value of $1$ means a contribution
of $\surplusunit$.  The squares above the dashed diagonal $\vb=\vs$, drawn as
empty gray squares, have $\vb<\vs$ and contribute nothing.  The squares next to the diagonal, with the
seller's value near $\qpoint{2\ell-1}$ and the buyer's value near
$\qpoint{2\ell+1}$, are the sets $\levelrect{2\ell-1}$ at level~$2$, and each
of them contributes about $\surplusunit$.  The other squares below the diagonal
have gains from trade, but they are unlikely
{. We choose the probabilities (Equation~\eqref{eq:ind-choice-of-R}) so that together they contribute at most $\surplusunit$, about as much as a single set $\levelrect{2\ell-1}$.}  As in
Figure~\ref{fig:ind-heatmap-probability}, the grid points are drawn evenly
spaced, not to scale.}
\label{fig:ind-heatmap-gft}
\end{figure}

\paragraph{Analysis.}
We first show that the grid in Equation~\eqref{eq:ind-grid} satisfies the grid condition of Definition~\ref{def:separation-grid}.

\begin{lemma}\label{lem:ind-grid}
The grid in Equation~\eqref{eq:ind-grid} satisfies the grid condition of Definition~\ref{def:separation-grid},
that is, Equation~\eqref{eq:grid-condition}.
\end{lemma}

The proof of Lemma~\ref{lem:ind-grid} is deferred to
Appendix~\ref{app:proof-ind-grid}.
\begin{proof}[Proof of Theorem~\ref{thm:wbb-sbb-unbounded-gap-ind-gft}]
Let $\distseller$, $\distbuyer$, and $\wbbmechanism$ be as above.  We apply
Lemma~\ref{lem:separation} with $I=\{2\ell-1:\ell\in[k-1]\}$, so $m=k-1$.  By
Lemma~\ref{lem:ind-grid}, the grid with slack $\slack$ is a separation grid.  Consecutive intervals
$[\qpoint{2\ell-1},\qpoint{2\ell+1}]$ share only an endpoint, so every price
belongs to at most two of them, that is, $\overlap\le2$.  It remains to verify the two conditions of
Lemma~\ref{lem:separation}.

For the first condition, fix $\ell\in[k-1]$ and let $i=2\ell-1$.  Since $\eta<\slack$, the profiles in $\sellerinterval{\ell}\times\buyerinterval{\ell}$ lie in $\levelrect{i}$, and no other profile in the support of $\distseller\times\distbuyer$ does.  Hence,
$\Pr\left[(\vs,\vb)\in\levelrect{i}\right]
=\Pr\left[(\vs,\vb)\in\sellerinterval{\ell}\times\buyerinterval{\ell}\right]
=\surplusunit/\profilegap{\ell}$. On $\sellerinterval{\ell}\times\buyerinterval{\ell}$, the expected gain from trade is $\profilegap{\ell}-\eta$, so the profiles in $\levelrect{i}$ contribute $\surplusunit(\profilegap{\ell}-\eta)/\profilegap{\ell}$ to the first-best gains-from-trade. This is between $\surplusunit/2$ and $\surplusunit$, since $\eta<\slack<\profilegap{\ell}/2$.

For the second condition, every profile in the support outside $\bigcup_{i\in I}\levelrect{i}$ lies in
$\sellerinterval{\ell}\times\buyerinterval{r}$ for some $\ell\neq r$.  If
$\ell>r$, then $\vb\le\vs$, so the gain from trade is $0$.  If $\ell<r$, the
gain from trade is at most $\qpoint{2r+1}-\qpoint{2\ell-1}$, and
$\Pr\left[(\vs,\vb)\in\sellerinterval{\ell}\times\buyerinterval{r}\right]=\surplusunit\,R^{\ell-r}/\sqrt{\profilegap{\ell}\profilegap{r}}$.
Summing over all pairs $\ell<r$, the first-best gains-from-trade outside these
sets is at most
\begin{align*}
    \surplusunit \cdot \sum_{1\le\ell<r\le k-1}
    R^{\ell-r}\,
    \frac{\qpoint{2r+1}-\qpoint{2\ell-1}}
         {\sqrt{\profilegap{\ell}\profilegap{r}}}
    \le\surplusunit,
\end{align*}
due to Equation~\eqref{eq:ind-choice-of-R}, so the second condition holds
with $\leftover=\surplusunit$.

By Lemma~\ref{lem:separation}, every SBB DSIC and IR
mechanism $\sbbmechanism$
satisfies
\begin{align*}
    \varepsilon\cdot\sgft_{\wbbmechanism}(\distseller\times\distbuyer)
    >
    \varepsilon\cdot\frac{(k-1)\surplusunit}{4}
    \ge
    3\surplusunit
    \ge
    \sgft_{\sbbmechanism}(\distseller\times\distbuyer),
\end{align*}
where the second inequality holds since $(k-1)/12\ge1/\varepsilon$, and the third since $\overlap\le2$ and $\leftover=\surplusunit$.  Since the
payments of an SBB mechanism cancel,
$\sgft_{\sbbmechanism}(\distseller\times\distbuyer)=\gft_{\sbbmechanism}(\distseller\times\distbuyer)$,
which proves the theorem.
\end{proof}

Recall that $\sgft^{b}_{\mechanism}$ and $\sgft^{s}_{\mechanism}$ denote the buyer's and the seller's surplus under a mechanism $\mechanism$, which sum to the traders' surplus $\sgft_{\mechanism}$ (Section~\ref{sec:objectives}).
\begin{remark}[Both traders benefit]\label{rem:both-traders-benefit} The separation of Theorem~\ref{thm:wbb-sbb-unbounded-gap-ind-gft} holds for each trader separately, that is, the benefit for \emph{each} of the agents from money burning can be arbitrarily large. For every $\varepsilon'>0$, there are seller and buyer distributions $\distseller$ and $\distbuyer$ and a dominant-strategy incentive-compatible, individually rational, and weakly budget-balanced mechanism $\wbbmechanism$ such that, for every dominant-strategy incentive-compatible, individually rational, and strongly budget-balanced mechanism $\sbbmechanism$ and every trader $a\in\{s,b\}$, \begin{align*} \varepsilon'\cdot\sgft^{a}_{\wbbmechanism}(\distseller\times\distbuyer) > \sgft_{\sbbmechanism}(\distseller\times\distbuyer) \ge \sgft^{a}_{\sbbmechanism}(\distseller\times\distbuyer). \end{align*} \end{remark}

The remark follows from the observation that in our construction, the two traders have the same expected surplus (Lemma~\ref{lem:equal-trader-surplus} below), so each of them obtains half of the traders' surplus. Apply Theorem~\ref{thm:wbb-sbb-unbounded-gap-ind-gft} with $\varepsilon=\varepsilon'/2$. Then for every SBB mechanism $\sbbmechanism$ and every trader $a\in\{s,b\}$, \begin{align*} \sgft^{a}_{\wbbmechanism}(\distseller\times\distbuyer) =\frac{1}{2}\,\sgft_{\wbbmechanism}(\distseller\times\distbuyer) >\frac{1}{\varepsilon'}\,\sgft_{\sbbmechanism}(\distseller\times\distbuyer) \ge\frac{1}{\varepsilon'}\,\sgft^{a}_{\sbbmechanism}(\distseller\times\distbuyer), \end{align*} so each trader alone obtains more than $1/\varepsilon'$ times the traders' surplus of $\sbbmechanism$, and in particular more than $1/\varepsilon'$ times the trader's own surplus under $\sbbmechanism$. The last inequality holds since, by individual rationality, the two traders' surpluses under $\sbbmechanism$ are nonnegative and sum to $\sgft_{\sbbmechanism}(\distseller\times\distbuyer)$.

It remains to show that the two traders have the same expected surplus. This follows from a symmetry of our construction: we match every profile with a \emph{mirror} profile that is equally likely, and in which the roles of the buyer and the seller are switched, so that the buyer's surplus at a profile equals the seller's surplus at its mirror.

\begin{lemma}\label{lem:equal-trader-surplus} In the construction of Theorem~\ref{thm:wbb-sbb-unbounded-gap-ind-gft}, $\sgft^{b}_{\wbbmechanism}(\distseller\times\distbuyer) =\sgft^{s}_{\wbbmechanism}(\distseller\times\distbuyer) =\frac{1}{2}\,\sgft_{\wbbmechanism}(\distseller\times\distbuyer)$. \end{lemma}

The proof of Lemma~\ref{lem:equal-trader-surplus} is deferred to Appendix~\ref{app:proof-equal-trader-surplus}.

\subsection{Proof of Theorem~\ref{thm:wbb-sbb-unbounded-gap-correlated-residual-welfare}}\label{sec:proof-wbb-sbb-seperation-welfare}


We prove the theorem by applying Lemma~\ref{lem:separation} to a specific grid
and a specific correlated distribution.  Throughout this subsection, fix
$\varepsilon>0$, and let $n\ge4$ be an integer such that
$(n-3)/16\ge1/\varepsilon$.

\paragraph{The grid.}
Let $\separation=1/(100n)$, and let the grid points be
\begin{align}\label{eq:corr-grid}
    \qpoint{i}
    =
    \frac{\separation^{\,n-i}-\separation^{\,n}}{1-\separation^{\,n}}
    \qquad\text{for every } i\in\{0,1,\ldots,n\}.
\end{align}
In particular, $\qpoint{0}=0$ and $\qpoint{n}=1$, and the gaps are
\begin{align*}
    \qlength{i}
    =\qpoint{i}-\qpoint{i-1}
    =\frac{(1-\separation)\,\separation^{\,n-i}}{1-\separation^{\,n}}
    \qquad\text{for every } i\in[n],
\end{align*}
so each gap is $1/\separation$ times the previous one.  Thus, the gaps grow
quickly from left to right.  In particular, for every $i\in[n]$, the gaps to
the left of $\qlength{i}$ sum to a small fraction of $\qlength{i}$, since
\begin{align*}
    \sum_{r=1}^{i-1}\qlength{r}
    =\qpoint{i-1}
    \le\frac{\separation^{\,n-i+1}}{1-\separation^{\,n}}
    =\frac{\separation}{1-\separation}\qlength{i}.
\end{align*}  Let $\slack>0$ be
small enough for the grid with slack $\slack$ to be a separation grid (this is possible by
Lemma~\ref{lem:corr-grid} below), and let $\wbbmechanism$ be the resulting
mechanism of Lemma~\ref{lem:general-wbb-step-construction}.

\paragraph{The distribution.}
Since the values may be correlated, we put all the probability on profiles in
$\levelrect{i-1}$, in which the seller's value is near $\qpoint{i-1}$ and the buyer's value is near
$\qpoint{i+1}$.  
We choose the
probabilities so that each $i$ contributes approximately the same amount to the
first-best social welfare.  At these profiles, trade is efficient, so the
first-best welfare is the buyer's value $\vb$, which equals the seller's value
$\vs$ plus the gain from trade $\vb-\vs$, where $\vs$ is close to
$\qpoint{i-1}$ and $\vb-\vs$ is close to $\qpoint{i+1}-\qpoint{i-1}$.
Since the gaps grow quickly, the seller's value $\qpoint{i-1}$ is small
compared to this gain.  Hence, as in
Section~\ref{sec:proof:wbb-sbb-unbounded-gap-ind-gft}, it suffices to make the
probability of each $i$ inversely proportional to $\qpoint{i+1}-\qpoint{i-1}$.

Formally, for every $i\in\{2,\ldots,n-2\}$, let
$\profilegap{i}=\qpoint{i+1}-\qpoint{i-1}$, fix
\begin{align*}
    0<\eta<\min\left\{\slack,\ \separation^2\min_{2\le i\le n-2}\profilegap{i}\right\},
\end{align*}
and let
\begin{align*}
    \sellerinterval{i}
    =[\qpoint{i-1},\qpoint{i-1}+\eta],
    \qquad
    \buyerinterval{i}
    =[\qpoint{i+1}-\eta,\qpoint{i+1}],
    \qquad
    \surplusunit
    =
    \frac{1}{\sum_{r=2}^{n-2}1/\profilegap{r}}.
\end{align*}
To draw the agents' values, we first draw an index $i\in\{2,\ldots,n-2\}$ with
probability $\surplusunit/\profilegap{i}$, and then draw the seller's value
uniformly from $\sellerinterval{i}$ and, independently, the buyer's value
uniformly from $\buyerinterval{i}$.  Let $\distjoint$ denote the resulting
joint distribution.  The two values are correlated, since the same index
selects both intervals.  Note that, for $(\vs,\vb)\sim\distjoint$ and every
$i\in\{2,\ldots,n-2\}$,
\begin{align*}
    \Pr\left[(\vs,\vb)\in\sellerinterval{i}\times\buyerinterval{i}\right]
    =\frac{\surplusunit}{\profilegap{i}}.
\end{align*}

\paragraph{Analysis.}
We first show that the grid in Equation~\eqref{eq:corr-grid} satisfies the grid condition of Definition~\ref{def:separation-grid}.

\begin{lemma}\label{lem:corr-grid}
The grid in Equation~\eqref{eq:corr-grid} satisfies the grid condition of Definition~\ref{def:separation-grid}, that is, Equation~\eqref{eq:grid-condition}.
\end{lemma}

To bound the traders' utility of an SBB mechanism,
we also need a bound on the expected seller's value.
We next show that it is small.

\begin{lemma}\label{lem:corr-seller-value}
$\Exp_{(\vs,\vb)\sim\distjoint}[\vs]\le\surplusunit$.
\end{lemma}

The proofs of Lemmas~\ref{lem:corr-grid} and~\ref{lem:corr-seller-value} are deferred to
Appendices~\ref{app:proof-corr-grid} and~\ref{app:proof-corr-seller-value}, respectively.

\begin{proof}[Proof of Theorem~\ref{thm:wbb-sbb-unbounded-gap-correlated-residual-welfare}]
Let $\distjoint$ and $\wbbmechanism$ be as above.  We apply
Lemma~\ref{lem:separation} with $I=\{1,\ldots,n-3\}$, so $m=n-3$.  By
Lemma~\ref{lem:corr-grid}, the grid with slack $\slack$ is a separation grid.  Every price belongs to at most three of
the intervals $[\qpoint{i},\qpoint{i+2}]$, $i\in I$, that is, $\overlap\le3$.  It remains to verify the
two conditions of Lemma~\ref{lem:separation}.

For the first condition, fix $i\in I$.  Since $\eta<\slack$, the profiles in
$\sellerinterval{i+1}\times\buyerinterval{i+1}$ lie in $\levelrect{i}$, and no
other profile in the support of $\distjoint$ does.  Hence,
$\Pr\left[(\vs,\vb)\in\levelrect{i}\right]
=\Pr\left[(\vs,\vb)\in\sellerinterval{i+1}\times\buyerinterval{i+1}\right]
=\surplusunit/\profilegap{i+1}$. On $\sellerinterval{i+1}\times\buyerinterval{i+1}$, the expected gain from trade is $\profilegap{i+1}-\eta$, so the profiles in $\levelrect{i}$ contribute $\surplusunit(\profilegap{i+1}-\eta)/\profilegap{i+1}$ to the first-best gains-from-trade. This is between $\surplusunit/2$ and $\surplusunit$, since $\eta<\slack<\profilegap{i+1}/2$.

For the second condition, all the probability is on $\bigcup_{i\in I}\levelrect{i}$, so it holds
with $\leftover=0$.  

By Lemma~\ref{lem:separation}, every SBB DSIC and IR
mechanism $\sbbmechanism$
satisfies
\begin{align*}
    \varepsilon\cdot\swelfare_{\wbbmechanism}(\distjoint)
    \ge
    \varepsilon\cdot\sgft_{\wbbmechanism}(\distjoint)
    >
    \varepsilon\cdot\frac{(n-3)\surplusunit}{4}
    \ge
    4\surplusunit
    \ge
    \swelfare_{\sbbmechanism}(\distjoint),
\end{align*}
where the first inequality holds since
$\swelfare_{\wbbmechanism}(\distjoint)=\Exp[\vs]+\sgft_{\wbbmechanism}(\distjoint)$,
and the third holds since $(n-3)/16\ge1/\varepsilon$. The last inequality holds since $\swelfare_{\sbbmechanism}(\distjoint)=\Exp[\vs]+\sgft_{\sbbmechanism}(\distjoint)$, where $\Exp[\vs]\le\surplusunit$ by Lemma~\ref{lem:corr-seller-value} and $\sgft_{\sbbmechanism}(\distjoint)\le3\surplusunit$ by Lemma~\ref{lem:separation}.  Since the
payments of an SBB mechanism cancel,
$\swelfare_{\sbbmechanism}(\distjoint)=\welfare_{\sbbmechanism}(\distjoint)$,
which proves the theorem.
\end{proof}

\begin{remark}\label{rem:constant-fraction}
On the distributions of Theorem~\ref{thm:wbb-sbb-unbounded-gap-ind-gft} and Theorem~\ref{thm:wbb-sbb-unbounded-gap-correlated-residual-welfare}, the mechanism $\wbbmechanism$ obtains a constant approximation to the first-best: its traders' surplus is more than $1/8$ of the first-best gains-from-trade for independent values, and its traders' utility is more than $1/8$ of the first-best social welfare for correlated values (see Appendix~\ref{app:proof-constant-fraction}).
\end{remark}




\section{The Limitations of Weakly Budget-Balanced Mechanisms}
\label{sec:wbb:limitations}

The separation results of Section~\ref{sec:separation-results} show that money burning can substantially improve the traders' surplus and the traders' utility.  Weakly budget-balanced mechanisms can obtain arbitrarily more traders' surplus and traders' utility than strongly budget-balanced ones, with independent and correlated values, respectively (Theorems~\ref{thm:wbb-sbb-unbounded-gap-ind-gft} and~\ref{thm:wbb-sbb-unbounded-gap-correlated-residual-welfare}).  This raises the natural question of whether this additional power is enough to overcome the known barriers to approximating the first-best with SBB DSIC mechanisms.  The following results show a sharp limit: even WBB DSIC mechanisms cannot guarantee a constant approximation to the first-best traders' surplus (for independent values) or the first-best traders' utility (for correlated values).  These impossibilities hold even if we ignore the fact that some money is burnt by the mechanism, and count it towards the gains-from-trade or the social welfare.

\begin{theorem}\label{thm:tie-wbb-no-constant-ind-gft}
For every $\varepsilon>0$, there are independent seller and buyer distributions $\distseller$ and $\distbuyer$ such that every dominant-strategy incentive-compatible, individually rational, and weakly budget-balanced mechanism $\wbbmechanism$ satisfies
\[
    \sgft_{\wbbmechanism}(\distseller\times\distbuyer)\le \gft_{\wbbmechanism}(\distseller\times\distbuyer)
    < \varepsilon\cdot\fbgft(\distseller\times\distbuyer)
    = \varepsilon\cdot\fbsgft(\distseller\times\distbuyer).
\]
\end{theorem}

\begin{theorem}\label{thm:wbb-tie-no-constant-correlated-welfare}
For every $\varepsilon>0$, there is a correlated distribution $\distjoint$ such that every dominant-strategy incentive-compatible, individually rational, and weakly budget-balanced mechanism $\wbbmechanism$ satisfies
\[
    \swelfare_{\wbbmechanism}(\distjoint)\le \welfare_{\wbbmechanism}(\distjoint)
    < \varepsilon\cdot\fbwelfare(\distjoint)
    = \varepsilon\cdot\fbswelfare(\distjoint).
\]
\end{theorem}

For independent values, Blumrosen and Dobzinski~\cite{BD14} showed that no SBB DSIC and IR mechanism guarantees a constant fraction of the first-best gains-from-trade.  Theorem~\ref{thm:tie-wbb-no-constant-ind-gft} establishes the same impossibility for the larger class of WBB DSIC and IR mechanisms, using the same family of distributions as in \cite{BD14}.

For correlated values, Dobzinski and Shaulker \cite{DS24} showed that no SBB mechanism in which both traders have a dominant strategy guarantees a constant approximation to first-best social welfare.  Theorem~\ref{thm:wbb-tie-no-constant-correlated-welfare} shows that allowing the randomized DSIC mechanism to be WBB is still not enough to overcome this barrier.

\subsection{Proof of Theorem~\ref{thm:tie-wbb-no-constant-ind-gft}}
\label{sec:tie-impossibility-gft}

To prove the theorem we consider the same family of independent distributions as in \cite{BD14}.

\paragraph{The distributions.}
For every $t\ge4$, let $\distseller^{t}$ and $\distbuyer^{t}$ be the seller's and the buyer's value distributions, respectively, both supported on the interval $[0,t]$.  We denote by $\cdfseller^{t}$ and $\pdfseller^{t}$ the cumulative distribution function and the probability density function of $\distseller^{t}$, and by $\cdfbuyer^{t}$ and $\pdfbuyer^{t}$ those of $\distbuyer^{t}$.  For every $z\in[0,t]$,
\begin{equation}\label{eq:lb-distributions}
    \begin{aligned}
        \cdfbuyer^{t}(z)&=\lambda(1-e^{-z}),
        &
        \pdfbuyer^{t}(z)&=\lambda e^{-z},\\
        \cdfseller^{t}(z)&=\lambda(e^{z-t}-e^{-t}),
        &
        \pdfseller^{t}(z)&=\lambda e^{z-t},
    \end{aligned}
    \qquad\text{where }\lambda=\frac{1}{1-e^{-t}}.
\end{equation}

\begin{proof}[Proof of Theorem~\ref{thm:tie-wbb-no-constant-ind-gft}]
Let $\varepsilon>0$ be as in the statement of the theorem, and let $t\ge4$ be such that
\begin{equation}\label{eq:lb-choice-of-t}
    \frac2t\left(1+\frac{11}{e}\left(\frac12+\log t\right)\right)<\varepsilon;
\end{equation}
such $t$ exists since the left-hand side tends to $0$ as $t\to\infty$.  Now that $t$ is fixed, we drop it from the notation and write $\distseller$, $\distbuyer$, $\cdfseller$, $\pdfseller$, $\cdfbuyer$, and $\pdfbuyer$ for simplicity.  Let $\mechanism=(\allocation,\paymentB,\paymentS)$ be a WBB DSIC and IR mechanism for $\distseller\times\distbuyer$.  For every $r\in[0,t]$, let
\begin{equation}\label{eq:lb-trade-mass}
    A(r)=\int_0^{t-r}\allocation(\vs,\vs+r)\d{\vs}.
\end{equation}
That is, $A(r)$ is the integral of the trade probability over the profiles with gain from trade $r$.

Lemma~\ref{lem:lb-gft-trade-mass} below expresses the gains-from-trade of $\mechanism$ in terms of $A$.  It uses the fact that, under $\distseller\times\distbuyer$, all the profiles with the same gain from trade $r$ have the same density.

\begin{lemma}[Gains-from-trade and trade probabilities]\label{lem:lb-gft-trade-mass}
The gains-from-trade of $\mechanism$ are
\[
    \gft_{\mechanism}(\distseller\times\distbuyer)
    =\lambda^2e^{-t}\int_0^tre^{-r}A(r)\d{r}.
\]
\end{lemma}

Lemma~\ref{lem:lb-trade-mass-bound} uses WBB to bound the expression for the gains-from-trade of $\mechanism$ given by Lemma~\ref{lem:lb-gft-trade-mass}.  Applying the inequality of Lemma~\ref{lem:wbb-allocation-inequality} at the profiles with gain from trade $r$ shows that trade at the profiles with a smaller gain from trade forces trade at the profiles with gain from trade $r$.  Since the trade probability is at most $1$, this limits the trade at the profiles with a small gain from trade, which carry most of the weight in this integral.

\begin{lemma}[Trade at profiles with a small gain from trade]\label{lem:lb-trade-mass-bound}
The function $A$ satisfies
\begin{equation}\label{eq:weighted-diagonal-mass}
    \int_0^tre^{-r}A(r)\d{r}\le1+\frac{11}{e}\left(\frac12+\log t\right).
\end{equation}
\end{lemma}

The following upper bound on the gains-from-trade of $\mechanism$ follows directly from Lemmas~\ref{lem:lb-gft-trade-mass} and~\ref{lem:lb-trade-mass-bound}, by substituting the bound of Lemma~\ref{lem:lb-trade-mass-bound} into the expression of Lemma~\ref{lem:lb-gft-trade-mass}.

\begin{corollary}\label{cor:lb-gft-bound}
The gains-from-trade of $\mechanism$ satisfy
\begin{equation}\label{eq:lb-gft-bound}
    \gft_{\mechanism}(\distseller\times\distbuyer)
    \le\lambda^2e^{-t}\left(1+\frac{11}{e}\left(\frac12+\log t\right)\right).
\end{equation}
\end{corollary}

Finally, Lemma~\ref{lem:lb-first-best} gives a lower bound on the first-best gains-from-trade.

\begin{lemma}[First-best gains-from-trade]\label{lem:lb-first-best}
The first-best gains-from-trade satisfy
\[
    \fbgft(\distseller\times\distbuyer)
    =\lambda^2\left((t-2)e^{-t}+(t+2)e^{-2t}\right)
    \ge\frac{\lambda^2te^{-t}}{2}.
\]
\end{lemma}

Combining Corollary~\ref{cor:lb-gft-bound} and Lemma~\ref{lem:lb-first-best} bounds the approximation ratio of $\mechanism$:
\[
    \frac{\gft_{\mechanism}(\distseller\times\distbuyer)}{\fbgft(\distseller\times\distbuyer)}
    \le\frac2t\left(1+\frac{11}{e}\left(\frac12+\log t\right)\right)
    <\varepsilon,
\]
where the last inequality holds due to Inequality~\eqref{eq:lb-choice-of-t}.  Moreover, since $\mechanism$ is WBB, the money it burns is nonnegative, so $\sgft_{\mechanism}(\distseller\times\distbuyer)\le\gft_{\mechanism}(\distseller\times\distbuyer)$.  Since $\mechanism$ is an arbitrary WBB DSIC and IR mechanism, this proves the theorem.
\end{proof}

The proofs of Lemmas~\ref{lem:lb-gft-trade-mass}, \ref{lem:lb-trade-mass-bound}, and~\ref{lem:lb-first-best} are deferred to Appendices~\ref{app:proof-lb-gft-trade-mass}, \ref{app:proof-lb-trade-mass-bound}, and~\ref{app:proof-lb-first-best}, respectively.

\subsection{Proof of Theorem~\ref{thm:wbb-tie-no-constant-correlated-welfare}}
\label{sec:wbb:corr:welfare}
\label{sec:wbb:corr:first-best-impossibility}

To prove the theorem we consider a family of correlated distributions supported on a geometric grid.

\paragraph{The distributions.}
For every integer $n\ge4$, let $\georatio=\sqrt n$, and let
\[
    \qpoint{i}=\georatio^{i-n},
    \qquad i=0,\ldots,n,
\]
so that $\qpoint{i}=\georatio\qpoint{i-1}$ for every $i\in[n]$, and $\qpoint{n}=1$.  Let $\distjoint^{n}$ be the joint distribution supported on the $n$ profiles $(\qpoint{i-1},\qpoint{i})$, $i\in[n]$,\footnote{This result easily extends to continuous distributions, by replacing each profile with a small square. 
} where the profile $(\qpoint{i-1},\qpoint{i})$ has probability
\begin{equation}\label{eq:lb-corr-distribution}
    \profileprob{i}=\frac{1}{\normalizer\qpoint{i}},
    \qquad\text{where }\normalizer=\sum_{r=1}^{n}\frac{1}{\qpoint{r}}.
\end{equation}

\begin{proof}[Proof of Theorem~\ref{thm:wbb-tie-no-constant-correlated-welfare}]
Let $\varepsilon>0$ be as in the statement of the theorem, and let $n\ge4$ be an integer such that $6/\sqrt n<\varepsilon$.  Now that $n$ is fixed, we drop it from the notation and write $\distjoint$ for $\distjoint^{n}$.  Let $\mechanism=(\allocation,\paymentB,\paymentS)$ be a WBB DSIC and IR mechanism for $\distjoint$.

Lemma~\ref{lem:lb-corr-welfare-trade} below expresses the social welfare of $\mechanism$ in terms of its trade probabilities at the profiles in the support.  It uses the fact that the probability of each profile in the support is inversely proportional to its buyer's value.

\begin{lemma}[Social welfare and trade probabilities]\label{lem:lb-corr-welfare-trade}
The social welfare of $\mechanism$ is
\[
    \welfare_{\mechanism}(\distjoint)
    =\frac{1}{\normalizer}
    \sum_{i=1}^{n}
    \left(
        \frac{1}{\georatio}
        +\left(1-\frac{1}{\georatio}\right)
        \allocation(\qpoint{i-1},\qpoint{i})
    \right).
\]
\end{lemma}

Lemma~\ref{lem:geometric-staircase-bound} bounds the sum of the trade probabilities in the expression of Lemma~\ref{lem:lb-corr-welfare-trade}.  The bound follows from WBB, through the inequality of Lemma~\ref{lem:wbb-allocation-inequality}, which combines WBB with Myerson's lemma.

\begin{lemma}[Allocation bound on a geometric grid]\label{lem:geometric-staircase-bound}
The trade probabilities of $\mechanism$ satisfy
\begin{equation}\label{eq:geometric-staircase-bound}
    \sum_{i=1}^{n}
    \allocation(\qpoint{i-1},\qpoint{i})
    \le 5\georatio.
\end{equation}
\end{lemma}

The following upper bound on the social welfare of $\mechanism$ follows directly from Lemmas~\ref{lem:lb-corr-welfare-trade} and~\ref{lem:geometric-staircase-bound}, by substituting the bound of Lemma~\ref{lem:geometric-staircase-bound} into the expression of Lemma~\ref{lem:lb-corr-welfare-trade}.

\begin{corollary}\label{cor:lb-corr-welfare-bound}
The social welfare of $\mechanism$ satisfies
\[
    \welfare_{\mechanism}(\distjoint)
    \le\frac{1}{\normalizer}\left(\frac{n}{\georatio}+5\georatio\right).
\]
\end{corollary}

Finally, Lemma~\ref{lem:lb-corr-first-best} gives the first-best social welfare.

\begin{lemma}[First-best social welfare]\label{lem:lb-corr-first-best}
The first-best social welfare is $\fbwelfare(\distjoint)=n/\normalizer$.
\end{lemma}

Combining Corollary~\ref{cor:lb-corr-welfare-bound} and Lemma~\ref{lem:lb-corr-first-best} bounds the approximation ratio of $\mechanism$:
\[
    \frac{\welfare_{\mechanism}(\distjoint)}{\fbwelfare(\distjoint)}
    \le
    \frac{1}{\georatio}+\frac{5\georatio}{n}
    =\frac{6}{\sqrt n}
    <\varepsilon.
\]
Moreover, since $\mechanism$ is WBB, the money it burns is nonnegative, so $\swelfare_{\mechanism}(\distjoint)\le\welfare_{\mechanism}(\distjoint)$.  Since $\mechanism$ is an arbitrary WBB DSIC and IR mechanism, this proves the theorem.
\end{proof}

The proofs of Lemmas~\ref{lem:lb-corr-welfare-trade}, \ref{lem:geometric-staircase-bound}, and~\ref{lem:lb-corr-first-best} are deferred to Appendices~\ref{app:proof-lb-corr-welfare-trade}, \ref{app:proof-geometric-staircase-bound}, and~\ref{app:proof-lb-corr-first-best}, respectively.

\section*{AI Disclosure}
{All ideas, technical content, proofs, and results are the authors' own. We used Codex and Claude to assist in drafting and editing the text throughout the paper, and in creating the figures. We  have also used AI tools to review the paper. The authors verified the correctness and originality of all content, including references, and take full responsibility for it.}





\bibliography{bilateral-trade}

\newcommand{\etalchar}[1]{$^{#1}$}
\begin{thebibliography}{CBCC{\etalchar{+}}24b}

\bibitem[AB20]{AmadorBagwell20}
Manuel Amador and Kyle Bagwell.
\newblock Money burning in the theory of delegation.
\newblock {\em Games and Economic Behavior}, 121:382--412, 2020.

\bibitem[ABDF24]{ABDF24}
Gagan Aggarwal, Ashwinkumar Badanidiyuru, Paul D{\"u}tting, and Federico Fusco.
\newblock Selling joint ads: A regret minimization perspective.
\newblock In {\em Proceedings of the 25th ACM Conference on Economics and Computation (EC)}, 2024.

\bibitem[AFF24]{regret2}
Yossi Azar, Amos Fiat, and Federico Fusco.
\newblock An {$\alpha$}-regret analysis of adversarial bilateral trade.
\newblock {\em Artificial Intelligence}, 337:104231, 2024.

\bibitem[Bai97]{Bailey97}
Martin~J. Bailey.
\newblock The demand revealing process: To distribute the surplus.
\newblock {\em Public Choice}, 91(2):107--126, 1997.

\bibitem[BCBCC24]{BCCC24fair}
Fran{\c{c}}ois Bachoc, Nicol{\`o} Cesa-Bianchi, Tommaso Cesari, and Roberto Colomboni.
\newblock Fair online bilateral trade.
\newblock In {\em Advances in Neural Information Processing Systems (NeurIPS)}, 2024.

\bibitem[BCC24]{BCC24brokerage}
Nata{\v{s}}a Boli{\'c}, Tommaso Cesari, and Roberto Colomboni.
\newblock An online learning theory of brokerage.
\newblock In {\em Proceedings of the 23rd International Conference on Autonomous Agents and Multiagent Systems (AAMAS)}, 2024.

\bibitem[BCC25]{BCC25contextual}
Fran{\c{c}}ois Bachoc, Tommaso Cesari, and Roberto Colomboni.
\newblock A tight regret analysis of non-parametric repeated contextual brokerage.
\newblock In {\em International Conference on Artificial Intelligence and Statistics (AISTATS)}, 2025.

\bibitem[BCCF24]{BCCF24}
Martino Bernasconi, Matteo Castiglioni, Andrea Celli, and Federico Fusco.
\newblock No-regret learning in bilateral trade via global budget balance.
\newblock In {\em Proceedings of the 56th Annual ACM Symposium on Theory of Computing (STOC)}, 2024.

\bibitem[BCGZ18]{BothWorlds}
Moshe Babaioff, Yang Cai, Yannai~A. Gonczarowski, and Mingfei Zhao.
\newblock The best of both worlds: Asymptotically efficient mechanisms with a guarantee on the expected gains-from-trade.
\newblock In {\em Proceedings of the 2018 ACM Conference on Economics and Computation}, EC '18, page 373, New York, NY, USA, 2018. Association for Computing Machinery.

\bibitem[BCWZ17]{Bru}
Johannes Brustle, Yang Cai, Fa~Wu, and Mingfei Zhao.
\newblock Approximating gains from trade in two-sided markets via simple mechanisms.
\newblock In {\em Proceedings of the 2017 ACM Conference on Economics and Computation}, EC '17, pages 589--590, New York, NY, USA, 2017. Association for Computing Machinery.

\bibitem[BD14]{BD14}
Liad Blumrosen and Shahar Dobzinski.
\newblock Reallocation mechanisms.
\newblock In {\em Proceedings of the Fifteenth ACM Conference on Economics and Computation}, EC '14, page 617, New York, NY, USA, 2014. Association for Computing Machinery.

\bibitem[BD21]{BD21}
Liad Blumrosen and Shahar Dobzinski.
\newblock {(Almost)} efficient mechanisms for bilateral trading.
\newblock {\em Games and Economic Behavior}, 130:369--383, 2021.

\bibitem[BDK21]{DKB}
Moshe Babaioff, Shahar Dobzinski, and Ron Kupfer.
\newblock A note on the gains from trade of the random-offerer mechanism, 2021.
\newblock arXiv:2111.07790.

\bibitem[BFN24]{BFN24}
Moshe Babaioff, Amitai Frey, and Noam Nisan.
\newblock Learning to maximize gains from trade in small markets.
\newblock In {\em Proceedings of the 25th ACM Conference on Economics and Computation (EC)}, 2024.

\bibitem[BLWZ26]{BLWZ26}
Xiaohui Bei, Bo~Li, Wenhao Wu, and Shengwei Zhou.
\newblock Second-best gains from trade in matching markets, 2026.
\newblock arXiv:2609.18724.

\bibitem[BM16]{BM}
Liad Blumrosen and Yehonatan Mizrahi.
\newblock Approximating gains-from-trade in bilateral trading.
\newblock In Yang Cai and Adrian Vetta, editors, {\em Web and Internet Economics}, pages 400--413, Berlin, Heidelberg, 2016. Springer Berlin Heidelberg.

\bibitem[BN04]{BabaioffNisan04}
Moshe Babaioff and Noam Nisan.
\newblock Concurrent auctions across the supply chain.
\newblock {\em Journal of Artificial Intelligence Research}, 21:595--629, 2004.

\bibitem[BN09]{BN09}
Tilman B{\"o}rgers and Peter Norman.
\newblock A note on budget balance under interim participation constraints: the case of independent types.
\newblock {\em Economic Theory}, 39(3):477--489, 2009.

\bibitem[BNP09]{BNP09}
Moshe Babaioff, Noam Nisan, and Elan Pavlov.
\newblock Mechanisms for a spatially distributed market.
\newblock {\em Games and Economic Behavior}, 66(2):660--684, 2009.

\bibitem[BRTW26]{BRTW26}
Moshe Babaioff, Aviad Rubinstein, Xizhi Tan, and Kangning Wang.
\newblock Approximating gains-from-trade in matching markets.
\newblock In {\em Proceedings of the 58th Annual ACM Symposium on Theory of Computing (STOC)}, 2026.

\bibitem[BW05]{BabaioffWalsh2005}
Moshe Babaioff and William~E. Walsh.
\newblock Incentive-compatible, budget-balanced, yet highly efficient auctions for supply chain formation.
\newblock {\em Decision Support Systems}, 39(1):123--149, 2005.

\bibitem[BWZ26]{BWZ26}
Xiaohui Bei, Wenhao Wu, and Shengwei Zhou.
\newblock The power of simple mechanisms: A tight $e$-approximation for gains from trade in matching markets, 2026.
\newblock arXiv:2609.37997.

\bibitem[Cav06]{Cavallo06}
Ruggiero Cavallo.
\newblock Optimal decision-making with minimal waste: Strategyproof redistribution of {VCG} payments.
\newblock In {\em Proceedings of the 5th International Joint Conference on Autonomous Agents and Multiagent Systems}, pages 882--889, 2006.

\bibitem[CBCC{\etalchar{+}}24a]{CCCFL24MOR}
Nicol{\`o} Cesa-Bianchi, Tommaso Cesari, Roberto Colomboni, Federico Fusco, and Stefano Leonardi.
\newblock Bilateral trade: A regret minimization perspective.
\newblock {\em Mathematics of Operations Research}, 49(1):171--203, 2024.

\bibitem[CBCC{\etalchar{+}}24b]{regret1}
Nicol{{\`o}} Cesa-Bianchi, Tommaso Cesari, Roberto Colomboni, Federico Fusco, and Stefano Leonardi.
\newblock Regret analysis of bilateral trade with a smoothed adversary.
\newblock {\em Journal of Machine Learning Research}, 25(234):1--36, 2024.

\bibitem[CBdKLT16]{italians}
Riccardo Colini-Baldeschi, Bart de~Keijzer, Stefano Leonardi, and Stefano Turchetta.
\newblock Approximately efficient double auctions with strong budget balance.
\newblock In {\em Proceedings of the Twenty-Seventh Annual ACM-SIAM Symposium on Discrete Algorithms}, SODA '16, pages 1424--1443, USA, 2016. Society for Industrial and Applied Mathematics.

\bibitem[CC25]{CC25volume}
Tommaso Cesari and Roberto Colomboni.
\newblock An online learning theory of trading-volume maximization.
\newblock In {\em International Conference on Learning Representations (ICLR)}, 2025.

\bibitem[CGLM26]{CGLM26}
Yang Cai, Vineet Gupta, Zun Li, and Aranyak Mehta.
\newblock A new lower bound for the random offerer mechanism in bilateral trade using {AI}-guided evolutionary search, 2026.
\newblock arXiv:2603.08679.

\bibitem[CGMZ21]{CGMZ21}
Yang Cai, Kira Goldner, Steven Ma, and Mingfei Zhao.
\newblock On multi-dimensional gains from trade maximization.
\newblock In {\em Proceedings of the 2021 ACM-SIAM Symposium on Discrete Algorithms (SODA)}, pages 1079--1098, 2021.

\bibitem[CJLZ26]{CJLZ26}
Houshuang Chen, Yaonan Jin, Pinyan Lu, and Chihao Zhang.
\newblock Tight regret bounds for fixed-price bilateral trade.
\newblock In {\em Proceedings of the 53rd International Colloquium on Automata, Languages, and Programming (ICALP)}, volume 374 of {\em LIPIcs}, pages 57:1--57:21, 2026.

\bibitem[CK13]{ChakravartyKaplan13}
Surajeet Chakravarty and Todd~R. Kaplan.
\newblock Optimal allocation without transfer payments.
\newblock {\em Games and Economic Behavior}, 77(1):1--20, 2013.

\bibitem[Con12]{Condorelli12}
Daniele Condorelli.
\newblock What money can't buy: Efficient mechanism design with costly signals.
\newblock {\em Games and Economic Behavior}, 75:613--624, 2012.

\bibitem[{\v{C}}P16]{CP16}
Jernej {\v{C}}opi{\v{c}} and Clara Ponsat{\'\i}.
\newblock Optimal robust bilateral trade: Risk neutrality.
\newblock {\em Journal of Economic Theory}, 163:276--287, 2016.

\bibitem[CW23]{CW-STOC}
Yang Cai and Jinzhao Wu.
\newblock On the optimal fixed-price mechanism in bilateral trade.
\newblock In {\em Proceedings of the 55th Annual ACM Symposium on Theory of Computing}, STOC 2023, pages 737--750, New York, NY, USA, 2023. Association for Computing Machinery.

\bibitem[dCNP{\etalchar{+}}14]{deClippel14}
Geoffroy de~Clippel, Victor Naroditskiy, Maria Polukarov, Amy Greenwald, and Nicholas~R. Jennings.
\newblock Destroy to save.
\newblock {\em Games and Economic Behavior}, 86:392--404, 2014.

\bibitem[DEG{\etalchar{+}}25]{DEGST25}
Shahar Dobzinski, Alon Eden, Kira Goldner, Ariel Shaulker, and Thodoris Tsilivis.
\newblock Bilateral trade with interdependent values: Information vs. approximation.
\newblock In {\em Proceedings of the 26th ACM Conference on Economics and Computation}, pages 641--665. Association for Computing Machinery, 2025.

\bibitem[DK15]{DK15}
Moritz Drexl and Andreas Kleiner.
\newblock Optimal private good allocation: The case for a balanced budget.
\newblock {\em Games and Economic Behavior}, 94:169--181, 2015.

\bibitem[DMSW22]{GDTCA}
Yuan Deng, Jieming Mao, Balasubramanian Sivan, and Kangning Wang.
\newblock Approximately efficient bilateral trade.
\newblock In {\em Proceedings of the 54th Annual ACM SIGACT Symposium on Theory of Computing}, STOC 2022, pages 718--721, New York, NY, USA, 2022. Association for Computing Machinery.

\bibitem[DS24]{DS24}
Shahar Dobzinski and Ariel Shaulker.
\newblock Bilateral trade with correlated values.
\newblock In {\em Proceedings of the 56th Annual ACM Symposium on Theory of Computing}, STOC 2024. Association for Computing Machinery, 2024.

\bibitem[DS26]{DS26}
Shahar Dobzinski and Ariel Shaulker.
\newblock Welfare maximization in bilateral trade: Improved approximation guarantees beyond the fixed price barrier, 2026.
\newblock arXiv:2606.04890.

\bibitem[EGR26]{EGR26}
Tomer Ezra, Aadityan Ganesh, and Aviad Rubinstein.
\newblock Welfare approximation in multilateral trade, 2026.
\newblock arXiv:2608.11351.

\bibitem[ESS25]{ESS25}
Tomer Ezra, Daniel Schoepflin, and Ariel Shaulker.
\newblock Multi-parameter mechanisms for consumer surplus maximization.
\newblock In {\em Proceedings of the 57th Annual ACM Symposium on Theory of Computing (STOC)}, 2025.

\bibitem[Ezr26]{Ezra26}
Tomer Ezra.
\newblock Money burning mechanism design: From welfare to surplus, 2026.
\newblock arXiv:2608.01693.

\bibitem[Fei22]{IGFT}
Yumou Fei.
\newblock Improved approximation to first-best gains-from-trade.
\newblock In Kristoffer~Arnsfelt Hansen, Tracy~Xiao Liu, and Azarakhsh Malekian, editors, {\em Web and Internet Economics}, pages 204--218, Cham, 2022. Springer International Publishing.

\bibitem[FTTZ16]{FTTZ16}
Dimitris Fotakis, Dimitris Tsipras, Christos Tzamos, and Emmanouil Zampetakis.
\newblock Efficient money burning in general domains.
\newblock {\em Theory of Computing Systems}, 59(4):619--640, 2016.

\bibitem[GBC{\etalchar{+}}25]{GBCCP25}
Solenne Gaucher, Martino Bernasconi, Matteo Castiglioni, Andrea Celli, and Vianney Perchet.
\newblock Feature-based online bilateral trade.
\newblock In {\em International Conference on Learning Representations (ICLR)}, 2025.

\bibitem[GC08]{GC08}
Mingyu Guo and Vincent Conitzer.
\newblock Better redistribution with inefficient allocation in multi-unit auctions with unit demand.
\newblock In {\em Proceedings of the 9th ACM Conference on Electronic Commerce}, pages 210--219, 2008.

\bibitem[GC09]{GC09}
Mingyu Guo and Vincent Conitzer.
\newblock Worst-case optimal redistribution of {VCG} payments in multi-unit auctions.
\newblock {\em Games and Economic Behavior}, 67(1):69--98, 2009.

\bibitem[Gd26]{improved-fp}
Giordano Giambartolomei and Bart {de Keijzer}.
\newblock On the approximation ratio of optimal fixed-price mechanisms for single and multi-unit bilateral trade.
\newblock In {\em Proceedings of the AAAI Conference on Artificial Intelligence}, volume~40, pages 16946--16953, 2026.

\bibitem[GL25]{GoldnerLundy25}
Kira Goldner and Taylor Lundy.
\newblock Multidimensional {B}ayesian utility maximization: Tight approximations to welfare.
\newblock In {\em Advances in Neural Information Processing Systems (NeurIPS)}, 2025.

\bibitem[GLT26]{GLT26}
Kira Goldner, Taylor Lundy, and Thodoris Tsilivis.
\newblock A constant-factor approximation to multidimensional consumer utility, 2026.
\newblock arXiv:2610.10516.

\bibitem[GMT26]{GMT26}
Kira Goldner, Divyarthi Mohan, and Thodoris Tsilivis.
\newblock Knowing who, not how much: Learning-augmented mechanisms for consumer utility maximization.
\newblock In {\em Proceedings of the 43rd International Conference on Machine Learning (ICML)}, 2026.

\bibitem[HR87]{HR87}
Kathleen~M Hagerty and William~P Rogerson.
\newblock Robust trading mechanisms.
\newblock {\em Journal of Economic Theory}, 42(1):94--107, 1987.

\bibitem[HR08]{HartlineRoughgarden08}
Jason~D. Hartline and Tim Roughgarden.
\newblock Optimal mechanism design and money burning.
\newblock In {\em Proceedings of the Fortieth Annual ACM Symposium on Theory of Computing}, pages 75--84, 2008.

\bibitem[JGC26]{JGC26}
Tao Jiang, Minbo Gao, and Shaowei Cai.
\newblock The exact approximation ratio of the optimal fixed-price mechanism in bilateral trade, 2026.
\newblock arXiv:2609.27878.

\bibitem[Jo26]{Jo26}
Sunghyeon Jo.
\newblock Tighter bounds for the random-offerer mechanism in bilateral trade, 2026.
\newblock arXiv:2607.13959.

\bibitem[KPV22]{KPV}
Zi~Yang Kang, Francisco Pernice, and Jan Vondrák.
\newblock Fixed-price approximations in bilateral trade.
\newblock In {\em Proceedings of the 2022 Annual ACM-SIAM Symposium on Discrete Algorithms (SODA)}, pages 2964--2985, Philadelphia, PA, 2022. Society for Industrial and Applied Mathematics.

\bibitem[LCM25a]{SublinearBudgetViolation25}
Anna Lunghi, Matteo Castiglioni, and Alberto Marchesi.
\newblock Better regret rates in bilateral trade via sublinear budget violation, 2025.
\newblock arXiv:2507.11419.

\bibitem[LCM25b]{LCM25}
Anna Lunghi, Matteo Castiglioni, and Alberto Marchesi.
\newblock Online two-sided markets: Many buyers enhance learning.
\newblock In {\em Advances in Neural Information Processing Systems (NeurIPS)}, 2025.
\newblock arXiv:2503.01529.

\bibitem[LCM26]{PerturbedMarkets26}
Anna Lunghi, Matteo Castiglioni, and Alberto Marchesi.
\newblock Regret minimization in bilateral trade with perturbed markets, 2026.
\newblock arXiv:2605.10475.

\bibitem[LQRW26]{LQRW26}
Zhengyang Liu, Ying Qin, Zeyu Ren, and Zihe Wang.
\newblock Second-best bilateral trade is $1/2$ efficient, 2026.
\newblock arXiv:2606.03849.

\bibitem[LQW26]{LQW26welfare}
Zhengyang Liu, Ying Qin, and Zihe Wang.
\newblock Sharp second-best welfare in bilateral and matching markets, 2026.
\newblock arXiv:2609.37366.

\bibitem[LRW23]{CSTOC}
Zhengyang Liu, Zeyu Ren, and Zihe Wang.
\newblock Improved approximation ratios of fixed-price mechanisms in bilateral trades.
\newblock In {\em Proceedings of the 55th Annual ACM Symposium on Theory of Computing}, STOC 2023, pages 751--760, New York, NY, USA, 2023. Association for Computing Machinery.

\bibitem[LSD{\etalchar{+}}26]{LSDKLYZ26}
Tingyi Lin, Yichen Shi, Ke~Dong, Jiazhuo Li, Shawn Yu, and Huanxi Zhang.
\newblock The exact welfare guarantee of fixed-price bilateral trade, 2026.
\newblock arXiv:2609.35888.

\bibitem[McA92]{TradeReduction}
R.~Preston McAfee.
\newblock A dominant strategy double auction.
\newblock {\em Journal of Economic Theory}, 56(2):434--450, 1992.

\bibitem[McA08]{mcaffee}
R.~Preston McAfee.
\newblock The gains from trade under fixed price mechanisms.
\newblock {\em Applied Economics Research Bulletin}, 1(1):1--10, 2008.

\bibitem[MM23]{MM23}
Mihai Manea and Eric Maskin.
\newblock Withholding and damage in {B}ayesian trade mechanisms.
\newblock {\em Games and Economic Behavior}, 142:243--265, 2023.

\bibitem[Mou09]{Moulin09}
Herv{\'e} Moulin.
\newblock Almost budget-balanced {VCG} mechanisms to assign multiple objects.
\newblock {\em Journal of Economic Theory}, 144(1):96--119, 2009.

\bibitem[MS83]{Myerson-Satterthwaite}
Roger~B Myerson and Mark~A Satterthwaite.
\newblock Efficient mechanisms for bilateral trading.
\newblock {\em Journal of Economic Theory}, 29(2):265--281, 1983.

\bibitem[RTZ26]{RTZ26}
Aviad Rubinstein, Xizhi Tan, and Zixin Zhou.
\newblock Approximately efficient multidimensional bilateral trade.
\newblock In {\em Proceedings of the 67th IEEE Symposium on Foundations of Computer Science (FOCS)}, 2026.
\newblock arXiv:2609.02872.

\bibitem[RW24]{RW-private}
Aviad Rubinstein and Kangning Wang.
\newblock Gains-from-trade approximations under correlation, 2024.
\newblock Unpublished manuscript.

\bibitem[SC26]{SC26}
Heyang Shang and Jing Chen.
\newblock An improved upper bound for the random-offerer mechanism in bilateral trade via recursive hard instances, 2026.
\newblock arXiv:2610.06547.

\bibitem[SHHA16]{SHA16}
Erel Segal-Halevi, Avinatan Hassidim, and Yonatan Aumann.
\newblock {SBBA}: A strongly-budget-balanced double-auction mechanism.
\newblock In {\em Algorithmic Game Theory (SAGT)}, 2016.

\bibitem[SW18]{SW18}
Jesse~A Schwartz and Quan Wen.
\newblock Robust trading mechanisms with budget surplus and partial trade.
\newblock {\em Economic Theory Bulletin}, 6(2):201--208, 2018.

\bibitem[SZ16]{ShaoZhou16}
Ran Shao and Lin Zhou.
\newblock Optimal allocation of an indivisible good.
\newblock {\em Games and Economic Behavior}, 100:95--112, 2016.

\bibitem[Yoo11]{Yoon11}
Kiho Yoon.
\newblock Optimal mechanism design when both allocative inefficiency and expenditure inefficiency matter.
\newblock {\em Journal of Mathematical Economics}, 47(6):670--676, 2011.

\end{thebibliography}
\bibliographystyle{alpha}
\appendix
\section{Omitted Proofs}
\label{app:proofs}

\subsection{Proof of Proposition~\ref{prop:prelim:dk-dominance}}
\label{app:proof-dk-dominance}

\begin{proof}[Proof of Proposition~\ref{prop:prelim:dk-dominance}]
Let $T$ be the set of profiles at which the given mechanism trades, that is, $T = \{ (\vs, \vb) \, |\, \allocation(\vs,\vb) = 1 \}$.  If $T$ is empty, the claim is immediate, so assume that it is not.  By individual rationality and weak budget balance, $\vs\le\vb$ for every $(\vs,\vb)\in T$.

We claim that $\vs'\le\vb$ for every two profiles $(\vs,\vb),(\vs',\vb')\in T$.  Suppose instead that $\vb<\vs'$.  Then $\vb<\vs'\le\vb'$, so by monotonicity the mechanism also trades at $(\vs,\vb')$.  By dominant-strategy incentive compatibility, the buyer's payment is the same at $(\vs,\vb')$ and at $(\vs,\vb)$, and by individual rationality at $(\vs,\vb)$ it is at most $\vb$.  Similarly, the amount that the seller receives is the same at $(\vs,\vb')$ and at $(\vs',\vb')$, and by individual rationality at $(\vs',\vb')$ it is at least $\vs'$.  Since $\vb<\vs'$, the buyer pays less than the seller receives at $(\vs,\vb')$, which violates weak budget balance.

Let $\price=\sup_{(\vs,\vb)\in T}\vs$.  By the claim, $\price$ is finite and $\vs\le\price\le\vb$ for every $(\vs,\vb)\in T$.  The mechanism that posts the price $\price$ is deterministic, dominant-strategy incentive-compatible, individually rational, and strongly budget-balanced, and it trades at every profile in $T$.
\end{proof}

\subsection{Proof of Lemma~\ref{lem:general-wbb-step-construction}}
\label{app:proof-wbb-construction}

\begin{proof}[Proof of Lemma~\ref{lem:general-wbb-step-construction}]
We first verify that the parameters of the construction are well defined.  The
restriction on $\slack$ guarantees that
$\sellerendpoint{i-1}<\buyerendpoint{i}$ for every $i\in[n]$.  It also
guarantees that every denominator in 
Equation~\eqref{eq:general-wbb-grid-condition} is positive: for every $1\le i<j\le n$,
the denominator is at least
\begin{align*}
    \qpoint{j}-\qpoint{i-1}-2\slack>0.
\end{align*}
Thus, $\growth$ is finite.

We next verify DSIC and IR.  In 
Equation~\eqref{eq:general-wbb-step-allocation}, the exponent of every positive
allocation probability lies between $0$ and $n-1$.  Hence, the normalization
by $\growth^{n-1}$ keeps all allocation probabilities in $[0,1]$, and the
smallest positive probability is $\growth^{-(n-1)}$.  Moreover, increasing the
seller's report can only increase her rounded index and therefore decrease the
allocation probability.  Similarly, increasing the buyer's report can only
increase his rounded index and therefore increase the allocation probability.
Thus, the allocation rule is nonincreasing in $\vs$ and nondecreasing in $\vb$.

Myerson's lemma therefore provides payments that make the mechanism DSIC and IR.  We use normalized
payments, which are zero whenever $\allocation(\vs,\vb)=0$.  For all agents' value profiles
$(\vs,\vb)$ with positive allocation probability, they are
\begin{align*}
    \paymentB(\vs,\vb)
    &=\vb\cdot\allocation(\vs,\vb)-\int_{\vs}^{\vb}\allocation(\vs,z)\d z,\\
    \paymentS(\vs,\vb)
    &=\vs\cdot\allocation(\vs,\vb)+\int_{\vs}^{\vb}\allocation(z,\vb)\d z.
\end{align*}
Thus, WBB holds immediately at all agents' profiles with zero
allocation probability.  At agents' profiles with positive allocation probability,
substituting these payments into
$\paymentB(\vs,\vb)\ge\paymentS(\vs,\vb)$ shows that WBB is equivalent to
\begin{align}\label{eq:constructed-rent-budget}
    \int_{\vs}^{\vb} \allocation(\vs,z)\d z
    +
    \int_{\vs}^{\vb} \allocation(z,\vb)\d z
    \le
    (\vb-\vs)\allocation(\vs,\vb).
\end{align}
It remains to verify this inequality.  Fix the agents' value profile $(\vs,\vb)$ with
positive allocation probability and write
\begin{align*}
    i=\sellerround{\vs}+1,
    \qquad
    j=\buyerround{\vb}.
\end{align*}
Partitioning the two integrals at the buyer and seller thresholds, and using
that $\allocation$ is constant between consecutive thresholds, gives
\begin{align*}
    \int_{\vs}^{\vb} \allocation(\vs,z)\d z
    &=\frac{1}{\growth^{n-1}}
    \left(
        \sum_{r=i}^{j-1}
        (\buyerendpoint{r+1}-\buyerendpoint{r})\growth^{r-i}
        +(\vb-\buyerendpoint{j})\growth^{j-i}
    \right),\\
    \int_{\vs}^{\vb} \allocation(z,\vb)\d z
    &=\frac{1}{\growth^{n-1}}
    \left(
        (\sellerendpoint{i-1}-\vs)\growth^{j-i}
        +\sum_{r=i+1}^{j}
        (\sellerendpoint{r-1}-\sellerendpoint{r-2})\growth^{j-r}
    \right).
\end{align*}
Substituting these expressions into
Equation~\eqref{eq:constructed-rent-budget} and rearranging the inequality yields
\begin{align}\label{eq:general-wbb-reduced-budget}
    (\buyerendpoint{j}-\sellerendpoint{i-1})\growth^{j-i}
    \ge
    \sum_{r=i}^{j-1}(\buyerendpoint{r+1}-\buyerendpoint{r})\growth^{r-i}
    +
    \sum_{r=i+1}^{j}(\sellerendpoint{r-1}-\sellerendpoint{r-2})\growth^{j-r}.
\end{align}
So we are left to show that this indeed holds. We consider two cases.
If $i=j$, both sums vanish, and the inequality follows immediately.  Suppose
instead that $i<j$.  Every exponent on the right-hand side is at most
$j-i-1$.  Moreover, by the definitions of the thresholds,
\begin{align*}
    \frac{(\buyerendpoint{j}-\buyerendpoint{i})+(\sellerendpoint{j-1}-\sellerendpoint{i-1})}
    {\buyerendpoint{j}-\sellerendpoint{i-1}}
    &=1+
    \frac{\qpoint{j-1}-\qpoint{i}+2\slack}{
        \qpoint{j}-\qpoint{i-1}
        -\slack\cdot\mathbbm{1}_{\{i>1\}}
        -\slack\cdot\mathbbm{1}_{\{j<n\}}}
    \le \growth,
\end{align*}
where the inequality follows from Equation~\eqref{eq:general-wbb-grid-condition}. We can therefore bound the right-hand
side of Equation~\eqref{eq:general-wbb-reduced-budget} as follows:
\begin{align*}
    &\sum_{r=i}^{j-1}(\buyerendpoint{r+1}-\buyerendpoint{r})\growth^{r-i}
    +
    \sum_{r=i+1}^{j}(\sellerendpoint{r-1}-\sellerendpoint{r-2})\growth^{j-r}\\
    &\quad\le \growth^{j-i-1}
    \left(
        (\buyerendpoint{j}-\buyerendpoint{i})
        +(\sellerendpoint{j-1}-\sellerendpoint{i-1})
    \right)\\
    &\quad\le \growth^{j-i}(\buyerendpoint{j}-\sellerendpoint{i-1}).
\end{align*}
Thus, inequality~\eqref{eq:general-wbb-reduced-budget} holds, and the
constructed mechanism is indeed WBB.
\end{proof}

%
\subsection{Proof of Lemma~\ref{lem:separation}}
\label{app:proof-separation}

\begin{proof}[Proof of Lemma~\ref{lem:separation}]
We first show two claims.  Claim~\ref{clm:grid-condition} shows that every
positive allocation probability of $\wbbmechanism$ is greater than $\frac{1}{2}$,
and Claim~\ref{clm:level-two-comparison} uses it to compare, at the profiles
in the sets $\levelrect{i}$, our mechanism with an SBB
mechanism.

\begin{claim}\label{clm:grid-condition}
The allocation rule $\allocation$ of $\wbbmechanism$ satisfies $\allocation(\vs,\vb)>\frac{1}{2}$
for every $(\vs,\vb)$ with $\sellerround{\vs}<\buyerround{\vb}$.
\end{claim}

\begin{proof}[Proof of Claim~\ref{clm:grid-condition}]
Let 
{$\separation$ and $L$ be as in}
Definition~\ref{def:separation-grid}.  By the definition,
$\slack\le\separation L/100<L/2$, which is the condition of
Lemma~\ref{lem:general-wbb-step-construction} in
Equation~\eqref{eq:general-wbb-slack-condition}.  Fix
$1\le i<j\le n$.  Since $i<j$, the difference
$\qpoint{j}-\qpoint{i-1}=\sum_{r=i}^{j}(\qpoint{r}-\qpoint{r-1})$ is a sum of
at least two gaps of the grid, so it is at least $2L$, and hence
$2\slack\le(\separation/100)\cdot 2L\le(\separation/100)(\qpoint{j}-\qpoint{i-1})$.  Therefore,
\begin{align*}
    \frac{\qpoint{j-1}-\qpoint{i}+2\slack}
         {\qpoint{j}-\qpoint{i-1}
          -\slack\cdot\mathbbm{1}_{\{i>1\}}
          -\slack\cdot\mathbbm{1}_{\{j<n\}}}
    &\le
    \frac{\qpoint{j-1}-\qpoint{i}+2\slack}
         {\qpoint{j}-\qpoint{i-1}-2\slack}
    \le
    \frac{(4\separation+\separation/100)(\qpoint{j}-\qpoint{i-1})}
         {(1-\separation/100)(\qpoint{j}-\qpoint{i-1})}\\
    &=
    \frac{4\separation+\separation/100}{1-\separation/100}
    \le 8\separation,
\end{align*}
where the first inequality holds since each indicator is at most $1$, the
second follows from Equation~\eqref{eq:grid-condition} for the first term of
the numerator and from $2\slack\le(\separation/100)(\qpoint{j}-\qpoint{i-1})$
for the remaining terms, and the last holds since $\separation\le 1/200$.
Since this holds for every $1\le i<j\le n$, the growth factor in
Equation~\eqref{eq:general-wbb-grid-condition} satisfies
$\growth\le 1+8\separation$, and
$\growth^{n-1}\le(1+8\separation)^{n-1}<e^{8\separation n}\le e^{0.08}<2$.
The claim follows since, by
Lemma~\ref{lem:general-wbb-step-construction}, $\allocation(\vs,\vb)\ge\growth^{-(n-1)}>\frac{1}{2}$
whenever $\sellerround{\vs}<\buyerround{\vb}$.
\end{proof}

%
\begin{claim}\label{clm:level-two-comparison}
Let $\sbbmechanism$ be a dominant-strategy incentive-compatible, individually
rational, and strongly budget-balanced mechanism.  By~\cite{CP16}, the
allocation rule of $\sbbmechanism$ satisfies $\sbballocation(\vs,\vb)\le\pricemeasure([\vs,\vb])$
for some distribution $\pricemeasure$ over posted prices.  Then for every
$1\le i\le n-3$ and every $(\vs,\vb)\in\levelrect{i}$,
\begin{align*}
    \sgft_{\wbbmechanism}(\vs,\vb)>\frac{\vb-\vs}{2}
    \qquad\text{and}\qquad
    \sgft_{\sbbmechanism}(\vs,\vb)\le(\vb-\vs)\cdot\pricemeasure([\qpoint{i},\qpoint{i+2}]).
\end{align*}
\end{claim}

\begin{proof}[Proof of Claim~\ref{clm:level-two-comparison}]
Since $1\le i$ and $i+2\le n-1$, we have
$\sellerendpoint{i-1}<\vs\le\sellerendpoint{i}$ and
$\buyerendpoint{i+2}\le \vb<\buyerendpoint{i+3}$, so $\sellerround{\vs}=i$ and
$\buyerround{\vb}=i+2$.  For the payments of
Lemma~\ref{lem:general-wbb-step-construction},
\begin{align*}
    \sgft_{\wbbmechanism}(\vs,\vb)
    &=
    \int_0^{\vb} \allocation(\vs,z)\d z
    +
    \int_{\vs}^1 \allocation(z,\vb)\d z
    \ge
    \int_{\buyerendpoint{i+1}}^{\buyerendpoint{i+2}} \allocation(\vs,z)\d z
    +
    \int_{\sellerendpoint{i}}^{\sellerendpoint{i+1}} \allocation(z,\vb)\d z\\
    &>
    \frac{1}{2}\left(\buyerendpoint{i+2}-\buyerendpoint{i+1}\right)
    +
    \frac{1}{2}\left(\sellerendpoint{i+1}-\sellerendpoint{i}\right)
    =
    \frac{1}{2}\left(\qpoint{i+2}-\qpoint{i+1}\right)
    +
    \frac{1}{2}\left(\qpoint{i+1}-\qpoint{i}\right)\\
    &=
    \frac{\qpoint{i+2}-\qpoint{i}}{2}
    \ge
    \frac{\vb-\vs}{2}.
\end{align*}
The first equality holds by Myerson's lemma, applied to the normalized
payments of Lemma~\ref{lem:general-wbb-step-construction}: the buyer's utility is
$\int_0^{\vb} \allocation(\vs,z)\d z$ and the seller's utility minus $\vs$ is
$\int_{\vs}^1 \allocation(z,\vb)\d z$.  The first inequality holds since the
allocation probabilities are nonnegative,
$[\buyerendpoint{i+1},\buyerendpoint{i+2}]\subseteq[0,\vb]$, and
$[\sellerendpoint{i},\sellerendpoint{i+1}]\subseteq[\vs,1]$.  The second
inequality holds by Claim~\ref{clm:grid-condition}, since
$\buyerround{z}=i+1>\sellerround{\vs}$ for
$z\in[\buyerendpoint{i+1},\buyerendpoint{i+2})$, and
$\sellerround{z}=i+1<\buyerround{\vb}$ for
$z\in(\sellerendpoint{i},\sellerendpoint{i+1}]$.  The second equality holds
since $\buyerendpoint{r}=\qpoint{r}-\slack$ and
$\sellerendpoint{r}=\qpoint{r}+\slack$ for every $1\le r\le n-1$, and the
indices $i$, $i+1$, and $i+2$ are all in this range, so $\slack$ cancels in
each difference. The third inequality holds since $\qpoint{i}\le\vs$ and $\vb\le\qpoint{i+2}$.

For the SBB mechanism, the payments cancel, so
\begin{align*}
    \sgft_{\sbbmechanism}(\vs,\vb)
    \le
    (\vb-\vs)\cdot\pricemeasure([\vs,\vb])
    \le
    (\vb-\vs)\cdot\pricemeasure([\qpoint{i},\qpoint{i+2}]),
\end{align*}
since $[\vs,\vb]\subseteq[\qpoint{i},\qpoint{i+2}]$.
\end{proof}

%

The sets $\levelrect{i}$, $i\in I$, are pairwise disjoint, since
$\slack$ is smaller than every gap of the grid.

By individual rationality, $\sgft_{\wbbmechanism}(\vs,\vb)\ge0$ at every
profile.  By Claim~\ref{clm:level-two-comparison}, for every
$(\vs,\vb)\in\levelrect{i}$, we have $\sgft_{\wbbmechanism}(\vs,\vb)>(\vb-\vs)/2$.  Hence, by the first condition,
\begin{align*}
    \sgft_{\wbbmechanism}(\distjoint)
    >
    \sum_{i\in I}\frac{1}{2}\Exp\left[(\vb-\vs)\cdot\indicator_{(\vs,\vb)\in\levelrect{i}}\right]
    \ge
    \sum_{i\in I}\frac{\surplusunit}{4}
    =\frac{m\,\surplusunit}{4}.
\end{align*}

For the SBB mechanism, let $\pricemeasure$ be the
distribution over posted prices from~\cite{CP16}.  Its payments cancel, so its
traders' surplus at $(\vs,\vb)$ is at most $(\vb-\vs)\,\pricemeasure([\vs,\vb])$, which is
at most $(\vb-\vs)\cdot\indicator_{\vs<\vb}$.  By Claim~\ref{clm:level-two-comparison}, it is at most
$(\vb-\vs)\cdot\pricemeasure([\qpoint{i},\qpoint{i+2}])$ for every
$(\vs,\vb)\in\levelrect{i}$.  Hence, by the two conditions,
\begin{align*}
    \sgft_{\sbbmechanism}(\distjoint)
    \le
    \surplusunit\sum_{i\in I}\pricemeasure([\qpoint{i},\qpoint{i+2}])
    +\leftover
    \le
    \overlap\,\surplusunit+\leftover,
\end{align*}
where the last inequality holds since every price belongs to at most $\overlap$ of
the intervals $[\qpoint{i},\qpoint{i+2}]$.
\end{proof}

\subsection{Proof of Lemma~\ref{lem:ind-grid}}
\label{app:proof-ind-grid}

\begin{proof}[Proof of Lemma~\ref{lem:ind-grid}]
Since $\qpoint{j-1}-\qpoint{i}=\sum_{r=i+1}^{j-1}\qlength{r}$ and
$\qlength{i}+\qlength{j}\le\qpoint{j}-\qpoint{i-1}$, it is enough to show that
\begin{align*}
    \sum_{r=i+1}^{j-1}\qlength{r}
    \le
    4\separation(\qlength{i}+\qlength{j})
    \qquad\text{for every } 1\le i<j\le n.
\end{align*}
We first bound, for every $1\le i\le k-1$, the sum of the gaps between
$\qpoint{i}$ and the middle point $\qpoint{k}=1/2$, that is,
$\sum_{r=i+1}^{k}\qlength{r}=\qpoint{k}-\qpoint{i}$.  For $i=1$,
\[
    \sum_{r=2}^{k}\qlength{r}=\qpoint{k}-\qpoint{1}=\separation \le 4\separation \qlength{1},
\]
because $\qlength{1}=\frac{1}{2}-\separation\ge\frac{1}{4}$. For $2\le i\le k-1$,
\[
    \sum_{r=i+1}^{k}\qlength{r}=\qpoint{k}-\qpoint{i}=\separation^i
    =\frac{\separation}{1-\separation}\qlength{i}
    \le 2\separation \qlength{i}
    \le 4\separation \qlength{i},
\]
where the first inequality holds since $1-\separation\ge\frac{1}{2}$.  Hence,
\begin{align}\label{eq:ind-left-bound}
    \sum_{r=i+1}^{k}\qlength{r}\le 4\separation\qlength{i}
    \qquad\text{for every } 1\le i\le k-1.
\end{align}
Symmetrically, we bound, for every $k+2\le j\le n$, the sum of the gaps
between the middle point $\qpoint{k}$ and $\qpoint{j-1}$, that is,
$\sum_{r=k+1}^{j-1}\qlength{r}=\qpoint{j-1}-\qpoint{k}$.  For $j=n$,
\[
    \sum_{r=k+1}^{n-1}\qlength{r}=\qpoint{n-1}-\qpoint{k}=\separation\le 4\separation \qlength{n},
\]
and for $k+2\le j\le n-1$,
\[
    \sum_{r=k+1}^{j-1}\qlength{r}=\qpoint{j-1}-\qpoint{k}=\separation^{n-j+1}
    =\frac{\separation}{1-\separation}\qlength{j}
    \le 2\separation \qlength{j}
    \le 4\separation \qlength{j},
\]
where the first inequality holds since $1-\separation\ge\frac{1}{2}$.  Hence,
\begin{align}\label{eq:ind-right-bound}
    \sum_{r=k+1}^{j-1}\qlength{r}\le 4\separation\qlength{j}
    \qquad\text{for every } k+2\le j\le n.
\end{align}
Now fix $1\le i<j\le n$.  We consider three cases.
\begin{itemize}
    \item If $j\le k$, then $i\le k-1$, and
    \begin{align*}
        \sum_{r=i+1}^{j-1}\qlength{r}
        \le
        \sum_{r=i+1}^{k}\qlength{r}
        \le
        4\separation\qlength{i},
    \end{align*}
    where the first inequality holds since $j-1\le k$ and the gaps are
    positive, and the second is Equation~\eqref{eq:ind-left-bound}.
    \item If $i\ge k+1$, then $j\ge k+2$, and
    \begin{align*}
        \sum_{r=i+1}^{j-1}\qlength{r}
        \le
        \sum_{r=k+1}^{j-1}\qlength{r}
        \le
        4\separation\qlength{j},
    \end{align*}
    where the first inequality holds since $i+1\ge k+1$ and the gaps are
    positive, and the second is Equation~\eqref{eq:ind-right-bound}.
    \item If $i\le k<j$, then
    \begin{align*}
        \sum_{r=i+1}^{j-1}\qlength{r}
        =
        \sum_{r=i+1}^{k}\qlength{r}
        +
        \sum_{r=k+1}^{j-1}\qlength{r}
        \le
        4\separation\qlength{i}+4\separation\qlength{j}.
    \end{align*}
    The first sum is $0$ if $i=k$, and is bounded by
    Equation~\eqref{eq:ind-left-bound} otherwise.  The second sum is $0$ if
    $j=k+1$, and is bounded by Equation~\eqref{eq:ind-right-bound} otherwise.
\end{itemize}
In all cases,
$\sum_{r=i+1}^{j-1}\qlength{r}\le 4\separation(\qlength{i}+\qlength{j})$.
\end{proof}

\subsection{Proof of Lemma~\ref{lem:equal-trader-surplus}}
\label{app:proof-equal-trader-surplus}

\begin{proof}[Proof of Lemma~\ref{lem:equal-trader-surplus}]
The mirror of a profile $(\vs,\vb)$ is the profile $(1-\vb,1-\vs)$, in which
the seller's value is $1-\vb$ and the buyer's value is $1-\vs$.  The mirror of
the mirror is the original profile.  We show that a profile and its mirror are
equally likely, and that the buyer's surplus at a profile equals the seller's
surplus at its mirror.

\emph{Equally likely.}  By Equation~\eqref{eq:ind-grid},
$1-\qpoint{i}=\qpoint{n-i}$ for every $i$.  Since $n=2k$, the mirror of
$\sellerinterval{\ell}$ is $\buyerinterval{k-\ell}$, and the mirror of
$\buyerinterval{r}$ is $\sellerinterval{k-r}$.  Hence, the mirror of a profile
in $\sellerinterval{\ell}\times\buyerinterval{r}$ is in
$\sellerinterval{k-r}\times\buyerinterval{k-\ell}$.  Moreover,
$\profilegap{k-j}=\profilegap{j}$ for every $j$, so
\begin{align*}
    \Pr\left[(\vs,\vb)\in\sellerinterval{\ell}\times\buyerinterval{r}\right]
    =\surplusunit\,\sellerweight{\ell}\,\buyerweight{r}
    =\surplusunit\,\frac{R^{\ell-r}}{\sqrt{\profilegap{\ell}\profilegap{r}}}
    =\surplusunit\,\sellerweight{k-r}\,\buyerweight{k-\ell}
    =\Pr\left[(\vs,\vb)\in\sellerinterval{k-r}\times\buyerinterval{k-\ell}\right].
\end{align*}
Since the values are uniform within these intervals, a profile and its mirror
are equally likely.

\emph{Switched surpluses.}  The mirror maps every seller threshold
$\sellerendpoint{i}$ to the buyer threshold $\buyerendpoint{n-i}$, so a
seller's value with rounded index $i$ becomes a buyer's value with rounded
index $n-i$, and vice versa.  Hence, a profile and its mirror are at the same
level, and since the allocation rule depends only on the level,
$\allocation(1-\vb,1-\vs)=\allocation(\vs,\vb)$.  With the payments of
Lemma~\ref{lem:general-wbb-step-construction}, the buyer's surplus at
$(\vs,\vb)$ is $\int_0^{\vb}\allocation(\vs,z)\d z$, and the seller's surplus
is $\int_{\vs}^1\allocation(z,\vb)\d z$.  Since $\allocation(1-z,1-\vs)=\allocation(\vs,z)$ for every $z$,
\begin{align*}
    \int_0^{\vb}\allocation(\vs,z)\d z
    =
    \int_0^{\vb}\allocation(1-z,1-\vs)\d z
    =
    \int_{1-\vb}^{1}\allocation(w,1-\vs)\d w,
\end{align*}
which is the seller's surplus at the mirror profile.

Since a profile and its mirror are equally likely, it follows that
$\sgft^{b}_{\wbbmechanism}(\distseller\times\distbuyer)
=\sgft^{s}_{\wbbmechanism}(\distseller\times\distbuyer)$.  The two sum to
$\sgft_{\wbbmechanism}(\distseller\times\distbuyer)$, so each of them is half
of it.
\end{proof}

\subsection{Proof of Lemma~\ref{lem:corr-grid}}
\label{app:proof-corr-grid}

\begin{proof}[Proof of Lemma~\ref{lem:corr-grid}]
Fix $1\le i<j\le n$.  Since every gap is $\separation$ times the next one,
$\qlength{r}=\separation^{\,j-r}\qlength{j}$ for every $r<j$.  Hence,
\begin{align*}
    \qpoint{j-1}-\qpoint{i}
    =
    \sum_{r=i+1}^{j-1}\qlength{r}
    =
    \qlength{j}\sum_{r=i+1}^{j-1}\separation^{\,j-r}
    \le
    \qlength{j}\sum_{t=1}^{\infty}\separation^{t}
    =
    \frac{\separation}{1-\separation}\qlength{j}
    \le
    2\separation\qlength{j}
    \le
    4\separation(\qpoint{j}-\qpoint{i-1}).
\end{align*}
The third equality is the sum of the geometric series,
$\sum_{t=1}^{\infty}\separation^{t}=\frac{\separation}{1-\separation}$ for
$0<\separation<1$.  The second inequality
holds since $\separation=1/(100n)\le\frac{1}{2}$, so
$\frac{1}{1-\separation}\le2$, and the last holds since
$\qpoint{j}-\qpoint{i-1}=\sum_{r=i}^{j}\qlength{r}\ge\qlength{j}$.
\end{proof}

\subsection{Proof of Lemma~\ref{lem:corr-seller-value}}
\label{app:proof-corr-seller-value}

\begin{proof}[Proof of Lemma~\ref{lem:corr-seller-value}]
Every $\vs\in\sellerinterval{i}$ satisfies $\vs\le\qpoint{i-1}+\eta$.  Since
every gap is $\separation$ times the next one,
$\qpoint{i-1}=\sum_{r=1}^{i-1}\qlength{r}\le\frac{\separation^2}{1-\separation}\qlength{i+1}\le\frac{\separation^2}{1-\separation}\profilegap{i}$,
and by the choice of $\eta$, $\eta\le\separation^2\profilegap{i}$.  Hence,
\begin{align*}
    \Exp[\vs]
    \le
    \sum_{i=2}^{n-2}\frac{\surplusunit}{\profilegap{i}}
    \left(\qpoint{i-1}+\eta\right)
    \le
    (n-3)\,\surplusunit
    \left(\frac{\separation^2}{1-\separation}+\separation^2\right)
    \le
    3(n-3)\separation^2\surplusunit
    \le
    \surplusunit.
\end{align*}
The first inequality holds since
$\Pr\left[(\vs,\vb)\in\sellerinterval{i}\times\buyerinterval{i}\right]=\surplusunit/\profilegap{i}$
and $\vs\le\qpoint{i-1}+\eta$ on $\sellerinterval{i}$.  The second holds since,
by the two bounds above, each of the $n-3$ terms satisfies
$\frac{\qpoint{i-1}}{\profilegap{i}}\le\frac{\separation^2}{1-\separation}$ and
$\frac{\eta}{\profilegap{i}}\le\separation^2$.  The third holds since
$\frac{1}{1-\separation}\le2$, and the last holds since
$3(n-3)\separation^2\le\frac{3n}{(100n)^2}<1$.
\end{proof}

\subsection{Proof of Remark~\ref{rem:constant-fraction}}
\label{app:proof-constant-fraction}

\begin{proof}[Proof of Remark~\ref{rem:constant-fraction}]
For independent values, consider the proof of Theorem~\ref{thm:wbb-sbb-unbounded-gap-ind-gft}.  Each of the $k-1$ sets $\levelrect{i}$, $i\in I$, contributes at most $\surplusunit$ to the first-best gains-from-trade, and the profiles outside these sets contribute at most $\surplusunit$, so $\fbgft(\distseller\times\distbuyer)\le k\surplusunit$.  By Lemma~\ref{lem:separation}, $\sgft_{\wbbmechanism}(\distseller\times\distbuyer)>(k-1)\surplusunit/4$.  Hence, the fraction is more than $(k-1)/(4k)\ge1/8$, since $k\ge2$.

For correlated values, consider the proof of Theorem~\ref{thm:wbb-sbb-unbounded-gap-correlated-residual-welfare}.  Each of the $n-3$ sets $\levelrect{i}$, $i\in I$, contributes at most $\surplusunit$ to the first-best gains-from-trade, and no other profile is in the support, so the first-best gains-from-trade is at most $(n-3)\surplusunit$.  The first-best social welfare is this quantity plus $\Exp[\vs]$, and $\Exp[\vs]\le\surplusunit$ by Lemma~\ref{lem:corr-seller-value}, so $\fbwelfare(\distjoint)\le(n-2)\surplusunit$.  By Lemma~\ref{lem:separation}, $\swelfare_{\wbbmechanism}(\distjoint)\ge\sgft_{\wbbmechanism}(\distjoint)>(n-3)\surplusunit/4$.  Hence, the fraction is more than $(n-3)/(4(n-2))\ge1/8$, since $n\ge4$.
\end{proof}

\subsection{Proof of Lemma~\ref{lem:lb-gft-trade-mass}}
\label{app:proof-lb-gft-trade-mass}

\begin{proof}[Proof of Lemma~\ref{lem:lb-gft-trade-mass}]
Let $t$, $\distseller$, $\distbuyer$, $\mechanism$, and $A$ be as in the proof of Theorem~\ref{thm:tie-wbb-no-constant-ind-gft}.  By Inequality~\eqref{ineq:wbb-ir}, IR and WBB imply that $\allocation(\vs,\vb)=0$ whenever $\vb<\vs$.  Hence, by Equation~\eqref{equ:def-gft}, only the profiles with $\vb\ge\vs$ contribute to the gains-from-trade.  By Equation~\eqref{eq:lb-distributions}, for every $r\in[0,t]$ and $\vs\in[0,t-r]$, the density of the profile $(\vs,\vs+r)$ is
\begin{equation}\label{eq:lb-diagonal-density}
    \pdfseller(\vs)\cdot\pdfbuyer(\vs+r)=\lambda^2e^{-t}e^{-r},
\end{equation}
which depends only on its gain from trade $r$.  Writing $\vb=\vs+r$ and using Equation~\eqref{eq:lb-diagonal-density},
\begin{align*}
    \gft_{\mechanism}(\distseller\times\distbuyer)
    &=\int_0^t\int_{\vs}^{t}(\vb-\vs)\allocation(\vs,\vb)\pdfseller(\vs)\pdfbuyer(\vb)\d{\vb}\d{\vs}\\
    &=\lambda^2e^{-t}\int_0^t\int_0^{t-\vs}re^{-r}\allocation(\vs,\vs+r)\d{r}\d{\vs}\\
    &=\lambda^2e^{-t}\int_0^tre^{-r}\int_0^{t-r}\allocation(\vs,\vs+r)\d{\vs}\d{r}
    =\lambda^2e^{-t}\int_0^tre^{-r}A(r)\d{r},
\end{align*}
where the third equality changes the order of integration over the pairs $(\vs,r)$ with $\vs,r\ge0$ and $\vs+r\le t$.
\end{proof}

\subsection{Proof of Lemma~\ref{lem:lb-trade-mass-bound}}
\label{app:proof-lb-trade-mass-bound}

\begin{proof}[Proof of Lemma~\ref{lem:lb-trade-mass-bound}]
Let $t$, $\mechanism$, and $A$ be as in the proof of Theorem~\ref{thm:tie-wbb-no-constant-ind-gft}.  For every $r\in[0,t]$, let $B(r)=\int_0^rA(s)\d{s}$.  We first show that, for every $r\in[0,t]$,
\begin{equation}\label{eq:diagonal-recursion}
    rA(r)\ge2B(r)-r^2.
\end{equation}
Fix $r\in[0,t]$ and $\vs\in[0,t-r]$.  Applying Lemma~\ref{lem:wbb-allocation-inequality} to the profile $(\vs,\vs+r)$ and substituting $z=\vs+s$ in both integrals gives
\[
    \int_0^r\allocation(\vs,\vs+s)\d{s}
    +\int_0^r\allocation(\vs+s,\vs+r)\d{s}
    \le r\allocation(\vs,\vs+r).
\]
The profile $(\vs,\vs+s)$ in the first integral has gain from trade $s$, and the profile $(\vs+s,\vs+r)$ in the second integral has gain from trade $r-s$.  Integrating over $\vs\in[0,t-r]$ and changing the order of integration,
\begin{equation}\label{eq:diagonal-recursion-integrated}
    rA(r)
    \ge
    \int_0^r\int_0^{t-r}\allocation(\vs,\vs+s)\d{\vs}\d{s}
    +\int_0^r\int_0^{t-r}\allocation(\vs+s,\vs+r)\d{\vs}\d{s}.
\end{equation}
We compare each inner integral to $A$.  Fix $s\in[0,r]$.  The first inner integral covers the profiles with gain from trade $s$, except for the seller's values in $[t-r,t-s]$, an interval of length $r-s$.  Since $\allocation\le1$,
\[
    \int_0^{t-r}\allocation(\vs,\vs+s)\d{\vs}\ge A(s)-(r-s).
\]
For the second inner integral, substituting $\ell=\vs+s$ gives
\[
    \int_0^{t-r}\allocation(\vs+s,\vs+r)\d{\vs}
    =\int_s^{t-(r-s)}\allocation(\ell,\ell+(r-s))\d{\ell},
\]
which covers the profiles with gain from trade $r-s$, except for the seller's values in $[0,s]$, an interval of length $s$.  Since $\allocation\le1$,
\[
    \int_0^{t-r}\allocation(\vs+s,\vs+r)\d{\vs}\ge A(r-s)-s.
\]
Substituting both bounds into Inequality~\eqref{eq:diagonal-recursion-integrated} gives
\[
    rA(r)
    \ge\int_0^r\bigl(A(s)-(r-s)\bigr)\d{s}+\int_0^r\bigl(A(r-s)-s\bigr)\d{s}
    =2B(r)-r^2,
\]
which proves Inequality~\eqref{eq:diagonal-recursion}.

We now use Inequality~\eqref{eq:diagonal-recursion} to bound $B$.  Since $0\le\allocation\le1$, we have $0\le A(u)\le t-u$ for every $u\in[0,t]$.  Hence, the function $B$ is Lipschitz, $B'(u)=A(u)$ for almost every $u$, and $B(t)\le\int_0^t(t-u)\d{u}=t^2/2$.  Fix $r\in(0,t]$, and let $G(u)=B(u)/u^2$ for $u\in[r,t]$.  The function $G$ is Lipschitz on $[r,t]$, and for almost every $u\in[r,t]$, by Inequality~\eqref{eq:diagonal-recursion},
\[
    G'(u)=\frac{uA(u)-2B(u)}{u^3}\ge-\frac{u^2}{u^3}=-\frac1u.
\]
Integrating over $[r,t]$ gives
\[
    G(t)-G(r)\ge-\int_r^t\frac{1}{u}\d{u}=-\log\frac tr.
\]
Since $B(t)\le t^2/2$, we have $G(t)\le1/2$, and therefore $G(r)\le\frac12+\log\frac tr$.  Multiplying by $r^2$ gives, for every $r\in(0,t]$,
\begin{equation}\label{eq:B-bound}
    B(r)\le r^2\left(\frac12+\log\frac tr\right).
\end{equation}
Finally, we use Inequality~\eqref{eq:B-bound} to bound the integral in the lemma.  Since $B(0)=0$ and $B'=A$ almost everywhere, integration by parts gives
\[
    \int_0^tre^{-r}A(r)\d{r}
    =te^{-t}B(t)+\int_0^t(r-1)e^{-r}B(r)\d{r}.
\]
For the first term, $B(t)\le t^2/2$ and $t^3e^{-t}\le27e^{-3}<2$, so $te^{-t}B(t)<1$.  For the second term, $(r-1)e^{-r}B(r)\le0$ for $r\in[0,1]$, and for $r\in[1,t]$, Inequality~\eqref{eq:B-bound} gives $B(r)\le r^2(\frac12+\log t)$.  Therefore,
\[
    \int_0^t(r-1)e^{-r}B(r)\d{r}
    \le\left(\frac12+\log t\right)\int_1^\infty(r-1)r^2e^{-r}\d{r}
    =\frac{11}{e}\left(\frac12+\log t\right).
\]
Combining the two bounds proves the lemma.
\end{proof}

\subsection{Proof of Lemma~\ref{lem:lb-first-best}}
\label{app:proof-lb-first-best}

\begin{proof}[Proof of Lemma~\ref{lem:lb-first-best}]
Let $t$, $\distseller$, and $\distbuyer$ be as in the proof of Theorem~\ref{thm:tie-wbb-no-constant-ind-gft}.  The first-best gains-from-trade are $\Exp[\max\{\vb-\vs,0\}]$, which are the gains-from-trade of the allocation that trades exactly at the profiles with $\vb\ge\vs$.  The computation in the proof of Lemma~\ref{lem:lb-gft-trade-mass} applies to this allocation, for which $A(r)=t-r$.  This gives
\[
    \fbgft(\distseller\times\distbuyer)
    =\lambda^2e^{-t}\int_0^t(t-r)re^{-r}\d{r}
    =\lambda^2\left((t-2)e^{-t}+(t+2)e^{-2t}\right).
\]
Since $t\ge4$, we have $t-2\ge t/2$, and the inequality follows.
\end{proof}

\subsection{Proof of Lemma~\ref{lem:lb-corr-welfare-trade}}
\label{app:proof-lb-corr-welfare-trade}

\begin{proof}[Proof of Lemma~\ref{lem:lb-corr-welfare-trade}]
Let $\georatio$, the grid $\qpoint{0},\ldots,\qpoint{n}$, $\distjoint$, and $\mechanism$ be as in the proof of Theorem~\ref{thm:wbb-tie-no-constant-correlated-welfare}.  Fix $i\in[n]$.  By Equation~\eqref{equ:def-welfare}, and since $\qpoint{i-1}=\qpoint{i}/\georatio$, the social welfare of $\mechanism$ at the profile $(\qpoint{i-1},\qpoint{i})$ is
\[
    \qpoint{i}\cdot\allocation(\qpoint{i-1},\qpoint{i})
    +\qpoint{i-1}\cdot\bigl(1-\allocation(\qpoint{i-1},\qpoint{i})\bigr)
    =\qpoint{i}\left(\frac{1}{\georatio}+\left(1-\frac{1}{\georatio}\right)\allocation(\qpoint{i-1},\qpoint{i})\right).
\]
Multiplying by $\profileprob{i}=1/(\normalizer\qpoint{i})$ and summing over $i\in[n]$ proves the lemma.
\end{proof}

\subsection{Proof of Lemma~\ref{lem:geometric-staircase-bound}}
\label{app:proof-geometric-staircase-bound}

\begin{proof}[Proof of Lemma~\ref{lem:geometric-staircase-bound}]
Let $\georatio$, the grid $\qpoint{0},\ldots,\qpoint{n}$, and $\mechanism$ be as in the proof of Theorem~\ref{thm:wbb-tie-no-constant-correlated-welfare}.  Recall that $\georatio=\sqrt n\ge2$ and $\qpoint{i}=\georatio\qpoint{i-1}$ for every $i\in[n]$.
Fix \(j\ge2\) and apply Lemma~\ref{lem:wbb-allocation-inequality} at
\((\qpoint{0},\qpoint{j})\).  We use the following two lower bounds on the
integrals in that lemma:
\begin{align*}
    \buyerintegralbound{j}
    &=
    \sum_{r=1}^{j-1}
    (\qpoint{r+1}-\qpoint{r})
    \cdot\allocation(\qpoint{0},\qpoint{r}),
    &
    \sellerintegralbound{j}
    &=
    \sum_{i=2}^{j}
    (\qpoint{i-1}-\qpoint{i-2})
    \cdot\allocation(\qpoint{i-1},\qpoint{i}).
\end{align*}
Indeed, on
\([\qpoint{r},\qpoint{r+1}]\), monotonicity in the buyer's value gives
\[
    \allocation(\qpoint{0},z)
    \ge \allocation(\qpoint{0},\qpoint{r}).
\]
Similarly, for \(i\le j\) and
\(z\in[\qpoint{i-2},\qpoint{i-1}]\), monotonicity in both reports gives
\[
    \allocation(z,\qpoint{j})
    \ge
    \allocation(\qpoint{i-1},\qpoint{j})
    \ge
    \allocation(\qpoint{i-1},\qpoint{i}).
\]
The first inequality follows from monotonicity in the seller's report, and the
second from monotonicity in the buyer's report.
The two integrals in the weak-budget-balance allocation inequality are therefore at least
\(\buyerintegralbound{j}\) and \(\sellerintegralbound{j}\), respectively.  Hence
\begin{equation}
\label{eq:geometric-row-wbb}
    \buyerintegralbound{j}+\sellerintegralbound{j}
    \le
    (\qpoint{j}-\qpoint{0})
    \cdot\allocation(\qpoint{0},\qpoint{j}).
\end{equation}

For the analysis, we extend the sequence to
\(\qpoint{n+1}=\georatio\qpoint{n}\).  By the definition of 
\(\buyerintegralbound{j}\),
\begin{align*}
    \buyerintegralbound{j+1}
    &=
    \buyerintegralbound{j}
    +(\qpoint{j+1}-\qpoint{j})
    \cdot\allocation(\qpoint{0},\qpoint{j})\\
    &=
    \buyerintegralbound{j}
    +(\georatio-1)\qpoint{j}
    \cdot\allocation(\qpoint{0},\qpoint{j}).
\end{align*}
Combining this identity with Equation~\eqref{eq:geometric-row-wbb} gives
\begin{align}
\label{eq:geometric-row-growth}
    \frac{\buyerintegralbound{j+1}}{\qpoint{j+1}}
    &\ge
    \frac{\buyerintegralbound{j}}{\georatio\qpoint{j}}
    +
    \frac{\georatio-1}{\georatio}
    \frac{\buyerintegralbound{j}+\sellerintegralbound{j}}
         {\qpoint{j}-\qpoint{0}}
    \notag\\
    &\ge
    \frac{\buyerintegralbound{j}}{\qpoint{j}}
    +
    \frac{\georatio-1}{\georatio}
    \frac{\sellerintegralbound{j}}
         {\qpoint{j}-\qpoint{0}}.
\end{align}
For the second inequality, we used
\(\buyerintegralbound{j}/(\qpoint{j}-\qpoint{0})
\ge \buyerintegralbound{j}/\qpoint{j}\).

We now verify the claimed upper bound on the final ratio.  Since every
allocation probability is at most one,
\begin{align}
\label{eq:buyer-integral-final-bound}
    \buyerintegralbound{n+1}
    &\le
    \sum_{r=1}^{n}(\qpoint{r+1}-\qpoint{r})
    =
    \qpoint{n+1}-\qpoint{1}
    <
    \qpoint{n+1}.
\end{align}
Summing Equation~\eqref{eq:geometric-row-growth} over
\(j=2,\ldots,n\) gives
\begin{align*}
    \sum_{j=2}^{n}
    \frac{\buyerintegralbound{j+1}}{\qpoint{j+1}}
    &\ge
    \sum_{j=2}^{n}
    \frac{\buyerintegralbound{j}}{\qpoint{j}}
    +
    \frac{\georatio-1}{\georatio}
    \sum_{j=2}^{n}
    \frac{\sellerintegralbound{j}}
         {\qpoint{j}-\qpoint{0}}.
\end{align*}
Canceling the common terms gives
\begin{align*}
    \frac{\buyerintegralbound{n+1}}{\qpoint{n+1}}
    &\ge
    \frac{\buyerintegralbound{2}}{\qpoint{2}}
    +
    \frac{\georatio-1}{\georatio}
    \sum_{j=2}^{n}
    \frac{\sellerintegralbound{j}}
         {\qpoint{j}-\qpoint{0}}\\
    &\ge
    \frac{\georatio-1}{\georatio}
    \sum_{j=2}^{n}
    \frac{\sellerintegralbound{j}}
         {\qpoint{j}-\qpoint{0}},
\end{align*}
where the last inequality uses \(\buyerintegralbound{2}\ge0\).  Together
with Equation~\eqref{eq:buyer-integral-final-bound}, this yields
\begin{equation}
\label{eq:geometric-total-charge}
    1
    >
    \frac{\buyerintegralbound{n+1}}{\qpoint{n+1}}
    \ge
    \frac{\georatio-1}{\georatio}
    \sum_{j=2}^{n}
    \frac{\sellerintegralbound{j}}
         {\qpoint{j}-\qpoint{0}}.
\end{equation}

The last term in \(\sellerintegralbound{j}\) is
\((\qpoint{j-1}-\qpoint{j-2})
\cdot\allocation(\qpoint{j-1},\qpoint{j})\).  Since
\(\qpoint{j}=\qpoint{0}\georatio^j\), we have
\begin{align*}
    \frac{\sellerintegralbound{j}}
         {\qpoint{j}-\qpoint{0}}
    &\ge
    \frac{
        (\qpoint{j-1}-\qpoint{j-2})
        \cdot\allocation(\qpoint{j-1},\qpoint{j})
    }{\qpoint{j}}\\
    &=
    \frac{\georatio-1}{\georatio^2}
    \cdot\allocation(\qpoint{j-1},\qpoint{j}).
\end{align*}
Substituting this bound into Equation~\eqref{eq:geometric-total-charge} gives
\[
    \sum_{j=2}^{n}\allocation(\qpoint{j-1},\qpoint{j})
    <
    \frac{\georatio^3}{(\georatio-1)^2}
    \le
    4\georatio,
\]
where the last inequality uses \(\georatio\ge2\).  Finally,
\(\allocation(\qpoint{0},\qpoint{1})\le1\le\georatio\), and therefore
\[
    \sum_{i=1}^{n}
    \allocation(\qpoint{i-1},\qpoint{i})
    \le5\georatio.
\]
\end{proof}

\subsection{Proof of Lemma~\ref{lem:lb-corr-first-best}}
\label{app:proof-lb-corr-first-best}

\begin{proof}[Proof of Lemma~\ref{lem:lb-corr-first-best}]
Let $\georatio$, the grid $\qpoint{0},\ldots,\qpoint{n}$, and $\distjoint$ be as in the proof of Theorem~\ref{thm:wbb-tie-no-constant-correlated-welfare}.  At every profile $(\qpoint{i-1},\qpoint{i})$ in the support, the buyer's value is larger than the seller's value, so the first-best trades and its social welfare is $\qpoint{i}$.  Hence,
\[
    \fbwelfare(\distjoint)
    =\sum_{i=1}^{n}\profileprob{i}\qpoint{i}
    =\sum_{i=1}^{n}\frac{1}{\normalizer}
    =\frac{n}{\normalizer}.
\]
\end{proof}

\subsection{From Traders' Surplus to Traders' Utility}
\label{app:ts-to-tu}


\begin{observation}\label{obs:ts-to-tu}
Let $\distjoint$ be a distribution over the buyer's and the seller's values such that the seller's expected value is bounded, and let $\mechanism$ and $\mechanism'$ be two mechanisms with finite objectives.  Then
\begin{align*}
    \sgft_{\mechanism}(\distjoint)>\sgft_{\mechanism'}(\distjoint)
    &\iff
    \swelfare_{\mechanism}(\distjoint)>\swelfare_{\mechanism'}(\distjoint),\\
    \gft_{\mechanism}(\distjoint)>\gft_{\mechanism'}(\distjoint)
    &\iff
    \welfare_{\mechanism}(\distjoint)>\welfare_{\mechanism'}(\distjoint).
\end{align*}
Moreover, for every $\alpha\ge1$,
\begin{align*}
    \swelfare_{\mechanism}(\distjoint)\ge\alpha\cdot\swelfare_{\mechanism'}(\distjoint)
    &\implies
    \sgft_{\mechanism}(\distjoint)\ge\alpha\cdot\sgft_{\mechanism'}(\distjoint),\\
    \welfare_{\mechanism}(\distjoint)\ge\alpha\cdot\welfare_{\mechanism'}(\distjoint)
    &\implies
    \gft_{\mechanism}(\distjoint)\ge\alpha\cdot\gft_{\mechanism'}(\distjoint).
\end{align*}
\end{observation}

\begin{proof}
By Equations~\eqref{equ:def-gft} and~\eqref{eq:residual-surplus-objective}, for every mechanism $\mechanism$,
\begin{align*}
    \swelfare_{\mechanism}(\distjoint)=\sgft_{\mechanism}(\distjoint)+\Exp_{(\vs,\vb)\sim\distjoint}[\vs],
    \qquad
    \welfare_{\mechanism}(\distjoint)=\gft_{\mechanism}(\distjoint)+\Exp_{(\vs,\vb)\sim\distjoint}[\vs].
\end{align*}
The term $\Exp_{(\vs,\vb)\sim\distjoint}[\vs]$ is non-negative and does not depend on the mechanism, so the first part follows.  For the second part, if $\swelfare_{\mechanism}(\distjoint)\ge\alpha\cdot\swelfare_{\mechanism'}(\distjoint)$, then $\sgft_{\mechanism}(\distjoint)\ge\alpha\cdot\swelfare_{\mechanism'}(\distjoint)-\Exp[\vs]=\alpha\cdot\sgft_{\mechanism'}(\distjoint)+(\alpha-1)\Exp[\vs]\ge\alpha\cdot\sgft_{\mechanism'}(\distjoint)$, and the same holds for social welfare and gains-from-trade.  Since this holds for every mechanism, for every class $\mathcal{C}$ of mechanisms,
\begin{align*}
    \sup_{\mechanism\in\mathcal{C}}\swelfare_{\mechanism}(\distjoint)&=\sup_{\mechanism\in\mathcal{C}}\sgft_{\mechanism}(\distjoint)+\Exp_{(\vs,\vb)\sim\distjoint}[\vs],\\
    \sup_{\mechanism\in\mathcal{C}}\welfare_{\mechanism}(\distjoint)&=\sup_{\mechanism\in\mathcal{C}}\gft_{\mechanism}(\distjoint)+\Exp_{(\vs,\vb)\sim\distjoint}[\vs],
\end{align*}
so the same statements hold between the best mechanisms in two classes, such as the best WBB and SBB mechanisms.
\end{proof}

\end{document}